\documentclass[a4paper,11pt]{article}
\usepackage[skins]{tcolorbox}
\usepackage{mathtools,slashed}
\usepackage[T1]{fontenc}
\usepackage{slashed,verbatim,subfigure}
\usepackage[numbers,sort&compress]{natbib}
\usepackage{amsmath}
\usepackage{mathrsfs}
\usepackage{amsbsy}
\usepackage{amssymb}
\usepackage{appendix}
\usepackage{graphicx}
\usepackage[final]{pdfpages}
\usepackage{dcolumn}
\usepackage{bm}
\usepackage{cancel}
\usepackage[numbers,sort&compress]{natbib}
\usepackage[colorlinks=true,linktocpage=true,
linkcolor=blue,citecolor=blue]{hyperref}
\usepackage{url}
\definecolor{db5}{cmyk}{0.5,0.5,0,0.5}
\definecolor{mauve}{cmyk}{0.3,0.7,0.1,0.3}
\definecolor{palemauve}{cmyk}{0.3,0.7,0.1,0.0}
\definecolor{pb}{cmyk}{0.4,0.1,0,0.1}
\definecolor{pgreen}{cmyk}{0.4,0.0,0.3,0.0}
\definecolor{pink}{cmyk}{0.0,0.5,0.3,0.0}
\newcommand{\bs}{\begin{slide}}
\newcommand{\es}{\end{slide}}
\newcommand{\tcb}{\textcolor {blue}}
\usepackage[a4paper]{geometry}
\catcode`@11
\def\seceqaa{\@addtoreset{equation}{section}
	\def\theequation{A\arabic{equation}}}
\def\seceqbb{\@addtoreset{equation}{section}
	\def\theequation{B\arabic{equation}}}
\def\seceqcc{\@addtoreset{equation}{section}
	\def\theequation{C\arabic{equation}}}
\def\seceqdd{\@addtoreset{equation}{section}
	\def\theequation{D\arabic{equation}}}
\def\seceqee{\@addtoreset{equation}{section}
	\def\theequation{E\arabic{equation}}}
\def\seceqff{\@addtoreset{equation}{section}
	\def\theequation{F\arabic{equation}}}	
\def\seceqgg{\@addtoreset{equation}{section}
	\def\theequation{G\arabic{equation}}}
\def\seceqhh{\@addtoreset{equation}{section}
	\def\theequation{H\arabic{equation}}}
\def\seceqii{\@addtoreset{equation}{section}
	\def\theequation{H\arabic{equation}}}	
\def\seceqjj{\@addtoreset{equation}{section}
	\def\theequation{H\arabic{equation}}}
\catcode`@11
\usepackage{authblk}
\newcommand{\be}{\begin{eqnarray}}
\newcommand{\ee}{\end{eqnarray}}
\begin{document}
\large
\title{Charge Transport in Magnetized Holographic $\mathcal{M}$-QGP}
\author{Shivam Singh Kushwah\footnote{email-shivams\_kushwah@ph.iitr.ac.in}~\\
Department of Physics,\\
Indian Institute of Technology Roorkee, Roorkee 247667, Uttarakhand, India}
\date{}
\maketitle
\begin{abstract}
We investigate DC transport in a top-down construction of thermal QGP-like theories using a holographic $\cal M$-theoretic background, incorporating quartic curvature corrections. The DC and Hall conductivities are computed from the Dirac–Born–Infeld (DBI) action of the corresponding type-IIA probe $D6$ flavor branes via the reality condition method proposed in\cite{OBannon:2007cex,Karch:2007pd}. We further analyze pair-production contributions in the presence and absence of an external magnetic field and work out regimes where thermal pair production dominates over charge carrier transport. These findings extend earlier AdS/CFT results to non-conformal, higher-derivative settings relevant to thermal QGP-like theories.
\end{abstract}
\newpage
\tableofcontents

\section{Introduction}
\label{Introduction}

The theory of strong interaction, known as quantum chromodynamics (QCD), exhibits several non-perturbative effects that are still challenging to examine with conventional theoretical methods. In particular, perturbative approaches are inapplicable in the strong-coupling limit; hence, an alternative method is required to handle such strongly coupled systems. One such method is the AdS/CFT correspondence \cite{Maldacena:1997re}, which connects some strongly coupled quantum field theory to a classically treatable one-dimensional higher gravitational system. By analysing related gravitational configurations, characteristics of highly interacting gauge theories can be investigated within this holographic framework. The phase structure of QCD itself is nontrivial: at low temperatures, it resides in a restricted phase with color-neutral hadrons, such as mesons and baryons, as its physical excitations. The system enters a deconfined phase with quark and gluon degrees of freedom as the temperature rises above a threshold value. These two regimes are linked to different gravitational backgrounds in holographic descriptions. The deconfined phase is associated with an AdS black hole spacetime, while the confined phase is usually represented by a thermal Anti–de Sitter (AdS) geometry. Therefore, studying the quark-gluon plasma characteristics in this duality framework is essentially equivalent to studying black hole physics in the gravitational theory. The fact that the Hawking–Page phase transition between hot AdS space and an AdS black hole on the gravity side is translated to the confinement–deconfinement transition in the gauge theory is a crucial finding establishing this correspondence\cite{Witten:1998zw}.

Transport phenomena play a major role in understanding the effective, long-wavelength behaviour of interacting many-body systems. In this sense, hydrodynamics, the effective low-energy description, provides a broadly applicable account of both near-equilibrium and significantly out-of-equilibrium dynamics. It is primarily utilised to explore a large class of transport coefficients. These transport coefficients capture significant aspects of the underlying microscopic theory and serve as an essential bridge between theoretical predictions and empirical facts.
Experimental findings from relativistic heavy-ion collisions (RHIC) reveal that the quark–gluon plasma exhibits fluid-like behavior with extremely low dissipation. In particular, the ratio of shear viscosity to entropy density is found to be unusually small, with phenomenological estimates placing it near $\frac{\eta}{s} \approx \frac{1}{4\pi}$\cite{Kovtun:2004de,Policastro:2001yc}. This observation has been investigated using a variety of theoretical approaches, including lattice QCD \cite{Meyer:2007ic,Astrakhantsev:2017nrs}, kinetic theory \cite{Arnold:2003zc}, as well as M-theoretic inspired quark–gluon plasma, where the ratios $\frac{\eta}{s}, \frac{\zeta}{s}$, and $\frac{\zeta}{\eta}$ have been analyzed \cite{Bulk-Viscosity-McGill-IIT-Roorkee} with the inclusion of higher-derivative corrections \cite{Kushwah:2024ngr}. Apart from there there are many other transport quantities which are of wide importance such as bulk viscosity ($\zeta$) more importantly bulk-viscosity-to-entropy-density-ratio $(\frac{\zeta}{s})$ (explored in \cite{Meyer:2008sn}), bulk viscosity-to-shear-viscosity ratio \cite{Bulk-Viscosity-McGill-IIT-Roorkee,Bluhm:2011xu,Kadam:2014cua,Yaresko:2014fia} for upto HD corrected lattice compatible ratio, see \cite{Kushwah:2024ngr}, thermal/AC/DC conductivity \cite{OBannon:2007cex,Karch:2007pd,Arnold:2003zc,Gupta:2003zh,Amato:2013naa}etc play crucial role in deciphering the nature and dynamics of the strongly coupled quark-gluon-plasma.

This highlights the central role of transport coefficients as key observables in the characterization of relativistic fluids. In this work, we focus on charge transport phenomena, for which the relevant transport coefficients are the electric and Hall conductivities. We compute the DC conductivity using a regulated form of Ohm’s law and subsequently extend this analysis to include an external magnetic field to investigate the Hall response of the top-down $\mathcal{M}$-QGP at intermediate coupling. The inclusion of a magnetic field is well motivated, as extremely strong magnetic fields are generated during heavy-ion collision experiments, with magnitudes reaching $\mathcal{O}(10^{18}\text{--}10^{19}G)$\cite{Skokov:2009qp}. The computation of DC and Hall conductivities has vast phenomenological relevance, with applications ranging from the electromagnetic response of the quark-gluon-plasma to charge diffusion and dilepton/photon production in the quark–gluon plasma; see \cite{Caron-Huot:2006pee,Kushwah:2025ymb} for more details. Hence, understanding the charge transport or conductivity in the presence of electromagnetic fields is a realistic description of QGP dynamics. Earlier explorations have addressed the charge transport in holographic setups designed to capture key features of QGP, with particular focus on DC conductivity and Hall response in both conformal and non-conformal plasma regimes\cite{OBannon:2007cex,Karch:2007pd,Iqbal:2008by}. However, a comprehensive and self-consistent analysis of DC and Hall conductivities within a top-down holographic framework that incorporates higher-derivative (HD) corrections is still absent. An important ingredient in such a study is the separate contributions originating from finite charge density and those induced by pair-production mechanisms under varying electromagnetic-field conditions, enabling a more precise understanding of transport behaviour in QGP -like matter. In this work, we aim to bridge this gap by systematically investigating charge transport in the $\mathcal{M}$-QGP at intermediate coupling with higher-derivative corrections, thereby advancing and complementing existing holographic analyses.

We use the M-theoretic uplift of the type IIA background generated by applying triple T-duality on the original type IIB setup developed by Dasgupta et al \cite{metrics,OR4} within the MQGP framework to calculate the conductivity in a domain relevant for quark–gluon plasma at intermediate coupling. We can work at finite string coupling $g_s$ and finite (though big) $N$ by using an $\mathcal{M}$-theoretic description, which is especially advantageous because it allows us to go beyond the tight large-$N$ constraint. Transport parameters like electrical conductivity may be significantly affected by the effects of quantum and stringy corrections, which are usually suppressed at conventional limits; see Sec.~(\ref{MQGP-review}) for more details. The uplift to eleven dimensions effectively incorporates these finite-$g_s$ effects into a geometric framework, within which charge transport can be consistently analyzed through probe brane embeddings and their world-volume fluctuations.  Furthermore, the ultraviolet geometry remains asymptotically AdS, ensuring the validity of standard holographic prescriptions for extracting boundary transport coefficients. Therefore, the M-theory construction provides a well-controlled and physically meaningful framework for studying conductivity in a strongly coupled plasma at finite coupling.

To compute charge transport in the $\mathcal{M}$-QGP framework, we use the proposal of O'Bannon et al. \cite{OBannon:2007cex,Karch:2007pd}. This approach is based on enforcing the reality condition on the Dirac–Born–Infeld (DBI) action, which ensures that the DBI action stays real by requiring the constraint equations that arise under the square root to vanish simultaneously at the shifted horizon, called the effective horizon. The conductivity tensor is then derived using Ohm's equation, $J_i=\sigma_{ij}E_j$. Transport coefficients are then calculated by evaluating the pertinent quantities at this point. The DC conductivity in this framework is written as 
\begin{equation}
\sigma_{\rm DC}=\sqrt{\sigma_{tpp}^{2}+\sigma_{\rm density}^{2}} , .
\end{equation}

A key feature of the O'Bannon proposal is that the charge-density contribution, $\sigma_{\rm density}$, vanishes for a neutral plasma, whereas the DC conductivity remains finite even in the absence of a net charge density. O'Bannon interpreted this density-independent contribution as arising from thermally generated charge-conjugate pairs in the finite-temperature plasma. This thermal-pair-production interpretation of $\sigma_{\rm tpp}$ has subsequently been corroborated in a variety of top-down and bottom-up holographic constructions, suggesting that it is a generic consequence of the DBI dynamics governing probe D-branes ( e.g., see.~\cite{OBannon:2007cex,Karch:2007pd,Ali-Akbari:2010xwz,Kim:2011zd,Lee:2010uy}). Motivated by this non-trivial structure of non-linear charge transport, we used this approach to study both DC and Hall conductivities in our holographic $\mathcal{M}$QGP model. In particular, we examine the impact of higher-derivative corrections on the distinct components of the DC conductivity as well as on the Hall response within a top-down holographic QGP setting, focusing on the $\mathcal{M}$-theory description of the quark–gluon plasma at intermediate coupling. For a comprehensive overview of the $\mathcal{M}$-QGP framework, see Sec.~(\ref{MQGP-review}).

The structure of the paper is as follows. In Sec .~(\ref{MQGP-review}), we outline the holographic framework and describe the finite-temperature background geometry relevant to the $\mathcal{M}$-QGP. In Sec.~(\ref{DCB0}), we evaluate the DC electrical conductivity in the absence of an external magnetic field using a regulated version of Ohm’s law. Sec.~(\ref{DCB}) focuses on charge transport in the presence of a magnetic field, where both longitudinal and Hall conductivities are studied in the regimes of weak and strong magnetic fields. In Sec.~(\ref{NDC}), we carry out a comprehensive numerical analysis of DC and Hall conductivities, considering cases with and without a magnetic field, and further examine the separate contributions from the charge-density and pair-production components as functions of temperature in the limit of a small electric field and an arbitrary magnetic field. Finally, Sec.~(\ref{Results}) provides a summary of our main results along with a discussion of potential future directions.

\section{$\mathcal{M}-$QGP Setup: A Short Review}
\label{MQGP-review}
This section details the uplift to M-theory originating from a Type IIA configuration. This specific framework is derived by applying the Strominger-Yau-Zaslow (SYZ) mirror symmetry prescription—executed through a sequence of three T-dualities—to the foundational Type IIB model established by Dasgupta and collaborators \cite{metrics}.

In this framework, the underlying Type IIB configuration is defined by $N$ spacetime-filling $D3$-branes situated at the apex of a warped resolved conifold. To this geometry, we add $M$ fractional $D5$-branes which wrap a vanishing $S^2$ at the North Pole of the resolved, squashed $S^2$ (characterized by the resolution parameter $a$); meanwhile, UV consistency is maintained by placing an equivalent number of anti-$D5$-branes at the South Pole. 

Flavor degrees of freedom are integrated into the model via $N_f$ $D7$-branes, which are embedded according to the Ouyang prescription. These branes wrap a vanishing squashed $S^3$ and extend into the IR up to a radial limit dictated by $|\mu_{\text{Ouyang}}|^{2/3}$, where $|\mu_{\text{Ouyang}}|$ denotes the modulus of the embedding parameter for the flavor $D7$-branes\cite{MQGP, ACMS}. We also include a matching set of $\overline{D7}$-branes in the UV to preserve conformality\cite{MQGP, ACMS}. This arrangement gives rise to a UV $SU(N_f) \times SU(N_f)$ symmetry which, in the IR, undergoes a breakdown to the diagonal $SU(N_f)$ subgroup—a process identified with chiral symmetry breaking\cite{MQGP, ACMS}.

From the bulk perspective \cite{metrics}, IR confinement in the dual gravity theory is driven by the deformation of the vanishing squashed $S^3$ of the conifold. Thermal effects are introduced by adopting thermal backgrounds for the regime $T < T_c$ and black-hole geometries for $T > T_c$. Furthermore, ensuring a finite separation between the $D5$ and $\overline{D5}$ branes necessitates an additional $S^2$ resolution of the conifold, which is also described by the parameter $a$. Once backreaction is accounted for, the resulting modifications to the warp factor and fluxes lead to a warped, resolved, and deformed conifold geometry in the large $N$ limit\cite{MQGP, ACMS}. This specific Type IIB dual of \cite{metrics} is particularly advantageous for exploring the intermediate-$N$ MQGP limit, as discussed in \cite{MQGP, ACMS}.

\begin{eqnarray}
\label{MQGP_limit}
& & g_s\equiv\frac{1}{{\cal O}(1)},\quad M, N_f \equiv {\cal O}(1),\quad
 N>1:\  \frac{\left(g_s M^2\right)^{m_1}\left(g_s N_f\right)^{m_2}}{N}<1,\ m_{1,2}\in\mathbb{Z}^+\cup\{0\},
\end{eqnarray}
 In the infrared, after the Seiberg-like duality cascade, one has $N_c = M$. From \cite{ACMS}, for the values of $g_s$, $M$, and $N_f$ listed in the table(\ref{MQGP_limit}).
\begin{table}[h]
\label{Parameters-real-QCD}
\begin{center}
\begin{tabular}{|c|c|c|c|} \hline
S. No. & Parameterc & Value chosen consistent with (\ref{MQGP_limit}) & Physics reason \\ \hline
1. & $g_s$ & 0.1 & QCD fine structure constant \\ 
&&& at EW scale \\ \hline
2. & $M$ & 3 & Number of colors after a \\ 
&&& Seiberg-like duality cascade \\
&&& to match real QCD \\ \hline
3. & $N_f$ & 2 or 3 & Number of light quarks in real QCD \\ \hline
\end{tabular}
\end{center}
\caption{QCD-Motivated values of $g_s, M, N_f$}
\end{table}
For the flavor sector, we study embeddings of D7-branes in the small Ouyang modulus limit 
 ($|\mu_{\rm Ouyang}|\ll 1$ as defined in \eqref{Ouyang-definition}), taking take $N_f=2$ or $3$, to model the light-quark sector \cite{Vikas+Gopal+Aalok}.

The $\mathcal{M}$-theory uplift of \cite{metrics} incorporates ${\cal O}(R^4)$  corrections in 11-dim. supergravity to probe the intermediate coupling regime of thermal QCD, as discussed in \cite{OR4}. The construction begins by obtaining the Type IIA Strominger–Yau–Zaslow (SYZ) mirror of the Type IIB background via triple T-duality along a local sLag 
 $T^3(x,y,z)$, where $(x,y,z)$ are actually the toroidal equivalents of $(\phi_1,\phi_2,\psi)$, in the large complex-structure limit \cite{MQGP,NPB}. Under T-duality, all color and flavor branes of type IIB are mapped to color and flavor $D6$-branes. The resulting background is then uplifted to M-theory, yielding a genuine $G_2$-structure consistent with the equations of motion, as in \cite{SYZ-free-delocalization}. The uplift is shown to remain valid even after removing the intermediate delocalization approximation. The analysis is further simplified by working near small $\theta_{1,2}$ and in the vanishing Ouyang modulus limit, effectively focusing on the light-quark sector (first-generation quarks[+s quark]) from 
 \begin{equation}
\label{Ouyang-definition}
\left(9 a^2 r^4 + r^6\right)^{1/4}e^{\frac{i}{2}(\psi - \phi_1-\phi_2)}\sin\frac{\theta_1}{2} \sin\frac{\theta_2}{2}=\mu_{\rm Ouyang}.
\end{equation}
This clearly indicates that one must operate in the regime where $\theta_{1,2}$ takes extremely small values. For instance, we work in the vicinity of
\begin{eqnarray}
\label{alpha_theta_12}
& & (\theta_1, \theta_2) = \left(\frac{\alpha_{\theta_1}}{N^{1/5}}, \frac{\alpha_{\theta_2}}{N^{3/10}}\right),\ \ \ \ \ \ \ \alpha_{\theta_{1,2}}\equiv{\cal O}(1).
\end{eqnarray}
Additionally, the somewhat differing powers of $N$ for the delocalized $\theta_{1,2}$ serve as a reminder that the resolved $S^2(\theta_2,\phi_2)$ and vanishing $S^2(\theta_1,\phi_1)$ in the pair of squashed $S^2$s do not sit at the ``same ground''.  From the on-shell action's perspective, by substituting $N^{1/5}\sin\theta_1$ or $N^{3/10}\sin\theta_2$ for the ${\cal O}(1)$ delocalization parameters $\alpha_{\theta_{1,2}}$, respectively, the outcomes up to ${\cal O}(\frac{1}{N})$ become independent of the delocalization (as shown in \cite{OR4}). Afterwards, one can select an alternative delocalization by substituting $\sin\theta_{1,2}$ with
\begin{eqnarray}
\label{alpha_theta_12_prime} 
& & \left(\frac{\tilde{\alpha}_{\theta_1}}{N^{\gamma_{\theta_1}}}, \frac{\tilde{\alpha}_{\theta_2}}{N^{\gamma_{\theta_2}}}\right), \gamma_{\theta_1}\neq\frac{1}{5}, \gamma_{\theta_2}\neq\frac{3}{10},
\ \ \ \ \ \ \ \ \tilde{\alpha}_{\theta_{1,2}}\equiv{\cal O}(1).
\end{eqnarray}
We define the ${\cal M}$-theory uplift (finite-but-large-$N$/intermediate coupling) metric of \cite{metrics} as follows \cite{MQGP}, \cite{OR4}, 
\begin{eqnarray}
\label{TypeIIA-from-M-theory-Witten-prescription-T>Tc}
\hskip -0.1in ds_{11}^2 & = & e^{-\frac{2\phi^{\rm IIA}}{3}}\Biggl[\frac{1}{\sqrt{h(r,\theta_{1,2})}}\left(-g(r) dt^2 + \left(dx^1\right)^2 +  \left(dx^2\right)^2 +\left(dx^3\right)^2 \right)
\nonumber\\
& & \hskip -0.1in+ \sqrt{h(r,\theta_{1,2})}\left(\frac{dr^2}{g(r)} + ds^2_{\rm IIA}(r,\theta_{1,2},\phi_{1,2},\psi)\right)
\Biggr] + e^{\frac{4\phi^{\rm IIA}}{3}}\left(dx^{11} + A_{\rm IIA}^{F_1^{\rm IIB} + F_3^{\rm IIB} + F_5^{\rm IIB}}\right)^2,
\end{eqnarray} 
where $A_{\rm IIA}^{F^{\rm IIB}_{i=1,3,5}}$ (type IIA RR 1-forms) are generated from type IIB  fluxes ($F_{1,3,5}^{\rm IIB}$) by the application of SYZ mirror to type IIB string dual \cite{metrics}, type IIA dilaton profile is $\phi^{\rm IIA}$, and $g(r) = 1 - \frac{r_h^4}{r^4}$. The thermal gravitational dual for low-temperature QCD, denoted as $T<T_c$, is defined as follows: 
\begin{eqnarray}
\label{TypeIIA-from-M-theory-Witten-prescription-T<Tc}
\hskip -0.1in ds_{11}^2 & = & e^{-\frac{2\phi^{\rm IIA}}{3}}\Biggl[\frac{1}{\sqrt{h(r,\theta_{1,2})}}\left(-dt^2 + \left(dx^1\right)^2 +  \left(dx^2\right)^2 + \tilde{g}(r)\left(dx^3\right)^2 \right)
\nonumber\\
& & \hskip -0.1in+ \sqrt{h(r,\theta_{1,2})}\left(\frac{dr^2}{\tilde{g}(r)} + ds^2_{\rm IIA}(r,\theta_{1,2},\phi_{1,2},\psi)\right)
\Biggr] + e^{\frac{4\phi^{\rm IIA}}{3}}\left(dx^{11} + A_{\rm IIA}^{F_1^{\rm IIB} + F_3^{\rm IIB} + F_5^{\rm IIB}}\right)^2,\nonumber\\
& & 
\end{eqnarray}
where $\tilde{g}(r) = 1 - \frac{r_0^4}{r^4}$. Observing that $t\rightarrow x^3,\ x^3\rightarrow t$ in (\ref{TypeIIA-from-M-theory-Witten-prescription-T>Tc}) followed by a double Wick rotation in the new $x^3, t$ coordinates yields (\ref{TypeIIA-from-M-theory-Witten-prescription-T<Tc}); the ten-dimensional warp factor is denoted by $h(r,\theta_{1,2})$ \cite{metrics, MQGP}. This may also be expressed as follows: $-g_{tt}^{\rm BH}(r_h\rightarrow r_0) = g_{x^3x^3}\ ^{\rm Thermal}(r_0),$ $ g_{x^3x^3}^{\rm BH}(r_h\rightarrow r_0) = -g_{tt}\ ^{\rm Themal}(r_0)$ in the outcomes of \cite{VA-Glueball-decay}, \cite{OR4} (Refer to \cite{Kruczenski et al-2003} regarding Euclidean/black $D4$-branes in type IIA). We are going to take the spatial component of the solitonic $M3$ brane [it, locally, might be seen as a homologous sum of $S^2_{\rm squashed}$ wrapping across a solitonic $M5$-brane \cite{DM-transport-2014}] in (\ref{TypeIIA-from-M-theory-Witten-prescription-T<Tc}), in which the extremely small $M_{\rm KK}$ is provided through $\frac{2r_0}{ L^2}\left[1 + {\cal O}\left(\frac{g_sM^2}{N}\right)\right]$ and their world volume provided by $\mathbb{R}^2(x^{1,2})\times S^1(x^3)$. The period of $S^1(x^3)$ is provided by an extremely big: $\frac{2\pi}{M_{\rm KK}}$, where $L = \left( 4\pi g_s N\right)^{\frac{1}{4}}$ and $r_0$ denote the extremely tiny IR cut-off that defines the thermal background (see also \cite{Armoni et al-2020}). Thus, 4D Physics is recovered by $\lim_{M_{\rm KK}\rightarrow0}\mathbb{R}^2(x^{1,2})\times S^1(x^3) = \mathbb{R}^3(x^{1,2,3})$. As for the thermal background corresponding to $T<T_c$, in (\ref{TypeIIA-from-M-theory-Witten-prescription-T<Tc}), $\tilde{g}(r)$ will be set to unity as the working metric.\par

{
The eleven-dimensional supergravity action used in \cite{OR4} that includes ${\cal O}(R^4)$ terms is the following:
\begin{eqnarray}
\label{D=11_O(l_p^6)}
& & \hskip -0.5in S = \frac{1}{2\kappa_{11}^2}\Biggl[\int_{M_{11}}\sqrt{g}R + \int_{\partial M_{11}}\sqrt{h}K -\frac{1}{2}\int_{M_{11}}\sqrt{g}G_4^2
-\frac{1}{6}\int_{M_{11}}C_3\wedge G_4\wedge G_4\nonumber\\
& & \hskip -0.5in  + \frac{\left(4\pi\kappa_{11}^2\right)^{\frac{2}{3}}}{{(2\pi)}^4 3^2.2^{13}}\Biggl(\int_{\cal{M}} d^{11}\!x \sqrt{g}\left(J_0-\frac{1}{2}E_8\right) + 3^2.2^{13}\int C_3 \wedge X_8 + \int t_8 t_8 G^2 R^3 + \cdot \cdot \Biggr)\Biggr] - {\cal S}^{\rm ct},\nonumber\\
\end{eqnarray}
where:
\begin{eqnarray}
\label{J0+E8-definitions}
& & \hskip -0.8inJ_0  =3\cdot 2^8 (R^{HMNK}R_{PMNQ}{R_H}^{RSP}{R^Q}_{RSK}+
{1\over 2} R^{HKMN}R_{PQMN}{R_H}^{RSP}{R^Q}_{RSK}),\nonumber\\
& & \hskip -0.8inE_8  ={ 1\over 3!} \epsilon^{ABCM_1 N_1 \dots M_4 N_4}
\epsilon_{ABCM_1' N_1' \dots M_4' N_4' }{R^{M_1'N_1'}}_{M_1 N_1} \dots
{R^{M_4' N_4'}}_{M_4 N_4},,\nonumber\\
& & \hskip -0.8in t_8t_8G^2R^3 = t_8^{M_1...M_8}t^8_{N_1....N_8}G_{M_1}\ ^{N_1 PQ}G_{M_2}\ ^{N_2}_{\ \ PQ}R_{M_3M_4}^{\ \ \ \ N_3N_4}R_{M_5M_6}^{\ \ \ \ N_5N_6}R_{M_7M_8}^{\ \ \ \ N_7N_8},
\nonumber\\
& & \hskip -0.8in X_8 = {1 \over 192} \left( {\rm tr}\ R^4 -
{1\over 4} ({\rm tr}\ R^2)^2\right),\nonumber\\
& & \hskip -0.8in\kappa_{11}^2 = \frac{(2\pi)^8 l_p^{9}}{2}.
\end{eqnarray}
$t_8$ symbol has the following structure \cite{OR4}:
{\footnotesize
\begin{eqnarray}
\label{t_8}
t_8^{N_1\dots N_8}   &=& \frac{1}{16} \big( -  2 \left(   g^{ N_1 N_3  }g^{  N_2  N_4  }g^{ N_5   N_7  }g^{ N_6 N_8  }
 + g^{ N_1 N_5  }g^{ N_2 N_6  }g^{ N_3   N_7  }g^{  N_4   N_8   }
 +  g^{ N_1 N_7  }g^{ N_2 N_8  }g^{ N_3   N_5  }g^{  N_4 N_6   }  \right) \nonumber \\
 & &  +
 8 \left(  g^{  N_2     N_3   }g^{ N_4    N_5  }g^{ N_6    N_7  }g^{ N_8   N_1   }
  +g^{  N_2     N_5   }g^{ N_6    N_3  }g^{ N_4    N_7  }g^{ N_8   N_1   }
  +   g^{  N_2     N_5   }g^{ N_6    N_7  }g^{ N_8    N_3  }g^{ N_4  N_1   }
\right) \nonumber \\
& &  - (N_1 \leftrightarrow  N_2) -( N_3 \leftrightarrow  N_4) - (N_5 \leftrightarrow  N_6) - (N_7 \leftrightarrow  N_8) \big),
\end{eqnarray}
}
where $g^{ N_i N_j}$ with ($i,j=1,2,..,8$) are in the inverse metric components. ${\cal S}^{\rm ct}$ denotes the counter terms required for holographic renormalization of (\ref{D=11_O(l_p^6)}).
The following are the equations of motion associated with the three-form potential $C$ and metric:
\begin{eqnarray}
\label{eoms}
& & {\rm EOM}_{\rm MN}:\ R_{MN} - \frac{1}{2}g_{MN}{\cal R} - \frac{1}{12}\left(G_{MPQR}G_N^{\ PQR} - \frac{g_{MN}}{8}G_{PQRS}G^{PQRS} \right)\nonumber\\
 & &  = - \beta\left[\frac{g_{MN}}{2}\left( J_0 - \frac{1}{2}E_8\right) + \frac{\delta}{\delta g^{MN}}\left( J_0 - \frac{1}{2}E_8\right)\right],\nonumber\\
& & d*G = \frac{1}{2} G\wedge G +3^22^{13} \left(2\pi\right)^{4}\beta X_8,\nonumber\\
& &
\end{eqnarray}
where \cite{Becker-sisters-O(R^4)}:
\begin{equation}
\label{beta-def}
\beta \equiv \frac{\left(2\pi^2\right)^{\frac{1}{3}}\left(\kappa_{11}^2\right)^{\frac{2}{3}}}{\left(2\pi\right)^43^22^{12}} \sim l_p^6.
\end{equation}
In (\ref{D=11_O(l_p^6)})/(\ref{eoms}), the eleven-dimensional Riemann curvature tensor, Ricci tensor, and Ricci scalar are denoted by the symbols $R_{MNPQ}, R_{MN}, {\cal R}$. The following was the ansatz constructed to solve (\ref{eoms}):
\begin{eqnarray}
\label{ansaetze}
& & \hskip -0.8ing_{MN} = g_{MN}^{\beta^0} +\beta g_{MN}^{\beta},\nonumber\\
& & \hskip -0.8inC_{MNP} = C^{(0)}_{MNP} + \beta C_{MNP}^\beta.
\end{eqnarray}
equations of motion corresponding to $C_{MNP}$ may be expressed symbolically as follows\footnote{{In (\ref{deltaC=0consistent}), $\epsilon_{11}\partial C^{\beta^0} \partial C^{\beta}$ is a schematic way of writing $\epsilon^{M_1...M_{11}}\partial_{[M_4}C_{M_5M_6M_7]}^{\beta^0}\partial_{[M_8}C_{M_9M_{10}M_{11}]}^{\beta}$ with $\epsilon_{11}$ denoting $\epsilon^{M_1...M_{11}}$.}}:
\begin{eqnarray}
\label{deltaC=0consistent}
& & \beta \partial\left(\sqrt{-g}\partial C^{\beta}\right) + \beta \partial\left[\left(\sqrt{-g}\right)^{\beta}\partial C^{\beta^0}\right] + \beta\epsilon_{11}\partial C^{\beta^0} \partial C^{\beta} = {\cal O}(\beta^2) \sim 0 [{\rm up\ to}\ {\cal O}(\beta)].
\nonumber\\
& & \end{eqnarray}
$C^{\beta}_{MNP}=0$ has been proven in \cite{OR4} to be a consistent truncation of ${\cal O}(\beta)$ corrections the ${\cal M}$-theory uplift of \cite{MQGP}, \cite{NPB} provided ${\cal C}_{zz} - 2 {\cal C}_{\theta_1z} = 0, |{\cal C}_{\theta_1x}|\ll1$ where $C_{MN}$ are the constants of integration appearing in the solutions to the equations of motion of $h_{MN}$ and the delocalized toroidal coordinates $T^3(x, y, z)$ corresponding to some $(r, \theta_1, \theta_2) = (\langle r\rangle, \langle\theta_1\rangle, \langle\theta_2\rangle)$ are defined as \cite{MQGP}: 
\begin{eqnarray}
\label{xyz-defs}
& & dx = \sqrt{\frac{1}{6} + {\cal O}\left(\frac{g_sM^2}{N}\right)}h^{\frac{1}{4}}\Bigl(\langle r\rangle, \langle \theta_{1,2}\rangle\Bigr)\langle r\rangle
\sin\langle\theta_{1}\rangle d\phi_1,\nonumber\\ 
& & dy = \sqrt{\frac{1}{6} + \frac{a^2}{r^2} + {\cal O}\left(\frac{g_sM^2}{N}\right)}h^{\frac{1}{4}}\Bigl(\langle r\rangle, \langle \theta_{1,2}\rangle\Bigr)
\langle r\rangle\sin\langle\theta_{2}\rangle d\phi_2,\nonumber\\  
& & dz =  \sqrt{\frac{1}{6} + {\cal O}\left(\frac{g_sM^2}{N}\right)}h^{\frac{1}{4}}\Bigl(\langle r\rangle, \langle \theta_{1,2}\rangle\Bigr)\langle r\rangle d\psi,
\end{eqnarray}
where the 10-D warp factor, which takes the following form
\begin{equation}
\label{h-def}
h(r,\theta_{1,2})
= \frac{L^4}{r^4}\left[1 + \frac{3g_sM_{\rm eff}^2}{N}\log r\left\{1 + \frac{3g_sN_f^{\rm eff}}{2\pi}\left(\log r + \frac{1}{2}\right) + \frac{g_sN_f^{\rm eff}}{4\pi}\log\left(\sin \frac{\theta_1}{2}\sin \frac{\theta_2}{2}\right)\right\}\right].
\end{equation}
 In light of this, ${\cal O}(R^4)$ corrections are applied exclusively to the metric and are defined as:
\begin{eqnarray}
\label{fMN-definitions}
\delta g_{MN} =g^{\beta}_{MN} = g_{MN}^{\beta^0} f_{MN}(r).
\end{eqnarray}
With ${\cal O}(R^4)$ corrections included, the ${\cal M}$ theory metric typically takes a particular form:
\begin{equation}
\label{fMN-def}
g_{MN} = g_{MN}^{\beta^0}\left(1+\beta f_{MN}(r)\right).
\end{equation}
where $f_{MN}(r)$ are given in \cite{OR4}. The metric components (\ref{fMN-def}) are worked out near the $\psi=2n\pi, \ n=0, 1, 2$-coordinate patches wherein $g_{rM}=0, M\neq r$ and the $M_5=(S^1 \times_w R^3) \times R_{>0}$ and the unwarped $\tilde{M}_6(S^1_{\cal M} \times_w T^{\rm NE})$ of $SU(3)$ structure wherein $S^1_{\cal M}$ is the ${\cal M}$-theory circle and $T^{\rm NE}$ is the non-Einsteinian deformation of $T^{1,1}$, decouple.

\section{DC Conductivity via Reality Condition of DBI Action\cite{OBannon:2007cex,Karch:2007pd}}
\label{DCB0}

In this section, we compute the components of the DC conductivity following the methodology developed by O’Bannon and collaborators \cite{OBannon:2007cex,Karch:2007pd}. Throughout this analysis, we restrict ourselves to the case of a vanishing magnetic field.
\\
Consider the DBI action for $N_f$ flavor $D6$-branes, 
\begin{equation}\label{D6DBI}
S_{D6}=-T_{D6}N_f\int d^{7}x~ e^{-\phi_{IIA}}\sqrt{-\det{\{i^*(g+B)+F\}}},
\end{equation}
with $2\pi\alpha^{\prime}=1$, $i:\Sigma_{D6}\hookrightarrow M_{10}$ defines the embedding of the $D6$-brane world volume in the ten-dimensional type IIA gravity dual involving a non-K\"ahler resolved conifold, and $t,x^1,x^2,x^3, Z,\theta_2,\tilde{y}$ are the coordinates of the $D6$-branes' worldvolume directions with $t,x^1 ,x^2,x^3 $ are standard Minkowski coordinates. In this case, the radial coordinate is redefined as $r=r_{h}e^{Z}$, where $r$ is the radial coordinate and $\tilde{y}$ and $\theta_{2}$ are the angular coordinates. $F_{\mu\nu}=\partial_{\mu} A_{\nu}-\partial_{\nu} A_{\mu}$ is the $U(1)$ gauge field strength, and $\phi_{IIA}$ is the type-IIA dilaton (triple T-dual of the type-IIB dilaton). The DBI action would be provided by
\cite{zeta_IITR+McGill}:
{\footnotesize
\begin{eqnarray}
\label{DBI-i}
& & \hskip -0.4in S_{\rm DBI} = -N_fT_{\rm D_6}\int_{\Sigma^{1,6}}\prod_{\mu=0}^3dr d\tilde{y}d\theta_2e^{-\phi^{\rm IIA}}\sqrt{i^*(g + B)^{\rm IIA} + F} \nonumber\\
& & \hskip -0.4in = -N_fT_{\rm D_6}\int_{\Sigma^{1,6}\times S^3}\prod_{\mu=0}^3dx^\mu dr d\theta_2 d\theta_1d\tilde{x}d\tilde{z}\delta\left(N^{1/5}\sin\theta_1 - \sqrt{2}N^{1/5}\sin\theta_2\right)\delta\left(\tilde{x}-c_{\tilde{x}}\right)\delta\left(\tilde{z}-c_{\tilde{z}}\right)\frac{\left(g_sN\right)^{1/4}\csc^2\theta_2d\psi}{6\sqrt{2}}\nonumber\\
& & \hskip 1.2in \times e^{-\phi^{\rm IIA}}\sqrt{i^*(g + B)^{\rm IIA} + F}\nonumber\\
& & \hskip -0.4in = - N_fT_{\rm D_6}\int_{\left(S^1\times\mathbb{R}^3\right)\times \mathbb{R}_{>0}\times S^2}\prod_{\mu=0}^3dx^\mu dr\mathcal{I}(\theta_2,\psi)e^{-\phi^{\rm IIA}}\sqrt{i^*(g + B)^{\rm IIA} + F},
\end{eqnarray}
}
with $\Theta$ being the Heaviside step function.
Define,
\begin{equation}
    \mathcal{I}(\theta_{2},\psi)=\frac{\left(g_sN\right)^{1/4}\csc^2\theta_2d\psi}{6\sqrt{2}N^{1/5}\sqrt{\cos(2\theta_2)}}d\theta_2 \Theta(\frac{\pi}{4}-\theta_2)\Theta(\theta_2)
\end{equation}
this integral diverges at $\theta_{2}=0$, and $\theta=\frac{\pi}{4}$, hence needs the angular regularization. After angular regularization, one obtains the following regularized multiplicative factor ($\mathcal{I}$) in the DBI action(\ref{DBI-i}).

\noindent

Incorporating the higher-derivative (HD) corrections to the background geometry and the worldvolume gauge field introduces the perturbative parameter $\beta$, which controls the strength of the higher-curvature contributions. At $\mathcal{O}(\beta)$, only the background metric and the worldvolume $U(1)$ gauge field receive nontrivial corrections, while the HD corrections to the remaining background fields are parametrically suppressed and do not contribute at this order. The justification for this approximation is presented in Appendix~C. We therefore expand the probe D6-brane $U(1)$ gauge field to linear order in $\beta$. The effect of a constant external electric field $E_{x^1}$ is incorporated through the spatial gauge field component along the $x^1$ direction, and we adopt the following ansatz:

\begin{equation}
\label{B0gaugefield}
    A_t = A_t^{(0)}(r) + \beta\, A_t^{(1)}(r), \qquad 
    A_{x^1} = -E_{x^{1}} t + A_{x^1}^{(0)}(r) + \beta\, A_{x^1}^{(1)}(r),
\end{equation}
where the superscripts $(0)$ and $(1)$ denote the leading-order and first-order corrections in the HD expansion, respectively.

\medskip

\noindent
Parallelly, the background metric $G_{ab}=g_{ab}+B^{NS-NS}_{ab}, where\,\, a,b\in (D6-brane\, directions)$, can be considered perturbatively up to higher-derivative (HD) order, and can be written in a compactified form that explicitly exhibits its dependence on the radial and angular coordinates. The corresponding corrected metric components are then given by\footnote{Throughout this work, we adopt the notation $( f_{11}^{(0)} = f_{x^{1}x^{1}}^{\beta^{0}} )$, and $( f_{tt}^{(0)} = \lvert f_{tt}^{\beta^{0}} \rvert )$. The superscripts in parentheses, $(0)$, and $(1)$ are used to denote contributions at order $(\mathcal{O}(\beta^{0}))$, and $(\mathcal{O}(\beta))$, respectively.}:
\begin{equation}
\label{compact-metric}
\begin{aligned}
G_{tt} &= f_{tt}^{(0)}(r) + \beta\, f_{tt}^{(1)}(r)\, f_{tt}^{(1)}(\theta_2), \qquad
G_{11} = f_{11}^{(0)}(r) + \beta\, f_{11}^{(1)}(r)\, f_{11}^{(1)}(\theta_2), \\
G_{rr} &= f_{rr}^{(0)}(r) + \beta\, f_{rr}^{(1)}(r), \qquad
G_{\tilde{Y}\tilde{Y}} = f_{\tilde{Y}\tilde{Y}}(\theta_2), \\
G_{\theta_2\theta_2} &= f_{\theta_2\theta_2}(\theta_2), \qquad
G_{\theta_2\tilde{Y}} = -G_{\tilde{Y}\theta_2} = f_{\theta_2\tilde{Y}}(\theta_2).
\end{aligned}
\end{equation}
where the first-order higher-derivative corrections are encoded by the functions $f_{ab}^{(1)}(r)$ and the leading-order background geometry is described by the functions $f_{ab}^{(0)}(r)$. The dependency on the internal coordinates of the compact manifold is captured by the angular functions $f_{ab}(\theta_2)$. Physical observables like the charge density and transport coefficients can be systematically expanded in powers of $\beta$ thanks to this perturbative framework, which also guarantees that the higher-derivative corrections stay under control.

Using eq(\ref{B0gaugefield}), and eq(\ref{compact-metric}), the DBI action at $\mathcal{O}(\beta^{0})$, can be written as
\begin{equation}
    S_{\text{DBI}}^{(0)} = -N_f T_{D6} \int d^7x\, e^{-\phi_{\text{IIA}}} 
   \sqrt{i^{\ast}B+g}|_{S^{2}} \sqrt{\, f_{11}^{(0)2} \left(
    -f_{11}^{(0)} A_t^{(0)'2} 
    + f_{tt}^{(0)} A_{x^1}^{(0)'2} 
    + f_{rr}^{(0)} \left(f_{11}^{(0)} f_{tt}^{(0)} - E_{x^{1}}^2\right)
    \right)}
\end{equation}
where, defining $\sqrt{i^{\ast}B+g}|_{S^{2}}=\sqrt{ f_{\theta_2\theta_2}(\theta_2)\, f_{\tilde{Y}\tilde{Y}}(\theta_2)-f_{\theta_2 \tilde{Y}}^2(\theta_2)}$, this quantity is purely of order $(\mathcal{O}(\beta^{0}))$, with the off-diagonal contributions originating from the $(B_{NS\text{-}NS})$ field. Considering $\mathcal{I}$ as the overall factor obtained after performing the angular regularization of the integral over $\theta_2$, along with the integral over the compact direction $\tilde{Y}$ (see Appendix~A for details), the DBI action takes the form:

\begin{equation}
    S_{\text{DBI}}^{(0)} = -N_f T_{D6} \int d^7x\, e^{-\phi_{\text{IIA}}} 
   \mathcal{I} \sqrt{\, f_{11}^{(0)2} \left(
    -f_{11}^{(0)} A_t^{(0)'2} 
    + f_{tt}^{(0)} A_{x^1}^{(0)'2} 
    + f_{rr}^{(0)} \left(f_{11}^{(0)} f_{tt}^{(0)} - E_{x^{1}}^2\right)
    \right)}
\end{equation}

 We absorb the multiplicative factor $(-N_{f}T_{D6})$ in the $\mathcal{I}$, and $\tilde{\mathcal{I}}$ (with $T_{D6}=\frac{1}{(2\pi)^{5/2}g_{s}}$ in the units of $2\pi\alpha^{\prime}=1$),  we will consider them separately while doing numerical analysis of the obtained results. The equations of motion for the gauge fields are obtained by varying the action with respect to the corresponding gauge fields, and are given by:

\begin{subequations}
\begin{align}
    -\frac{\mathcal{I}\, e^{-\Phi}\, f_{11}^{(0)3}\, A_t^{(0)'}}{\sqrt{ f_{11}^{(0)2} \left(
    -f_{11}^{(0)} A_t^{(0)'2} 
    + f_{tt}^{(0)} A_{x^1}^{(0)'2} 
    + f_{rr}^{(0)} \left(f_{11}^{(0)} f_{tt}^{(0)} - E_{x^{1}}^2\right)
    \right)}} &= \Pi_t^{(0)},
    \label{eq:EOMAt}
    \\[8pt]
    \frac{ \mathcal{I}\,e^{-\Phi}\, f_{11}^{(0)2}\, f_{tt}^{(0)}\, A_{x^1}^{(0)'}}{\sqrt{ f_{11}^{(0)2} \left(
    -f_{11}^{(0)} A_t^{(0)'2} 
    + f_{tt}^{(0)} A_{x^1}^{(0)'2} 
    + f_{rr}^{(0)} \left(f_{11}^{(0)} f_{tt}^{(0)} - E_{x^{1}}^2\right)
    \right)}} &= \Pi_{x^1}^{(0)},
    \label{eq:EOMAx}
\end{align}
\end{subequations}

where $\Pi_{t}^{(0)}$, and $\Pi_{x^{0}}^{(1)}$ are the constants of integration appearing in the EOMs of the respective gauge fields. From (\ref{eq:EOMAt}), and (\ref{eq:EOMAx}), one gets 
\begin{equation}
\label{constraint1}
    -\frac{f_{11}^{(0)}\, A_t^{(0)'}}{f_{tt}^{(0)}\, A_{x^1}^{(0)'}} = \frac{\Pi_t^{(0)}}{\Pi_{x^1}^{(0)}}
\end{equation}

By utilizing eq(\ref{constraint1}), the radial derivatives of the gauge fields may be expressed in the following form:
 
\begin{subequations}
\label{eq:gauge_fields_beta_B0}
\begin{align}
    \partial_r A_t^{(0)} &= -\frac{J_t^{(0)} \sqrt{f_{rr}^{(0)}} \sqrt{f_{tt}^{(0)}} 
    \sqrt{f_{11}^{(0)} f_{tt}^{(0)} - E_{x^{1}}^2}}
    {\sqrt{f_{11}^{(0)}} \sqrt{\mathcal{I}^{2}\, e^{-2\Phi}\, f_{11}^{(0)3} f_{tt}^{(0)} 
    - J_{x^1}^{(0)2} f_{11}^{(0)} + \Pi_t^{(0)2} f_{tt}^{(0)}}},
    \label{eq:At_beta_B0}
    \\[8pt]
    \partial_r A_{x^1}^{(0)} &= \frac{\Pi_{x^1}^{(0)} \sqrt{f_{11}^{(0)}} \sqrt{f_{rr}^{(0)}} 
    \sqrt{f_{11}^{(0)} f_{tt}^{(0)} - E_{x^{1}}^2}}
    {\sqrt{f_{tt}^{(0)}} \sqrt{\mathcal{I}^{2}\, e^{-2\Phi}\, f_{11}^{(0)3} f_{tt}^{(0)} 
    - \Pi_{x^1}^{(0)2} f_{11}^{(0)} + \Pi_t^{(0)2} f_{tt}^{(0)}}},
    \label{eq:Ax1_beta_B0}
\end{align}
\end{subequations}

Upon substituting the radial derivatives of the gauge fields, the DBI action takes the following form:

\begin{equation}
    S_{\text{DBI}} = -\int d^7x\, e^{-2\Phi} \,
   \mathcal{I}^{2}\, \sqrt{\frac{f_{11}^{(0)5}\, f_{rr}^{(0)}\, f_{tt}^{(0)}
    \left(f_{11}^{(0)} f_{tt}^{(0)} - E_{x^{1}}^2\right)}
    {\mathcal{I}^{2}\, e^{-2\phi}\, f_{11}^{(0)3} f_{tt}^{(0)} 
    - \Pi_{x^1}^{(0)2} f_{11}^{(0)} + \Pi_t^{(0)2} f_{tt}^{(0)}}}
\end{equation}
where the $T_{D6}$ is included in $\mathcal{I}$. Following \cite{OBannon:2007cex, Karch:2007pd}, the $f_{tt}=0$ at the horizon, resulting in both numerator and denominator negative, and going towards the UV boundary, both numerator and denominator turn out to be positive. Hence, for the reality of DBI action, the numerator and denominator must vanish simultaneously at the same point called the effective horizon($r_{\ast}$), i.e,
\begin{align}
\label{constraints1}
\xi_{B=0}^{(0)} 
&= f_{11}^{(0)} f_{tt}^{(0)} - E_{x^{1}}^2 
= 0, \\[6pt]
\chi_{B=0}^{(0)}
&= \mathcal{I}^{2}\, e^{-2\Phi}\, f_{11}^{(0)3}\, f_{tt}^{(0)} 
- \Pi_{x^1}^{(0)2}\, f_{11}^{(0)} 
+ \Pi_t^{(0)2}\, f_{tt}^{(0)} 
= 0.
\end{align}
where, the root of $\xi^{(0)}=0$, turns out to be $r_{\ast}^{(0)4}=r_{h}^{4}+4\pi g_{s}N E_{{x}^{1}}^{2}$, which is the effective horizon for $B=0$ scenario.

To implement the renormalized boundary current, we define the renormalized boundary current as
\begin{equation}
J_{\mu}
=
\lim_{r \to r_{\rm UV}}
\frac{r_{\rm UV}^{\,a}}{\sqrt{-\gamma}}\,
\Pi_{\mu},
\label{RenBoundaryCurrent}
\end{equation}
where $\Pi_{\mu}$ denotes the canonical momentum conjugate to the worldvolume gauge field, and $\gamma_{i\Pi}$ is the induced metric on the UV cutoff hypersurface. The determinant of the induced metric evaluates to
\begin{equation}
\sqrt{-\gamma} \equiv \sqrt{-\det(\gamma_{ij})}
=
\frac{r^4}{6 \sqrt[3]{3}\, \pi^{3/4}\, g_s^{3/4}\, N^{3/20}\, }
\equiv \mathcal{Z}.
\label{GammaDeterminant}
\end{equation}
For notational convenience, we define the renormalization factor, $\mathcal{Z} \equiv \sqrt{-\gamma}$. Following the holographic renormalization prescription of 
O’Bannon \cite{OBannon:2007cex} and Karch \emph{et al.} \cite{Karch:2007pd}, the bulk currents are related to the renormalized boundary currents via $J_{t}^{(0)}
=
\lim_{r \to r_{\rm UV}}
\frac{r_{\rm UV}^{\,a}}{\sqrt{-\gamma}}\,
\Pi_{t}^{(0)}$, and $J_{x^{1}}^{(0)}=\lim_{r \to r_{\rm UV}}
\frac{r_{\rm UV}^{\,a}}{\sqrt{-\gamma}}\,
\Pi_{x^1}^{(0)}$.

Imposing Ohm's law for the boundary theory, $J_{x^{1}} =\sigma_{xx} E_{x^{1}}$, one can directly read off the DC conductivity as, $\sigma_{xx}=\left. \frac{J_{x^{1}}}{E_{x^{1}}} \right|_{r = r_{\ast}}$, where all background quantities are evaluated at the effective horizon $r_{\ast}$, as required by the reality condition of the DBI action.

\begin{equation}
    J_{x^1}^{(0)} = \frac{E_{x^{1}} \sqrt{\mathcal{Z}^{-2}\mathcal{I}^{2}\, e^{-2\Phi}\, f_{11}^{(0)3} + J_{t}^{(0)2}}}{f_{11}^{(0)}}
\end{equation}
The resulting DC conductivity will be,

\begin{equation}
\label{conductivityxxE0}
    \sigma_{xx}^{B=0} = \frac{\sqrt{\mathcal{Z}^{-2}\mathcal{I}^{2}\, e^{-2\Phi}\, f_{11}^{(0)3} + J_t^{(0)2}}}{f_{11}^{(0)}}
\end{equation}

which is in agreement with \cite{OBannon:2007cex,Karch:2007pd}}.

\subsection{At $\mathcal{O}(\beta)$ for $B=0$ case:}

In this subsection, we compute the higher-derivative (HD) correction to the conductivity obtained in eq(\ref{conductivityxxE0}). To this end, we consider the metric decomposition up to $\mathcal{O}(\beta)$, as introduced in eq(\ref{compact-metric}), together with the gauge field ansatz given in eq(\ref{B0gaugefield}).

Using eq(\ref{compact-metric}) and (\ref{B0gaugefield}), we expand the DBI action perturbatively up to $\mathcal{O}(\beta)$. The resulting first-order HD corrected action then takes the form:

\begin{align}
    S^{(1)}_{\text{DBI}} = -N_f T_{D6} \int d^7 x\, \frac{\beta\, e^{-\Phi}\, f_{11}^{(0)}}{2\sqrt{f_{11}^{(0)2} \left(
    -f_{11}^{(0)} A_t^{(0)'2} 
    + f_{tt}^{(0)} A_{x^1}^{(0)'2} 
    + f_{rr}^{(0)} \left(f_{11}^{(0)} f_{tt}^{(0)} - E_{x^{1}}^2\right)
    \right)}}
    \notag \\
    \times\Biggl[
        \sqrt{i^{\ast}B+g}|_{S^{2}} \Bigl(
            2 f_{11}^{(0)} f_{tt}^{(0)} A_{x^1}^{(1)'} A_{x^1}^{(0)'}
            - 2 f_{11}^{(0)2} A_t^{(1)'} A_t^{(0)'}
        \Bigr)
        \notag \\
        + \mathcal{F} \Biggl(
            f_{11}^{(1)} \Bigl(
                -3 f_{11}^{(0)} A_t^{(0)'2}
                + 2 f_{tt}^{(0)} A_{x^1}^{(0)'2}
                + f_{rr}^{(0)} \left(3 f_{11}^{(0)} f_{tt}^{(0)} - 2 E_{x^{1}}^2\right)
            \Bigr)
            \notag \\
            + f_{11}^{(0)} f_{tt}^{(1)} A_{x^1}^{(0)'2}
            + f_{11}^{(0)2} f_{rr}^{(0)} f_{tt}^{(1)}
        \Biggr)
    \Biggr]
\end{align}

The angular factor $\sqrt{i^{\ast}B+g}|_{S^{2}} = \sqrt{ f_{\theta_2\theta_2}(\theta_2) f_{\tilde{Y}\tilde{Y}}(\theta_2) - f_{\theta_2 \tilde{Y}}^2(\theta_2) }$, as previously established, represents the factors needed the angular regularization for the action at $\mathcal{O}(\beta^0)$. However, at $\mathcal{O}(\beta)$, an additional term requiring angular regularization emerges, defined by the factor $\mathcal{F} = f_{11}^{(1)}(\theta_2) \sqrt{ f_{\theta_2\theta_2}(\theta_2) f_{\tilde{Y}\tilde{Y}}(\theta_2) - f_{\theta_2 \tilde{Y}}^2(\theta_2) }$. Let $\mathcal{I}$ and $\tilde{\mathcal{I}}$ denote the finite multiplicative constants obtained after performing the complete angular regularization of the integrals over $\theta_2$ and $\tilde{Y}$ for the terms $\sqrt{i^{\ast}B+g}|_{S^{2}}$ and $\mathcal{F}$, respectively. Under this framework, the DBI action at $\mathcal{O}(\beta)$ is expressed as:

\begin{align}
    S^{(1)}_{\text{DBI}} = -N_f T_{D6} \int d^7 x\, \frac{\beta\, e^{-\Phi}\, f_{11}^{(0)}}{2\sqrt{f_{11}^{(0)2} \left(
    -f_{11}^{(0)} A_t^{(0)'2} 
    + f_{tt}^{(0)} A_{x^1}^{(0)'2} 
    + f_{rr}^{(0)} \left(f_{11}^{(0)} f_{tt}^{(0)} - E_{x^{1}}^2\right)
    \right)}}
    \notag \\
    \times\Biggl[
        \mathcal{I} \Bigl(
            2 f_{11}^{(0)} f_{tt}^{(0)} A_{x^1}^{(1)'} A_{x^1}^{(0)'}
            - 2 f_{11}^{(0)2} A_t^{(1)'} A_t^{(0)'}
        \Bigr)
        \notag \\
        + \mathcal{\tilde{I}} \Biggl(
            f_{11}^{(1)} \Bigl(
                -3 f_{11}^{(0)} A_t^{(0)'2}
                + 2 f_{tt}^{(0)} A_{x^1}^{(0)'2}
                + f_{rr}^{(0)} \left(3 f_{11}^{(0)} f_{tt}^{(0)} - 2 E_{x^{1}}^2\right)
            \Bigr)
            \notag \\
            + f_{11}^{(0)} f_{tt}^{(1)} A_{x^1}^{(0)'2}
            + f_{11}^{(0)2} f_{rr}^{(0)} f_{tt}^{(1)}
        \Biggr)
    \Biggr]
\end{align}
where as previously stated, now at $\mathcal{O}(\beta)$, $\mathcal{I}$, and $\tilde{\mathcal{I}}$ represent the factors arising from the angular regularization procedure at $\mathcal{O}(\beta)$. The equations of motion (EOMs) for the corresponding $U(1)$ gauge fields then take the form:

\begin{subequations}
\begin{align}
\Pi_t^{(1)} &= -\frac{\beta\, e^{-\Phi}\, f_{11}^{(0)4}}{2\,X_{(0)}^{3/2}}
\Biggl[
    -f_{11}^{(0)} \Bigl(
        A_{x^1}^{(0)'} \bigl(
            2\mathcal{I} f_{tt}^{(0)} (A_t^{(0)'} A_{x^1}^{(1)'} - A_t^{(1)'} A_{x^1}^{(0)'})
            + \mathcal{\tilde{I}} f_{tt}^{(1)} A_t^{(0)'} A_{x^1}^{(0)'}
        \bigr)
        \notag \\
        &\quad + f_{rr}^{(0)} \bigl(
            2\mathcal{I} E_{x^{1}}^2 A_t^{(1)'}
            - 3\mathcal{\tilde{I}} f_{11}^{(1)} f_{tt}^{(0)} A_t^{(0)'}
        \bigr)
        + 3\mathcal{\tilde{I}} f_{11}^{(1)} A_t^{(0)'3}
    \Bigr)
    \notag \\
    &\quad + f_{11}^{(0)2} f_{rr}^{(0)} \bigl(
        2\mathcal{I} f_{tt}^{(0)} A_t^{(1)'}
        - \mathcal{\tilde{I}} f_{tt}^{(1)} A_t^{(0)'}
    \bigr)
    \notag \\
    &\quad + 4\mathcal{\tilde{I}} f_{11}^{(1)} A_t^{(0)'} \bigl(
        f_{tt}^{(0)} A_{x^1}^{(0)'2} - E_{x^{1}}^2 f_{rr}^{(0)}
    \bigr)
\Biggr],
\label{eq:Pi_t1}
\\[10pt]
\Pi_{x^1}^{(1)} &= \frac{\beta\, e^{-\Phi}\, f_{11}^{(0)3}}{2\,X_{(0)}^{3/2}}
\Biggl[
    f_{11}^{(0)2} \Bigl(
        f_{rr}^{(0)} f_{tt}^{(0)} \bigl(
            2\mathcal{I} f_{tt}^{(0)} A_{x^1}^{(1)'}
            + \mathcal{\tilde{I}} f_{tt}^{(1)} A_{x^1}^{(0)'}
        \bigr)
        \notag \\
        &\quad - 2 A_t^{(0)'} \bigl(
            \mathcal{I} f_{tt}^{(0)} (A_t^{(0)'} A_{x^1}^{(1)'} - A_t^{(1)'} A_{x^1}^{(0)'})
            + \mathcal{\tilde{I}} f_{tt}^{(1)} A_t^{(0)'} A_{x^1}^{(0)'}
        \bigr)
    \Bigr)
    \notag \\
    &\quad + f_{11}^{(0)} \Bigl(
        \mathcal{\tilde{I}} f_{tt}^{(0)} A_{x^1}^{(0)'} \bigl(
            f_{tt}^{(1)} A_{x^1}^{(0)'2} - f_{11}^{(1)} A_t^{(0)'2}
        \bigr)
        \notag \\
        &\quad + f_{rr}^{(0)} \bigl(
            -2\mathcal{I} E_{x^{1}}^2 f_{tt}^{(0)} A_{x^1}^{(1)'}
            - 2\mathcal{\tilde{I}} E_{x^{1}}^2 f_{tt}^{(1)} A_{x^1}^{(0)'}
            + \mathcal{\tilde{I}} f_{11}^{(1)} f_{tt}^{(0)2} A_{x^1}^{(0)'}
        \bigr)
    \Bigr)
    \notag \\
    &\quad + 2\mathcal{\tilde{I}} f_{11}^{(1)} f_{tt}^{(0)} A_{x^1}^{(0)'} \bigl(
        f_{tt}^{(0)} A_{x^1}^{(0)'2} - E_{x^{1}}^2 f_{rr}^{(0)}
    \bigr)
\Biggr],
\label{eq:Pi_x1}
\end{align}
\end{subequations}

where,

\begin{align}
X_{(0)} &= f_{11}^{(0)}\Bigl[
    f_{11}^{(0)2}\bigl(f_{rr}^{(0)} f_{tt}^{(0)} - A_t^{(0)'2}\bigr)
    + f_{11}^{(0)} f_{tt}^{(0)} A_{x^1}^{(0)'2}
    - E_{x^{1}}^2 f_{11}^{(0)} f_{rr}^{(0)}
\Bigr],
\label{eq:calD0}
\\[6pt]
\mathcal{N}_{(0)} &\equiv \mathcal{\tilde{I}}\Bigl[
    -3f_{11}^{(0)2} f_{11}^{(1)}\bigl(A_t^{(0)'2} - f_{rr}^{(0)} f_{tt}^{(0)}\bigr)
    + 2f_{11}^{(0)} f_{11}^{(1)}\bigl(f_{tt}^{(0)} A_{x^1}^{(0)'2} - E_{x^{1}}^2 f_{rr}^{(0)}\bigr)
    + f_{11}^{(0)2} f_{tt}^{(1)}\bigl(A_{x^1}^{(0)'2} + f_{11}^{(0)2} f_{rr}^{(0)}\bigr)
\Bigr] \notag \\
&\quad - 2\mathcal{I}\, f_{11}^{(0)}\Bigl[
    A_t^{(0)'} A_t^{(1)'} f_{11}^{(0)2}
    - f_{11}^{(0)} f_{tt}^{(0)} A_{x^1}^{(0)'} A_{x^1}^{(1)'}
\Bigr].
\label{eq:calN0}
\end{align}

$\Pi_{t}^{(1)}$ and $\Pi_{x^{1}}^{(1)}$ are constants of integration that can be identified with radially conserved currents via the gauge/gravity correspondence, wherein the appropriately regularized integration constants are mapped to conserved currents at $\mathcal{O}(\beta)$ in the dual field theory.

Then the radial derivative of gauge fields at $\mathcal{O}(\beta)$ can be written in terms of the constants of integration as

\begin{align}
A_t^{(1)'} &= \frac{1}{2\mathcal{I}\, e^{-\Phi}\, f_{11}^{(0)3}\, 
    f_{rr}^{(0)}\left(f_{11}^{(0)} f_{tt}^{(0)} - E_{x^{1}}^2\right)}
\notag \\
&\quad \times\Biggl\{
    e^{-\Phi}\, \mathcal{\tilde{I}}\, f_{11}^{(1)}\, f_{11}^{(0)2}\, A_t^{(0)'}
    \Bigl(
        3 f_{11}^{(0)} A_t^{(0)'2}
        - 2 f_{tt}^{(0)} A_{x^1}^{(0)'2}
        + f_{rr}^{(0)} \bigl(4E_{x^{1}}^2 - 3 f_{11}^{(0)} f_{tt}^{(0)}\bigr)
    \Bigr)
    \notag \\
    &\qquad - 2 f_{11}^{(0)}\, \sqrt{X_{(0)}}
    \Bigl(
        \Pi_t^{(1)} f_{rr}^{(0)} f_{tt}^{(0)}
        - A_t^{(0)'}\bigl(\Pi_t^{(1)} A_t^{(0)'} + \Pi_{x^1}^{(1)} A_{x^1}^{(0)'}\bigr)
    \Bigr)
    \notag \\
    &\qquad + 2E_{x^{1}}^2\, \Pi_t^{(1)}\, f_{rr}^{(0)}\, \sqrt{X_{(0)}}
    \notag \\
    &\qquad - e^{-\Phi}\, \mathcal{\tilde{I}}\, f_{11}^{(0)3}\, f_{tt}^{(1)}\, 
    A_t^{(0)'} A_{x^1}^{(0)'2}
    \notag \\
    &\qquad + e^{-\Phi}\, \mathcal{\tilde{I}}\, f_{11}^{(0)4}\, f_{rr}^{(0)}\, f_{tt}^{(1)}\, A_t^{(0)'}
\Biggr\},
\label{eq:At1prime}
\\[10pt]
A_{x^1}^{(1)'} &= -\frac{1}{2\mathcal{I}\, e^{-\Phi}\, f_{11}^{(0)2}\, 
    f_{rr}^{(0)}\, f_{tt}^{(0)}\left(f_{11}^{(0)} f_{tt}^{(0)} - E_{x^{1}}^2\right)}
\notag \\
&\quad \times\Biggl\{
    e^{-\Phi}\, \mathcal{\tilde{I}}\, f_{11}^{(0)2}\, A_{x^1}^{(0)'}
    \Bigl(
        f_{tt}^{(0)}\bigl(f_{tt}^{(1)} A_{x^1}^{(0)'2} - 3 f_{11}^{(1)} A_t^{(0)'2}\bigr)
        + f_{rr}^{(0)}\bigl(f_{11}^{(1)} f_{tt}^{(0)2} - 2E_{x^{1}}^2 f_{tt}^{(1)}\bigr)
    \Bigr)
    \notag \\
    &\qquad - 2 f_{11}^{(0)}\, f_{tt}^{(0)}
    \Bigl(
        \Pi_{x^1}^{(1)} f_{rr}^{(0)}\, \sqrt{X_{(0)}}
        + e^{-\Phi}\, \mathcal{\tilde{I}}\, f_{11}^{(1)}\, A_{x^1}^{(0)'}
        \bigl(E_{x^{1}}^2 f_{rr}^{(0)} - f_{tt}^{(0)} A_{x^1}^{(0)'2}\bigr)
    \Bigr)
    \notag \\
    &\qquad + 2\sqrt{X_{(0)}}
    \Bigl(
        E_{x^{1}}^2\, \Pi_{x^1}^{(1)}\, f_{rr}^{(0)}
        - f_{tt}^{(0)}\, A_{x^1}^{(0)'}
        \bigl(\Pi_t^{(1)} A_t^{(0)'} + \Pi_{x^1}^{(1)} A_{x^1}^{(0)'}\bigr)
    \Bigr)
    \notag \\
    &\qquad + e^{-\Phi}\, \mathcal{\tilde{I}}\, f_{11}^{(0)3}\, f_{rr}^{(0)}\, 
    f_{tt}^{(1)}\, f_{tt}^{(0)}\, A_{x^1}^{(0)'}
\Biggr\},
\label{eq:Ax1prime}
\end{align}

Using these radial derivatives of gauge fields in eqs(\ref{eq:At0},\ref{eq:At1},\ref{eq:At1prime},\ref{eq:At1prime}), the resulting DBI Lagrangian can be written as at $\mathcal{O}(\beta)$:

\begin{align}
L_{DBI}^{(1)} 
&= \frac{\beta}{2\sqrt{f_{11}^{(0)}(r)\,f_{tt}^{(0)}(r)-E_{x^1}^{2}}}
\left(
\mathcal{I}^2 e^{-2\Phi_{IIA}}(r)\bigl(f_{11}^{(0)}(r)\bigr)^3 f_{tt}^{(0)}(r)
-(\Pi_{x}^{(0)})^2 f_{11}^{(0)}(r)
+(\Pi_{t}^{(0)})^2 f_{tt}^{(0)}(r)
\right)^2
\nonumber\\[4pt]
&\quad \times 
\sqrt{
\frac{
\mathcal{I}^2 e^{-2\Phi_{IIA}}(r)\bigl(f_{11}^{(0)}(r)\bigr)^5 f_{rr}^{(0)}(r) f_{tt}^{(0)}(r)
\left(f_{11}^{(0)}(r)f_{tt}^{(0)}(r)-E_{x^1}^{2}\right)
}{
\mathcal{I}^2 e^{-2\Phi_{IIA}}(r)\bigl(f_{11}^{(0)}(r)\bigr)^3 f_{tt}^{(0)}(r)
-(\Pi_{x}^{(0)})^2 f_{11}^{(0)}(r)
+(\Pi_{t}^{(0)})^2 f_{tt}^{(0)}(r)
}
}
\nonumber\\[6pt]
&\quad \times \Biggl[
\mathcal{I}^4 \tilde{\mathcal{I}} e^{-5\Phi_{IIA}}(r)
\bigl(f_{11}^{(0)}(r)\bigr)^7 f_{rr}^{(0)}(r) \bigl(f_{tt}^{(0)}(r)\bigr)^2
\sqrt{f_{11}^{(0)}(r)f_{tt}^{(0)}(r)-E_{x^1}^{2}}
\nonumber\\
&\qquad \times 
\left(
f_{11}^{(1)}(r)\left(3 f_{11}^{(0)}(r)f_{tt}^{(0)}(r)-2E_{x^1}^{2}\right)
+\bigl(f_{11}^{(0)}(r)\bigr)^2 f_{tt}^{(1)}(r)
\right)
\nonumber\\[4pt]
&\quad + \mathcal{I}^2 \tilde{\mathcal{I}} e^{-3\Phi_{IIA}}(r)
\bigl(f_{11}^{(0)}(r)\bigr)^4 f_{rr}^{(0)}(r)
\sqrt{f_{11}^{(0)}(r)f_{tt}^{(0)}(r)-E_{x^1}^{2}}
\nonumber\\
&\qquad \times \Bigl[
\bigl(f_{11}^{(0)}(r)\bigr)^2
\Bigl(
f_{tt}^{(1)}(r)\bigl(E_{x^1}^{2}(\Pi_{x}^{(0)})^2
+(\Pi_{t}^{(0)})^2 (f_{tt}^{(0)}(r))^2\bigr)
\nonumber\\
&\qquad\qquad
-5(\Pi_{x}^{(0)})^2 f_{11}^{(1)}(r) (f_{tt}^{(0)}(r))^2
\Bigr)
\nonumber\\
&\qquad\quad
+2 f_{11}^{(1)}(r) f_{11}^{(0)}(r) f_{tt}^{(0)}(r)
\Bigl(2E_{x^1}^{2}(\Pi_{x}^{(0)})^2
+3(\Pi_{t}^{(0)})^2 (f_{tt}^{(0)}(r))^2\Bigr)
\nonumber\\
&\qquad\quad
-5E_{x^1}^{2}(\Pi_{t}^{(0)})^2 f_{11}^{(1)}(r) (f_{tt}^{(0)}(r))^2
\nonumber\\
&\qquad\quad
-2(\Pi_{x}^{(0)})^2 (f_{11}^{(0)}(r))^3 f_{tt}^{(1)}(r) f_{tt}^{(0)}(r)
\Bigr]
\nonumber\\[4pt]
&\quad + 
\frac{
2\left(
\mathcal{I}^2 e^{-2\Phi_{IIA}}(r)\bigl(f_{11}^{(0)}(r)\bigr)^3 f_{tt}^{(0)}(r)
-(\Pi_{x}^{(0)})^2 f_{11}^{(0)}(r)
+(\Pi_{t}^{(0)})^2 f_{tt}^{(0)}(r)
\right)^{3/2}
}{(f_{11}^{(0)}(r))^{5/2}\sqrt{f_{rr}^{(0)}(r)}\sqrt{f_{tt}^{(0)}(r)}}
\nonumber\\
&\qquad \times
\left(\Pi_{x}^{(1)}\Pi_{x}^{(0)} f_{11}^{(0)}(r)
-\Pi_{t}^{(1)}\Pi_{t}^{(0)} f_{tt}^{(0)}(r)\right)
\nonumber\\
&\qquad \times
\left(
\frac{
\mathcal{I}^2 e^{-2\Phi_{IIA}}(r)\bigl(f_{11}^{(0)}(r)\bigr)^5 f_{rr}^{(0)}(r) f_{tt}^{(0)}(r)
\left(f_{11}^{(0)}(r)f_{tt}^{(0)}(r)-E_{x^1}^{2}\right)
}{
\mathcal{I}^2 e^{-2\Phi_{IIA}}(r)\bigl(f_{11}^{(0)}(r)\bigr)^3 f_{tt}^{(0)}(r)
-(\Pi_{x}^{(0)})^2 f_{11}^{(0)}(r)
+(\Pi_{t}^{(0)})^2 f_{tt}^{(0)}(r)
}
\right)^{3/2}
\Biggr]
\end{align}

We note that at the leading-order effective horizon $r=r_{\ast}^{\beta^0}$, the O'Bannon reality
conditions require $\xi_{B=0}^{(0)}(r_{\ast}^{\beta^0})=\chi_{B=0}^{(0)}(r_{\ast}^{\beta^0})=0$, so
that the ratio $\chi_{B=0}^{(0)}/\xi_{B=0}^{(0)}$ appearing implicitly in the DBI action is formally
indeterminate at $r=r_{\ast}^{\beta^0}$. Since the $\mathcal{O}(\beta)$ expansion is carried out in a
small neighborhood of the effective horizon, both functions admit regular Taylor expansions in
$(r-r_{\ast}^{\beta^0})$,
\begin{equation}
    \xi_{B=0}^{(0)}(r) = \partial_r\xi_{B=0}^{(0)}\Big|_{r=r_{\ast}^{\beta^0}}(r-r_{\ast}^{\beta^0}) + \mathcal{O}\big((r-r_{\ast}^{\beta^0})^2\big),
\end{equation}
\begin{equation}
    \chi_{B=0}^{(0)}(r) = \partial_r\chi_{B=0}^{(0)}\Big|_{r=r_{\ast}^{\beta^0}}(r-r_{\ast}^{\beta^0}) + \mathcal{O}\big((r-r_{\ast}^{\beta^0})^2\big),
\end{equation}
Consequently, in the near-horizon limit, $\frac{\chi_{B=0}^{(0)}}{\xi_{B=0}^{(0)}}\sim\frac{\partial_r\chi_{B=0}^{(0)}}{\partial_r\xi_{B=0}^{(0)}}$,
which remains finite, with its value being fixed by the integration constants of the $\mathcal{O}(\beta^{0})$ equations of motion for the $U(1)$ gauge fields.

Using the constraints obtained from the $\mathcal{O}(\beta^{0})$ part of eq(\ref{constraint1}) expanded near effective horizon $r=r_{\ast}$, we first impose the condition, $
\xi_{B=0}^{(0)}=f_{11}^{(0)}\, f_{tt}^{(0)} - E_{x^{1}}^{2} \equiv \epsilon $,
evaluated at the zeroth-order effective horizon $r_{\ast}^{\beta^{0}}$. Expanding the higher-derivative (HD) corrected part of the action in the limit $\epsilon \to 0$, the numerator of the expanded DBI contribution takes the form

\begin{equation}
    \mathcal{A}_{B=0} = \beta\, \tilde{\mathcal{I}}\, E_{x^{1}}^2\, e^{-3\Phi}\, 
    \mathcal{I}^{2}\, f_{11}^{(0)3}\, f_{rr}^{(0)}
    \Bigl(E_{x^{1}}^2 f_{11}^{(1)} + f_{11}^{(0)2} f_{tt}^{(1)}\Bigr)
    \partial_{r}\chi^{(0)}_{r=r_{\ast}^{B=0}}\epsilon
\end{equation}
and,
\begin{equation}
\mathcal{B}_{B=0}
=
2\,\mathcal{I}\,
\left.\partial_{r}\xi^{(0)}_{B=0}\right|_{r=r_{\ast}^{B=0}}
\left(
\left.\partial_{r}\chi^{(0)}_{B=0}\right|_{r=r_{\ast}^{B=0}}\epsilon
\right)^{3/2}
E_{x^{1}}\,
e^{-\Phi}\,
\left(f_{11}^{(0)}\right)^{2}
\sqrt{f_{rr}^{(0)}}.
\end{equation}
then the resulting DBI action at $\mathcal{O}(\beta)$ takes the form

\begin{eqnarray}
\frac{S_{DBI}^{\beta}}{N_{f}T_{D6}}&=&-\int d^{7}x \frac{\mathcal{I}\,E_{x^1}\,\tilde{\mathcal{I}}\,f_{11}^{(0)}\,e^{-2\Phi_{IIA}}(r)\sqrt{f_{rr}^{(0)}(r)}\left(E_{x^1}^{2}\,f_{11}^{(1)}(r)+\bigl(f_{11}^{(0)}(r)\bigr)^2 f_{tt}^{(1)}(r)\right)}{2\epsilon\sqrt{\left(\partial_{r}\xi_{B=0}^{(0)}|_{r=r_{\ast}^{B=0}}\right) \,\left(\partial_{r}\chi_{B=0}^{(0)}|_{r=r_{\ast}^{B=0}}\right)}}\nonumber\\
\end{eqnarray}

Since the $\mathcal{O}(\beta)$ DBI action behaves as $S_{\rm DBI}^{\beta}\sim 1/\epsilon$ in the vicinity of the effective horizon ($\epsilon\equiv r-r_{\ast}^{B=0}\rightarrow0$), the higher-derivative contribution is potentially singular. Therefore, for the perturbative higher-derivative expansion of the DBI action to remain well-defined, the numerator must vanish sufficiently rapidly to compensate for the divergence arising from the denominator. This regularity condition requires that $\xi_{B=0}^{(1)}$ vanish, thereby determining the $\mathcal{O}(\beta)$ correction to the effective horizon. We further note that the higher-derivative correction $\xi_{B=0}^{(1)}$ is independent of the conserved current, so that the location of the corrected effective horizon is fixed solely by the condition $\xi_{B=0}^{(1)}=0$. The resulting condition at $\mathcal{O}(\beta)$ can therefore be written as follows:

\begin{equation}
\xi^{(1)}_{B=0}
=
E_{x^{1}}^{2}\, f_{11}^{(1)}(r)
+
f_{11}^{(0)}(r)^{2}\, f_{tt}^{(1)}(r)
\end{equation}

Using the metric components, $\xi^{(1)}_{B=0}=0$ , one gets
\begin{equation}
    4\pi E_{x^{1}}^2 g_s N - r^4 + r_h^4 = 0,
\end{equation}
This relation is automatically satisfied at $r = r_{\ast}^{\beta^0}$, which shows that the effective horizon does not acquire any higher-derivative correction. As a result, the conductivity obtained in the previous section remains unchanged. In other words, when the magnetic field is absent, the DC conductivity does not receive corrections from $\mathcal{O}(R^4)$.
\section{Conductivity in the presence of an electromagnetic setup}
\label{DCB}
In this section, we present a detailed analysis extending the results of the previous section by additionally turning on the magnetic field along the $x^{2}$-direction. This setup enables us to compute both the longitudinal (DC) conductivity and the Hall conductivity in the presence of a background magnetic field.

We now employ the following ansatz for the worldvolume gauge field to introduce a constant magnetic field oriented along the $x^{2}$-direction:

\begin{eqnarray}
A_t=A_{t}^{(0)}(r)+\beta A_{t}^{(1)}(r) \quad\quad
A_x = -E_{x^{1}} t + A_{x^{1}}^{(0)}(r)+\beta A_{x^{1}}^{(1)} (r) \quad\quad
A_y = B x^{1} + A_{x^{2}}^{(0)}(r)+\beta A_{x^{2}}^{(1)} (r)\nonumber\\
\end{eqnarray}

The resulting DBI action for the flavor D6-brane in the presence of an electromagnetic setup takes the following form,

\begin{align}
    S_{\text{DBI},B}^{(0)} &= -N_f T_{D6} \int d^7x\, \mathcal{I}_{B}e^{-\Phi}
    \notag \\
    &\quad \times\sqrt{f_{11}^{(0)}
    \left(
    \begin{aligned}
    &-\left(B A_t^{(0)'} - E_{x^{1}} A_{x^2}^{(0)'}\right)^2
    + f_{11}^{(0)2}\left(f_{rr}^{(0)} f_{tt}^{(0)} - A_t^{(0)'2}\right)\\
    &+ f_{11}^{(0)}\left(f_{tt}^{(0)}\left(A_{x^1}^{(0)'2} + A_{x^2}^{(0)'2}\right) 
    - E_{x^{1}}^2 f_{rr}^{(0)}\right)
    + B^2 f_{rr}^{(0)} f_{tt}^{(0)}
    \end{aligned}
    \right)}
\end{align}

where $\mathcal{I}$ is the multiplicative factor that appeared post angular regularization for the term involving $\sqrt{i^{\ast}B+g}|_{S^{2}}$. Assuming,

\begin{equation}
\label{eq:Xgaugefields}
X=\sqrt{f_{11}^{(0)}
    \left(
    \begin{aligned}
    &-\left(B A_t^{(0)'} - E_{x^{1}} A_{x^2}^{(0)'}\right)^2
    + f_{11}^{(0)2}\left(f_{rr}^{(0)} f_{tt}^{(0)} - A_t^{(0)'2}\right)\\
    &+ f_{11}^{(0)}\left(f_{tt}^{(0)}\left(A_{x^1}^{(0)'2} + A_{x^2}^{(0)'2}\right) 
    - E_{x^{1}}^2 f_{rr}^{(0)}\right)
    + B^2 f_{rr}^{(0)} f_{tt}^{(0)}
    \end{aligned}
    \right)}
\end{equation}
from which the the EOMs for the $U(1)$ gauge fields will be

\begin{subequations}
\label{eq:gauge_fields_zeroth_beta}
\begin{align}
    A_{t}^{(0)'} &= -\frac{X \left(
        -B E_{x^{1}} \Pi_{x^2}^{(0)} 
        - E_{x^{1}}^2 \Pi_t^{(0)} 
        + \Pi_t^{(0)} f_{11}^{(0)} f_{tt}^{(0)}
    \right)}{\mathcal{I}\, e^{-\Phi}\, f_{11}^{(0)2} 
    \left(B^2 f_{tt}^{(0)} - E_{x^{1}}^2 f_{11}^{(0)} + f_{11}^{(0)2} f_{tt}^{(0)}\right)},
    \label{eq:At0_beta}
    \\[8pt]
    A_{x^1}^{(0)'} &= \frac{\Pi_{x^1}^{(0)}\, X}{\mathcal{I}\, e^{-\Phi}\, 
    f_{11}^{(0)2}\, f_{tt}^{(0)}},
    \label{eq:Ax0_beta}
    \\[8pt]
    A_{x^2}^{(0)'} &= \frac{X \left(
        B^2 \Pi_{x^2}^{(0)} 
        + B E_{x^{1}} \Pi_t^{(0)} 
        + \Pi_{x^2}^{(0)} f_{11}^{(0)2}
    \right)}{\mathcal{I}\, e^{-\Phi}\, f_{11}^{(0)2} 
    \left(B^2 f_{tt}^{(0)} - E_{x^{1}}^2 f_{11}^{(0)} + f_{11}^{(0)2} f_{tt}^{(0)}\right)},
    \label{eq:Ay0_beta}
\end{align}
\end{subequations}

Using eq(\ref{eq:Xgaugefields}) and eq(\ref{eq:gauge_fields_zeroth_beta}), the quantity $X$ can be written entirely in terms of the constants of 
integration $\Pi_t,\Pi_{x^{1}}$, and $\Pi_{x^{2}}$, together with the external fields $E_{x^{1}}$, and $B$, as
\begin{equation}
\resizebox{\textwidth}{!}{$\displaystyle
    X^{2} = \frac{\mathcal{I}^2 e^{-2\Phi} f_{11}^{(0)3} f_{rr}^{(0)} f_{tt}^{(0)} 
    \left(E_{x^{1}}^2 f_{11}^{(0)} - f_{tt}^{(0)} \left(B^2 + f_{11}^{(0)2}\right)\right)^2}{
    \mathcal{I} e^{-2\Phi} f_{11}^{(0)2} f_{tt}^{(0)} 
    \left(f_{tt}^{(0)}\left(B^2 + f_{11}^{(0)2}\right) - E_{x^{1}}^2 f_{11}^{(0)}\right)
    - f_{tt}^{(0)}\left(B^2 + f_{11}^{(0)2}\right)\left(\Pi_{x^1}^{(0)2} + \Pi_{x^2}^{(0)2}\right)
    + f_{11}^{(0)} f_{tt}^{(0)2} \Pi_t^{(0)2}
    - 2B E_{x^{1}} f_{tt}^{(0)} \Pi_t^{(0)} \Pi_{x^2}^{(0)}
    + E_{x^{1}}^2 \left(f_{11}^{(0)} \Pi_{x^1}^{(0)2} - f_{tt}^{(0)} \Pi_t^{(0)2}\right)}
$}
\end{equation}
Using this expression, the gauge fields can be rewritten completely in terms of $E_{x^{1}}$, $B$, currents, and metric components. The compact form of which can be written as 

\begin{subequations}
\begin{align}
    A_{t}^{(0)'} &= -\frac{\sqrt{f_{rr}^{(0)}f_{tt}^{(0)}}\, 
    \bigl(\Pi_t^{(0)}\, \xi^{(0)} - \mathfrak{\alpha}^{(0)} B\bigr)}
    {\sqrt{\chi^{(0)}\, \xi^{(0)} - \mathfrak{\alpha}^{(0)2}}},
    \label{eq:At0}
    \\[8pt]
    A_{x^1}^{(0)'} &= \frac{\sqrt{f_{rr}^{(0)}}\, \Pi_{x^1}^{(0)}\, \xi^{(0)}}
    {\sqrt{f_{tt}^{(0)}}\,\sqrt{\chi^{(0)}\, \xi^{(0)} - \alpha^{(0)2}}},
    \label{eq:Ax0}
    \\[8pt]
    A_{x^2}^{(0)'} &= \frac{\sqrt{f_{rr}^{(0)}}\, 
    \bigl(\mathfrak{\alpha}^{(0)} E_{x^{1}} + \Pi_{x^2}^{(0)}\, \xi^{(0)}\bigr)}
    {\sqrt{f_{tt}^{(0)}}\,\sqrt{\chi^{(0)}\, \xi^{(0)} - \mathfrak{\alpha}^{(0)2}}},
    \label{eq:Ay0}
\end{align}
\end{subequations}
where,
\begin{subequations}
\label{eq:definitions_zeroth_order}
\begin{align}
    \xi^{(0)} &= B^2 f_{tt}^{(0)} - E_{x^{1}}^2 f_{11}^{(0)} + f_{11}^{(0)2} f_{tt}^{(0)},
    \label{eq:xi0}
    \\[8pt]
    \chi^{(0)} &= \mathcal{I}^{2}\, e^{-2\Phi}\, f_{11}^{(0)3} f_{tt}^{(0)} 
    - f_{11}^{(0)} \bigl(\Pi_{x^1}^{(0)2} + \Pi_{x^2}^{(0)2}\bigr) 
    + f_{tt}^{(0)} \Pi_t^{(0)2},
    \label{eq:chi0}
    \\[8pt]
    \alpha^{(0)} &= B f_{tt}^{(0)} \Pi_t^{(0)} + E_{x^{1}} f_{11}^{(0)} \Pi_{x^2}^{(0)},
    \label{eq:alpha0}
\end{align}
\end{subequations}

The complete on-shell DBI action in the compact form can be written as:
\begin{equation}
S_{DBI}^{on-shell,\beta^{0}}=- \int d^{7}x \frac{e^{-2\phi_{IIA}} \mathcal{I}^{2}\sqrt{f_{tt}^{(0)}} f_{11}^{(0)} \sqrt{f_{rr}^{(0)}}  \xi_{B=0}^{(0)}}{\sqrt{\chi^{(0)} \xi^{(0)}-\alpha^{(0)^2}}}
\end{equation}

Similar to the $B=0$ case, as $|f_{tt}|$ vanishes at the horizon, $\xi^{(0)} < 0$, while toward the boundary, $\xi^{(0)} > 0$. Also, at the horizon $\chi^{(0)} < 0$, and at the UV boundary $\chi^{(0)} > 0$; therefore, to satisfy $\xi^{(0)}\chi^{(0)} \geq 0$, both $\xi^{(0)}$ and $\chi^{(0)}$ must share the same zero. The root of $\xi^{(0)}$ provides the effective horizon ($r_{\ast}^{(0)}$) for this case. Since $\xi^{(0)}$ and $\chi^{(0)}$ both vanish at the same point $r_{\ast}$, the reality of the DBI action requires that $\alpha^{(0)}$ must share the same root as $\xi^{(0)}$ and $\chi^{(0)}$. In conclusion, for the DBI action to remain real, all three constraints---$\xi^{(0)}$, $\chi^{(0)}$, and $\alpha^{(0)}$---must vanish simultaneously at the same point, known as the effective horizon.

From these constraints, one may proceed exactly as in the $B=0$ case and relate the conserved bulk quantities in the gauge-field equations of motion, namely $(\Pi_t^{(0)}), (\Pi_{x^1}^{(0)})$, and $(\Pi_{x^2}^{(0)})$, to the regularized boundary currents $(J_t^{(0)}, J_{x^1}^{(0)}, J_{x^2}^{(0)})$. The DC and Ohmic conductivities then follow from the regularized Ohm’s law, $J_i=\sigma_{ij}E_j$. In this notation, $\sigma_{ii}$ corresponds to the DC conductivities, whereas $\sigma_{ij}$ with $i\neq j$ describes the Hall response. One therefore finds the DC conductivity to be

\begin{equation}
\label{OhmicBbeta0}
    \sigma_{xx} = \frac{f_{11}^{(0)} \sqrt{\mathcal{I}^{2}\, e^{-2\Phi}\,\mathcal{Z}^{-2}\, f_{11}^{(0)}\bigl(B^2 + f_{11}^{(0)2}\bigr) + J_t^{(0)2}}}{B^2 + f_{11}^{(0)2}}
\end{equation}

and the Hall conductivity as,
\begin{equation}
\label{HallBbeta0}
    \sigma_{xy}^{(0)} = \frac{B  J_t^{(0)}}{B^2 + f_{11}^{(0)2}}
\end{equation}
From \cite{OBannon:2007cex, Karch:2007pd}, the resulting effective horizon in the presence of an arbitrary electromagnetic set-up can be obtained from the vanishing of Eq~(\ref{eq:xi0}), which turns out to be,

{\footnotesize
\begin{eqnarray}
    r_{\ast}^{4}=\frac{r_{h}^{4}-4 \pi  B^2 {g_s} N+4 \pi 
   E_{x^{1}}^2 {g_s} N+\sqrt{\left(r_{h}^{4}-4 \pi  B^2 {g_s} N+4 \pi 
   E_{x^{1}}^2 {g_s} N\right)^2+16 \pi  B^2 {g_s} N r_{h}^4}}{2}
\end{eqnarray}
}

\subsection{At $\mathcal{O}(\beta)$:}
In this subsection, we discuss the $\mathcal{O}(\beta)$ corrections to the conductivity components obtained in eq(\ref{OhmicBbeta0}) and (\ref{HallBbeta0}).

 The DBI action in the electromagnetic setup, including contributions up to $\mathcal{O}(\beta)$, can be written as

\begin{align}
    \mathcal{S}_{\text{DBI}}^{\beta} &= -N_{f}T_{D6}\int d^{7}x\frac{e^{-\Phi}}{2\sqrt{X}}
    \Biggl(
        \mathcal{\tilde{I}} \Bigl[
            -3 f_{11}^{(0)2} f_{11}^{(1)} \bigl(A_{t}^{(0)'2} - f_{rr}^{(0)} f_{tt}^{(0)}\bigr)
            + f_{11}^{(1)} \bigl(B^2 f_{rr}^{(0)} f_{tt}^{(0)} - (A_{t}^{(0)'} B - A_{x^{2}}^{(0)'} E_{x^{1}})^2\bigr)
            \notag \\
            &\qquad + 2 f_{11}^{(0)} f_{11}^{(1)} \bigl(f_{tt}^{(0)}(A_{x}^{(0)'2} + A_{y}^{(0)'2}) - E_{x^{1}}^2 f_{rr}^{(0)}\bigr)
            + f_{11}^{(0)2} f_{tt}^{(1)} \bigl(A_{x}^{(0)'2} + A_{y}^{(0)'2} + B^2 f_{rr}^{(0)} + f_{11}^{(0)2} f_{rr}^{(0)}\bigr)
        \Bigr]
        \notag \\
        &\quad - 2\mathcal{I}\, f_{11}^{(0)} \Bigl[
            A_{t}^{(0)'} A_{t}^{(1)'} \bigl(B^2 + f_{11}^{(0)2}\bigr)
            - B E_{x^{1}} \bigl(A_{t}^{(0)'} A_{y}^{(1)'} + A_{t}^{(1)'} A_{x^{2}}^{(0)'}\bigr)
            \notag \\
            &\qquad - f_{11}^{(0)} f_{tt}^{(0)} \bigl(A_{x}^{(0)'} A_{x}^{(1)'} + A_{x^{2}}^{(0)'} A_{y}^{(1)'}\bigr)
            + A_{x^{2}}^{(0)'} A_{y}^{(1)'} E_{x^{1}}^2
        \Bigr]
    \Biggr)
\end{align}

The equations of motion for the gauge fields at $\mathcal{O}(\beta)$, take the form
\begin{subequations}
\label{eq:currents_first_order}
\begin{align}
\Pi_{t}^{(1)} &= \frac{e^{-\Phi}}{2\sqrt{X}}
    \left[
        2\, \mathcal{\tilde{I}}\, f_{11}^{(1)} \left( A_{x^{2}}^{(0)'} B E_{x^{1}} - A_{t}^{(0)'} \bigl(B^2 + 3 f_{11}^{(0)2}\bigr) \right)
        - 2\mathcal{I}\, f_{11}^{(0)}
        \left( A_{t}^{(1)'} \bigl(B^2 + f_{11}^{(0)2}\bigr) - A_{y}^{(1)'} B E_{x^{1}} \right)
    \right] \notag \\
    &\quad - \frac{f_{11}^{(0)}\, e^{-\Phi}}{4\,X^{3/2}}
    \left( 2 A_{x^{2}}^{(0)'} B E_{x^{1}} - 2 A_{t}^{(0)'} \bigl(B^2 + f_{11}^{(0)2}\bigr) \right) \mathcal{N},
    \label{eq:Jt}
\\[8pt]
\Pi_{x^1}^{(1)} &= \frac{f_{11}^{(0)}\, e^{-\Phi}}{\sqrt{X}}
    \left[
        A_{x}^{(0)'}\, \mathcal{\tilde{I}} \bigl(f_{11}^{(0)} f_{tt}^{(1)} + 2 f_{11}^{(1)} f_{tt}^{(0)}\bigr)
        + A_{x}^{(1)'}\mathcal{I}\, f_{11}^{(0)} f_{tt}^{(0)}
    \right] \notag \\
    &\quad - \frac{A_{x}^{(0)'} f_{11}^{(0)2} f_{tt}^{(0)}\, e^{-\Phi}}{2\,X^{3/2}}\, \mathcal{N},
    \label{eq:Jx1}
\\[8pt]
\Pi_{x^2}^{(1)} &= \frac{e^{-\Phi}}{2\sqrt{X}}
    \left[
        \mathcal{\tilde{I}} \Bigl(
            2 f_{11}^{(1)} E_{x^{1}} \bigl(A_{t}^{(0)'} B - A_{x^{2}}^{(0)'} E_{x^{1}}\bigr)
            + 2 A_{x^{2}}^{(0)'} f_{11}^{(0)2} f_{tt}^{(1)}
            + 4 A_{x^{2}}^{(0)'} f_{11}^{(0)} f_{11}^{(1)} f_{tt}^{(0)}
        \Bigr) \right. \notag \\
    &\qquad \left.
        + 2\mathcal{I}\, f_{11}^{(0)}
        \Bigl( A_{t}^{(1)'} B E_{x^{1}} - A_{y}^{(1)'} E_{x^{1}}^2 + A_{y}^{(1)'} f_{11}^{(0)} f_{tt}^{(0)} \Bigr)
    \right] \notag \\
    &\quad - \frac{f_{11}^{(0)}\, e^{-\Phi}}{4\,X^{3/2}}
    \Bigl( 2 E_{x^{1}} \bigl(A_{t}^{(0)'} B - A_{x^{2}}^{(0)'} E_{x^{1}}\bigr)
    + 2 A_{x^{2}}^{(0)'} f_{11}^{(0)} f_{tt}^{(0)} \Bigr)\, \mathcal{N},
    \label{eq:Jx2}
\end{align}
\end{subequations}

where $\mathcal{N}$ is defined as,

\begin{align}
\mathcal{N} &\equiv \mathcal{\tilde{I}} \Bigl[
    -3 f_{11}^{(0)2} f_{11}^{(1)} \bigl(A_{t}^{(0)'2} - f_{rr}^{(0)} f_{tt}^{(0)}\bigr)
    + f_{11}^{(1)} \bigl(B^2 f_{rr}^{(0)} f_{tt}^{(0)} - (A_{t}^{(0)'} B - A_{x^{2}}^{(0)'} E_{x^{1}})^2\bigr)
    \notag \\
    &\qquad + 2 f_{11}^{(0)} f_{11}^{(1)} \bigl(f_{tt}^{(0)}(A_{x}^{(0)'2} + A_{y}^{(0)'2}) - E_{x^{1}}^2 f_{rr}^{(0)}\bigr)
    + f_{11}^{(0)2} f_{tt}^{(1)} \bigl(A_{x}^{(0)'2} + A_{y}^{(0)'2}
    + B^2 f_{rr}^{(0)} + f_{11}^{(0)2} f_{rr}^{(0)}\bigr)
\Bigr] \notag \\
&\quad - 2\mathcal{I}\, f_{11}^{(0)} \Bigl[
    A_{t}^{(0)'} A_{t}^{(1)'} \bigl(B^2 + f_{11}^{(0)2}\bigr)
    - B E_{x^{1}} \bigl(A_{t}^{(0)'} A_{y}^{(1)'} + A_{t}^{(1)'} A_{x^{2}}^{(0)'}\bigr)
    \notag \\
    &\qquad - f_{11}^{(0)} f_{tt}^{(0)} \bigl(A_{x}^{(0)'} A_{x}^{(1)'} + A_{x^{2}}^{(0)'} A_{y}^{(1)'}\bigr)
    + A_{x^{2}}^{(0)'} A_{y}^{(1)'} E_{x^{1}}^2
\Bigr].
\label{eq:calN}
\end{align}
Similar to the $\mathcal{O}(\beta^{0})$ case, the radial derivatives of the gauge-field components at $\mathcal{O}(\beta)$ can be written as
{\footnotesize
\begin{subequations}
\label{eq:gauge_fields_first_order}
\begin{align}
A_{t}^{(1)'} &= \frac{2X\left(
        A_{t}^{(0)'2} f_{11}^{(0)} \Pi_t^{(1)}
        + A_{t}^{(0)'} f_{11}^{(0)} \bigl(A_{x}^{(0)'} \Pi_{x^1}^{(1)} + A_{x^{2}}^{(0)'} \Pi_{x^2}^{(1)}\bigr)
        + f_{rr}^{(0)} \bigl(B E_{x^{1}} \Pi_{x^2}^{(1)} + E_{x^{1}}^2 \Pi_t^{(1)} - f_{11}^{(0)} f_{tt}^{(0)} \Pi_t^{(1)}\bigr)
    \right)}{2\mathcal{I}\, f_{11}^{(0)2} f_{rr}^{(0)}\, e^{-\Phi}
    \left(f_{tt}^{(0)}\bigl(B^2 + f_{11}^{(0)2}\bigr) - E_{x^{1}}^2 f_{11}^{(0)}\right)}
    \notag \\
    &\quad + \frac{\mathcal{\tilde{I}}\, f_{11}^{(0)} \Bigl(
        A_{t}^{(0)'3} f_{11}^{(1)}\bigl(B^2 + 3 f_{11}^{(0)2}\bigr)
        - 2 A_{t}^{(0)'2} A_{x^{2}}^{(0)'} B E_{x^{1}} f_{11}^{(1)}
        + A_{t}^{(0)'} \,\mathcal{P}_t
        - 2 A_{x^{2}}^{(0)'} B E_{x^{1}} f_{rr}^{(0)}\bigl(f_{11}^{(0)} f_{tt}^{(1)} + f_{11}^{(1)} f_{tt}^{(0)}\bigr)
    \Bigr)}{2\mathcal{I}\, f_{11}^{(0)2} f_{rr}^{(0)}
    \left(f_{tt}^{(0)}\bigl(B^2 + f_{11}^{(0)2}\bigr) - E_{x^{1}}^2 f_{11}^{(0)}\right)},
    \label{eq:At1}
\\[10pt]
A_{x^1}^{(1)'} &= \frac{2X\left(
        A_{t}^{(0)'} A_{x}^{(0)'} f_{11}^{(0)} f_{tt}^{(0)} \Pi_t^{(1)}
        + f_{tt}^{(0)}\bigl(
            A_{x}^{(0)'2} f_{11}^{(0)} \Pi_{x^1}^{(1)}
            + A_{x}^{(0)'} A_{x^{2}}^{(0)'} f_{11}^{(0)} \Pi_{x^2}^{(1)}
            + f_{rr}^{(0)} \Pi_{x^1}^{(1)}\bigl(B^2 + f_{11}^{(0)2}\bigr)
        \bigr)
        - E_{x^{1}}^2 f_{11}^{(0)} f_{rr}^{(0)} \Pi_{x^1}^{(1)}
    \right)}{2\mathcal{I}\, f_{11}^{(0)2} f_{rr}^{(0)} f_{tt}^{(0)}\, e^{-\Phi}
    \left(f_{tt}^{(0)}\bigl(B^2 + f_{11}^{(0)2}\bigr) - E_{x^{1}}^2 f_{11}^{(0)}\right)}
    \notag \\
    &\quad - \frac{\mathcal{\tilde{I}}\, A_{x}^{(0)'} f_{11}^{(0)}\, \mathcal{Q}}{2\mathcal{I}\,
    f_{11}^{(0)2} f_{rr}^{(0)} f_{tt}^{(0)}
    \left(f_{tt}^{(0)}\bigl(B^2 + f_{11}^{(0)2}\bigr) - E_{x^{1}}^2 f_{11}^{(0)}\right)},
    \label{eq:Ax1}
\\[10pt]
A_{x^2}^{(1)'} &= \frac{2X\left(
        A_{t}^{(0)'} A_{x^{2}}^{(0)'} f_{11}^{(0)} \Pi_t^{(1)}
        + f_{11}^{(0)}\bigl(
            A_{x}^{(0)'} A_{x^{2}}^{(0)'} \Pi_{x^1}^{(1)}
            + A_{y}^{(0)'2} \Pi_{x^2}^{(1)}
            + f_{11}^{(0)} f_{rr}^{(0)} \Pi_{x^2}^{(1)}
        \bigr)
        + B^2 f_{rr}^{(0)} \Pi_{x^2}^{(1)}
        + B E_{x^{1}} f_{rr}^{(0)} \Pi_t^{(1)}
    \right)}{2\mathcal{I}\, f_{11}^{(0)2} f_{rr}^{(0)}\, e^{-\Phi}
    \left(f_{tt}^{(0)}\bigl(B^2 + f_{11}^{(0)2}\bigr) - E_{x^{1}}^2 f_{11}^{(0)}\right)}
    \notag \\
    &\quad - \frac{\mathcal{\tilde{I}}\, f_{11}^{(0)}\, \mathcal{R}}{2\mathcal{I}\,
    f_{11}^{(0)2} f_{rr}^{(0)}
    \left(f_{tt}^{(0)}\bigl(B^2 + f_{11}^{(0)2}\bigr) - E_{x^{1}}^2 f_{11}^{(0)}\right)},
    \label{eq:Ay1}
\end{align}
\end{subequations}

}
where the shorthand symbols $\mathcal{P}_t \, \mathcal{Q}$, and  $\mathcal{R} $ are defined as 

\begin{align}
\mathcal{P}_t &\equiv
    - A_{x}^{(0)'2} f_{11}^{(0)2} f_{tt}^{(1)}
    - 2 A_{x}^{(0)'2} f_{11}^{(0)} f_{11}^{(1)} f_{tt}^{(0)}
    + A_{y}^{(0)'2}\bigl(E_{x^{1}}^2 f_{11}^{(1)} - f_{11}^{(0)}\bigl(f_{11}^{(0)} f_{tt}^{(1)} + 2 f_{11}^{(1)} f_{tt}^{(0)}\bigr)\bigr)
    \notag \\
    &\quad + B^2 f_{11}^{(0)} f_{rr}^{(0)} f_{tt}^{(1)}
    - B^2 f_{11}^{(1)} f_{rr}^{(0)} f_{tt}^{(0)}
    + 4 E_{x^{1}}^2 f_{11}^{(0)} f_{11}^{(1)} f_{rr}^{(0)}
    + f_{11}^{(0)3} f_{rr}^{(0)} f_{tt}^{(1)}
    - 3 f_{11}^{(0)2} f_{11}^{(1)} f_{rr}^{(0)} f_{tt}^{(0)},
\label{eq:calPt}
\\[6pt]
\mathcal{Q} &\equiv
    - A_{t}^{(0)'2} f_{11}^{(1)} f_{tt}^{(0)}\bigl(B^2 + 3 f_{11}^{(0)2}\bigr)
    + 2 A_{t}^{(0)'} A_{x^{2}}^{(0)'} B E_{x^{1}} f_{11}^{(1)} f_{tt}^{(0)}
    + A_{x}^{(0)'2} f_{11}^{(0)2} f_{tt}^{(0)} f_{tt}^{(1)}
    + 2 A_{x}^{(0)'2} f_{11}^{(0)} f_{11}^{(1)} f_{tt}^{(0)2}
    \notag \\
    &\quad + A_{y}^{(0)'2} f_{tt}^{(0)}\bigl(
        - E_{x^{1}}^2 f_{11}^{(1)}
        + f_{11}^{(0)2} f_{tt}^{(1)}
        + 2 f_{11}^{(0)} f_{11}^{(1)} f_{tt}^{(0)}
    \bigr)
    + B^2 f_{11}^{(0)} f_{rr}^{(0)} f_{tt}^{(0)} f_{tt}^{(1)}
    + 3 B^2 f_{11}^{(1)} f_{rr}^{(0)} f_{tt}^{(0)2}
    \notag \\
    &\quad - 2 E_{x^{1}}^2 f_{11}^{(0)2} f_{rr}^{(0)} f_{tt}^{(1)}
    - 2 E_{x^{1}}^2 f_{11}^{(0)} f_{11}^{(1)} f_{rr}^{(0)} f_{tt}^{(0)}
    + f_{11}^{(0)3} f_{rr}^{(0)} f_{tt}^{(0)} f_{tt}^{(1)}
    + f_{11}^{(0)2} f_{11}^{(1)} f_{rr}^{(0)} f_{tt}^{(0)2},
\label{eq:calQ}
\\[6pt]
\mathcal{R} &\equiv
    - A_{t}^{(0)'2} A_{x^{2}}^{(0)'} f_{11}^{(1)}\bigl(B^2 + 3 f_{11}^{(0)2}\bigr)
    + 2 A_{t}^{(0)'} B E_{x^{1}} f_{11}^{(1)}\bigl(A_{y}^{(0)'2} - 2 f_{11}^{(0)} f_{rr}^{(0)}\bigr)
    \notag \\
    &\quad + A_{x^{2}}^{(0)'}\Bigl[
        A_{x}^{(0)'2} f_{11}^{(0)}\bigl(f_{11}^{(0)} f_{tt}^{(1)} + 2 f_{11}^{(1)} f_{tt}^{(0)}\bigr)
        + A_{y}^{(0)'2}\bigl(f_{11}^{(0)}\bigl(f_{11}^{(0)} f_{tt}^{(1)} + 2 f_{11}^{(1)} f_{tt}^{(0)}\bigr) - E_{x^{1}}^2 f_{11}^{(1)}\bigr)
        \notag \\
    &\qquad + B^2 f_{rr}^{(0)}\bigl(f_{11}^{(0)} f_{tt}^{(1)} + 3 f_{11}^{(1)} f_{tt}^{(0)}\bigr)
        + f_{11}^{(0)2} f_{rr}^{(0)}\bigl(f_{11}^{(0)} f_{tt}^{(1)} + f_{11}^{(1)} f_{tt}^{(0)}\bigr)
    \Bigr].
\label{eq:calR}
\end{align}

Similar to the $B=0$ case, the constraint equations obtained at $\mathcal{O}(\beta^{0})$ vanish at the effective horizon $r=r_{\ast}$, they can be used to perturbatively expand the $\mathcal{O}(\beta)$ Lagrangian in the vicinity of $r_{\ast}$. In particular, one takes $\chi^{(0)},\xi^{(0)}\sim C\epsilon$, where $\epsilon\equiv r-r_{\ast}\to 0$ and $C$ denotes a finite constant determined by the corresponding first radial derivative evaluated at the effective horizon, namely $\partial_r\chi^{(0)}|_{r=r_{\ast}}$ or $\partial_r\xi^{(0)}|_{r=r_{\ast}}$. This follows directly from the Taylor expansions $\chi^{(0)}(r)=\chi^{(0)}(r_{\ast})+\partial_r\chi^{(0)}|_{r=r_{\ast}}(r-r_{\ast})+\mathcal{O}((r-r_{\ast})^2)$ and, similarly, $\xi^{(0)}(r)=\xi^{(0)}(r_{\ast})+\partial_r\xi^{(0)}|_{r=r_{\ast}}(r-r_{\ast})+\mathcal{O}((r-r_{\ast})^2)$. Since the leading-order constraint equations imply $\chi^{(0)}(r_{\ast})=\xi^{(0)}(r_{\ast})=0$, both functions ($\chi^{(0)}$, and $\xi^{(0)}$) scale linearly with $\epsilon$ near the effective horizon at the leading order $\mathcal{O}(r-r_{\ast})$. Consequently, the numerator and denominator of the DBI Lagrangian at $\mathcal{O}(\beta)$ are
{\footnotesize
\begin{align}
    \mathcal{A} &= (\partial_{r}\chi^{(0)}|_{r=r_{\ast}}\epsilon)^{3/2}\,\sqrt{\partial_{r}\xi^{(0)}|_{r=r_{\ast}}\epsilon}\, E_{x^{1}}^4\, f_{11}^{(0)2}\,  \mathcal{\tilde{I}}\, 
     e^{-\Phi}
    \left(
        B^4 f_{tt}^{(1)}
        + B^2 \left(2 f_{11}^{(0)} f_{tt}^{(1)} - E_{x^{1}}^2 f_{11}^{(1)}\right)
        + E_{x^{1}}^2 f_{11}^{(0)2} f_{11}^{(1)}
        + f_{11}^{(0)4} f_{tt}^{(1)}
    \right)
    \notag\\
    &\quad \times\sqrt{f_{rr}^{(0)} \left( f_{11}^{(0)2} \Pi_{x^1}^{(0)2}-B^2 \Pi_{x^2}^{(0)2}- 2 B E_{x^{1}} f_{11}^{(0)} \Pi_t^{(0)} \Pi_{x^2}^{(0)}- E_{x^{1}}^2 f_{11}^{(0)2} \Pi_t^{(0)2}\right)}
\end{align}
}
and 
\begin{equation}
    \mathcal{B}=-2\,(\partial_{r}\chi^{(0)}|_{r=r_{\ast}}\epsilon)\, f_{tt}^{(0)}\,(\partial_{r}\xi^{(0)}|_{r=r_{\ast}}\epsilon)^{2}
\end{equation}

Hence, the DBI Lagrangian at $\mathcal{O}(\beta)$, can be written as
{\footnotesize
\begin{eqnarray}
\label{finalldbibeta}
\frac{-\mathcal{L}^{(1)}_{DBI}}{N_{f}T_{D6}}
&=&\Biggl(\frac{\partial_r\chi^{(0)}\Big|_{r=r_{\ast}}}{
\partial_r\xi^{(0)}\Big|_{r=r_{\ast}}\,}\Biggr)^{3/2}\Biggl(\frac{\xi^{(1)} E_{x^{1}}^4
f_{11}^{(0)2}\,
\tilde{\mathcal{I}}\,
\,e^{-\Phi}\sqrt{f_{rr}^{(0)} \left( f_{11}^{(0)2} \Pi_{x^1}^{(0)2}-B^2 \Pi_{x^2}^{(0)2}- 2 B E_{x^{1}} f_{11}^{(0)} \Pi_t^{(0)} \Pi_{x^2}^{(0)}- E_{x^{1}}^2 f_{11}^{(0)2} \Pi_t^{(0)2}\right)}}{-2\,
f_{tt}^{(0)}\left(\partial_r\chi^{(0)}\Big|_{r=r_{\ast}}\right)\epsilon}\Biggr).\nonumber\\
\end{eqnarray}
}
where,
\begin{equation}
\label{xibeta}
\xi^{(1)}=\left(
B^4f_{tt}^{(1)}
+B^2\left(2f_{11}^{(0)}f_{tt}^{(1)}
-E_{x^{1}}^2f_{11}^{(1)}\right)
+E_{x^{1}}^2f_{11}^{(0)2}f_{11}^{(1)}
+f_{11}^{(0)4}f_{tt}^{(1)}
\right)
\end{equation}

We note that at the leading-order effective horizon $r=r_*$, the O'Bannon reality conditions require $\xi^{(0)}(r_*)=\chi^{(0)}(r_*)=0$, so that the ratio $\chi^{(0)}/\xi^{(0)}$ is formally indeterminate at $r=r_*$. Since the $\mathcal{O}(\beta)$ expansion is carried out in a small neighborhood of the effective horizon, both functions admit regular Taylor expansions in $(r-r_*)$, and the ratio $\xi^{(0)}/\chi^{(0)}$ consequently approaches a finite limit, $\partial_r\chi^{(0)}|_{r_*}/\partial_r\xi^{(0)}|_{r_*}$, provided the regularity, i.e.\ $\partial_r\xi^{(0)}|_{r_*}\neq 0$. This has a direct bearing on the structure of the DBI Lagrangian, writing $r_*-r=\epsilon$, the relevant term takes the schematic form $L_{\rm DBI}^{\beta}\sim\Biggl({\frac{\partial_{r}\chi^{(0)}|_{r=r_{\ast}^{(0)}}}{\partial_{r}\xi^{(0)}|_{r=r_{\ast}^{(0)}}}}\,\Biggr)^{3/2}\frac{\xi^{(1)}\mathcal{T}(r,E_{x^{1}})}{(\partial_{r}\xi^{(0)}|_{r=r_{\ast}^{(0)}})\epsilon}$,(where $\mathcal{T}(r,E_{x^{1}})$ includes remaining multiplicative terms of Eq(\ref{finalldbibeta})) which develops a potential $1/\epsilon$ divergence as $r\to r_*$ unless the numerator vanishes correspondingly. Demanding that $L_{\rm DBI}$ admit a convergent, non-singular expansion near the effective horizon therefore requires $\xi^{(1)}\big(r_{\ast}^{(0)}+\beta\, r_{\ast}^{\beta}\big)=0$.

Inspection of eq. (\ref{xibeta}) shows that this quantity must vanish at least perturbatively at $\mathcal{O}(\beta)$. The effective horizon may therefore be written as
$r = r_{\ast}^{(0)} + \beta, r_{\ast}^{(1)}$.
As in the $B=0$ case, we now work with an arbitrary magnetic field $B$ and a small electric field $E_{x^{1}}$ to determine the corrected horizon at this order. The resulting shift in the horizon then produces higher-derivative corrections to the conductivity, which take the form $
    \sigma(r) = \sigma^{(0)}\!\left(r_{\ast}^{(0)}\right) + \delta r \cdot \partial_r \sigma^{(0)}\!\left(r\right)\bigg|_{r = r_{\ast}^{(0)}} $, where $\delta r=\beta r^{\beta}_{\ast}$.

Now, from eq(\ref{xibeta}), one obtains the following expression
\begin{align}
    \xi^{(1)} &= -\frac{\mathcal{C}\,(r - 2r_h)}{r}
    \Bigl(
        16\pi^2 B^4 g_s^2 N^2 \left(r^4 - r_h^4\right)
        \notag \\
        &\quad + 8\pi B^2 g_s N r^4 \left(2\pi E_{x^{1}}^2 g_s N + r^4 - r_h^4\right)
        \notag \\
        &\quad + r^8\left(-4\pi E_{x^{1}}^2 g_s N + r^4 - r_h^4\right)
    \Bigr)
\end{align}
where $\mathcal{C}$ denotes the overall multiplicative factor defined as,
\begin{equation}
    \mathcal{C}=-\frac{27\, b^{10} \left(9b^2+1\right)^4 M \left(\frac{1}{N}\right)^{15/4} 
    \left(6a^2 + r_h^2\right) \log^3(r_h)}{64\, \pi^{7/2} \left(3b^2-1\right)^5 
    \left(6b^2+1\right)^4 g_s^{5/2}\, \log(N)^4\, N_f\, r_h^4 \left(9a^2 + r_h^2\right)}
\end{equation}

and, the radial derivative of $\xi^{\beta}$ is

\begin{align}
    \partial_r \xi^{(1)} &= \frac{2\mathcal{C}}{r^2}
    \Bigl(
        16\pi^2 B^4 g_s^2 N^2 \left(-2r^5 + 3r^4 r_h + r_h^5\right)
        \notag \\
        &\quad - 8\pi B^2 g_s N r^4 \Bigl(
            2\pi E_{x^{1}}^2 g_s N (2r - 3r_h)
            + 4r^5 - 7r^4 r_h - 2r\, r_h^4 + 3r_h^5
        \Bigr)
        \notag \\
        &\quad + r^8 \Bigl(
            4\pi E_{x^{1}}^2 g_s N (4r - 7r_h)
            - 6r^5 + 11r^4 r_h + 4r\, r_h^4 - 7r_h^5
        \Bigr)
    \Bigr)
\end{align}
from, $r_{\ast}^{(\beta)}=\delta r^{1}=-\frac{\xi^{(\beta)}}{\partial_{r}\xi^{(\beta)}}$, one obtains the $\mathcal{O}(\beta)$ correction in the effective horizon, which in the weak electric field limit takes the form, 

\begin{equation}
\label{rstarbeta}
    r^{\beta}_{\ast} = -\frac{8\pi^2 B^2 E_{x^{1}}^2 g_s^2 N^2 r_h}{\left(4\pi B^2 g_s N + r_h^4\right)^2}
\end{equation}
The correction is found to be purely magnetic, and thus remains consistent with the result obtained in the $B=0$ limit, wherein the effective horizon does not receive any $\mathcal{O}(\beta)$ corrections.

\section{Numerical Investigations}
\label{NDC}
\begin{itemize}

\item \textbf{Inverse Magnetic Catalysis}:
The expressions obtained for the effective horizons, in the presence and absence of a magnetic field, respectively, are:
\begin{equation}
\label{effhorizon}
r_{\ast}^{(0)}(B) = r_h \left( 1 + \frac{E_{x^{1}}^2}{4 g_s N \pi B^2 + 9 \pi^5 g_s N T^4} \right), \quad r_{\ast}^{(0)}(0) = r_h \left( 1 + \frac{E_{x^{1}}^2}{9 \pi^5 g_s N T^4} \right) .
\end{equation}
where, $T=\frac{r_{h}}{\sqrt{3}\pi^{3/2}\sqrt{gs_{s}}\sqrt{N}}$ \cite{Yadav:2022qcl}. 
The effective horizon $r_\ast$ describes an effective temperature shift within the non-equilibrium state, just as the black hole horizon $r_h$ is connected to the Hawking temperature $T$ in the dual field theory. In the weak electric field limit, expanding Eq(\ref{effhorizon}) yields:
\begin{equation}
\label{IMC}
T_{\text{eff}}(B) = T_{\text{eff}}(B=0) \left( 1 - \frac{4 B^2 E_{x^{1}}^2}{81 \pi^{10} g_s^2 N^2 T^8} \right) ,
\end{equation} 
This finding shows that as the magnetic field $B$ increases, the dual field theory's effective temperature drops. This magnetic field-induced suppression of the de-confinement temperature is a transport manifestation of the phenomenon called \textit{Inverse Magnetic Catalysis}.

\item \textbf{Density part contribution:} The charge density-dependent part contribution to the conductivity is given by :

\begin{equation}    
\sigma_{\text{density}}^{\beta^{0}}\Big|_{\text{small}\,E_{x^1},\,\text{arbitrary}\,B} = 
    \frac{6\pi^{5/2} \sqrt{g_s}\, J_t^{(0)} \sqrt{N}\, T^2}{4B^2 + 9\pi^5 g_s N T^4}+\frac{12\pi^{5/2} E_{x^{1}}^2 \sqrt{g_s}\, J_t^{(0)} \sqrt{N} \left(4B^2 T^2 - 9\pi^5 g_s N T^6\right)}
    {\left(4B^2 + 9\pi^5 g_s N T^4\right)^3}
\end{equation}

    Considering the small $E_{x^{1}}$ expansion of the same, then the small B expansion, one gets 

\begin{equation}
\label{U1num}
\begin{aligned}
\sigma_{\text{density}}^{\beta^{0}} &= \frac{2 J_{t}^{(0)}}{3 \pi^{5/2} \sqrt{g_s N} T^{2}} - \frac{4 J_{t}^{(0)} E_{x^{1}}^{2}}{27 \pi^{15/2} (g_s N)^{3/2} T^{6}} \\
&\quad + \left[ -\frac{8 J_{t}^{(0)}}{27 \pi^{15/2} (g_s N)^{3/2} T^{6}} + \frac{64 J_{t}^{(0)} E_{x^{1}}^{2}}{243 \pi^{25/2} (g_s N)^{5/2} T^{10}} \right] B^{2} + \mathcal{O}(E_{x^{1}}^4, B^4) \, .
\end{aligned}
\end{equation}

In this expression, the leading-order term represents the Ohmic (or linear) contribution to the conductivity. This term is independent of both the electric and magnetic fields and exhibits a temperature dependence of $T^{-2}$. This is analogous to Drude-like behavior, where conductivity decreases with increasing temperature due to the enhancement of thermal fluctuations and momentum relaxation among charge carriers. 

The next-to-leading order (NLO) correction, appearing at $\mathcal{O}(E_{x^{1}}^2)$, serves to decrease the overall conductivity. This suppression can be understood through the expression for the effective horizon Eq~(\ref{effhorizon}): the application of the external electric field $E_{x^{1}}$ "stretches" the worldvolume horizon. This geometric deformation promotes a shift toward a higher effective temperature, which increases the drag force and consequently reduces the mobility of the charge carriers. Furthermore, the $B^2$ terms demonstrate that the magnetic field modulates this non-linear response, a behavior consistent with the dynamics of inverse magnetic catalysis; see eq(\ref{IMC}). Finally, we note that these non-linear corrections are suppressed by higher powers of $N$, implying they become negligible in the strict large-$N$ limit.

At $\mathcal{O}(\beta)$, the conductivity in the limit of weak electric and magnetic fields is given by:

\begin{equation}
\sigma_{\text{density}}^{\beta} = \frac{32 \, B^{2} E_{x^{1}}^{2}  J_{t}^{(0)}}{243 \, \pi^{25/2} (g_s N)^{5/2} T^{10}} + \mathcal{O}(E_{x^{1}}^4, B^4) \, .
\end{equation}

The $\mathcal{O}(\beta)$ contribution to the density-dependent conductivity is significantly suppressed in the weak-field regime. This suppression is governed by a $(\sqrt{N} T^{2})^{-5}$ power-law scaling, which indicates that both the large-$N$ limit and thermal fluctuations effectively dampen the non-linear transport corrections at this order. Physically, this suggests that the non-linear response is a sub-leading effect that is rapidly washed out as the temperature increases or as the system approaches the strict planar limit.

\begin{figure}[h]
    \centering
    \subfigure[]{
        \includegraphics[width=0.45\textwidth]{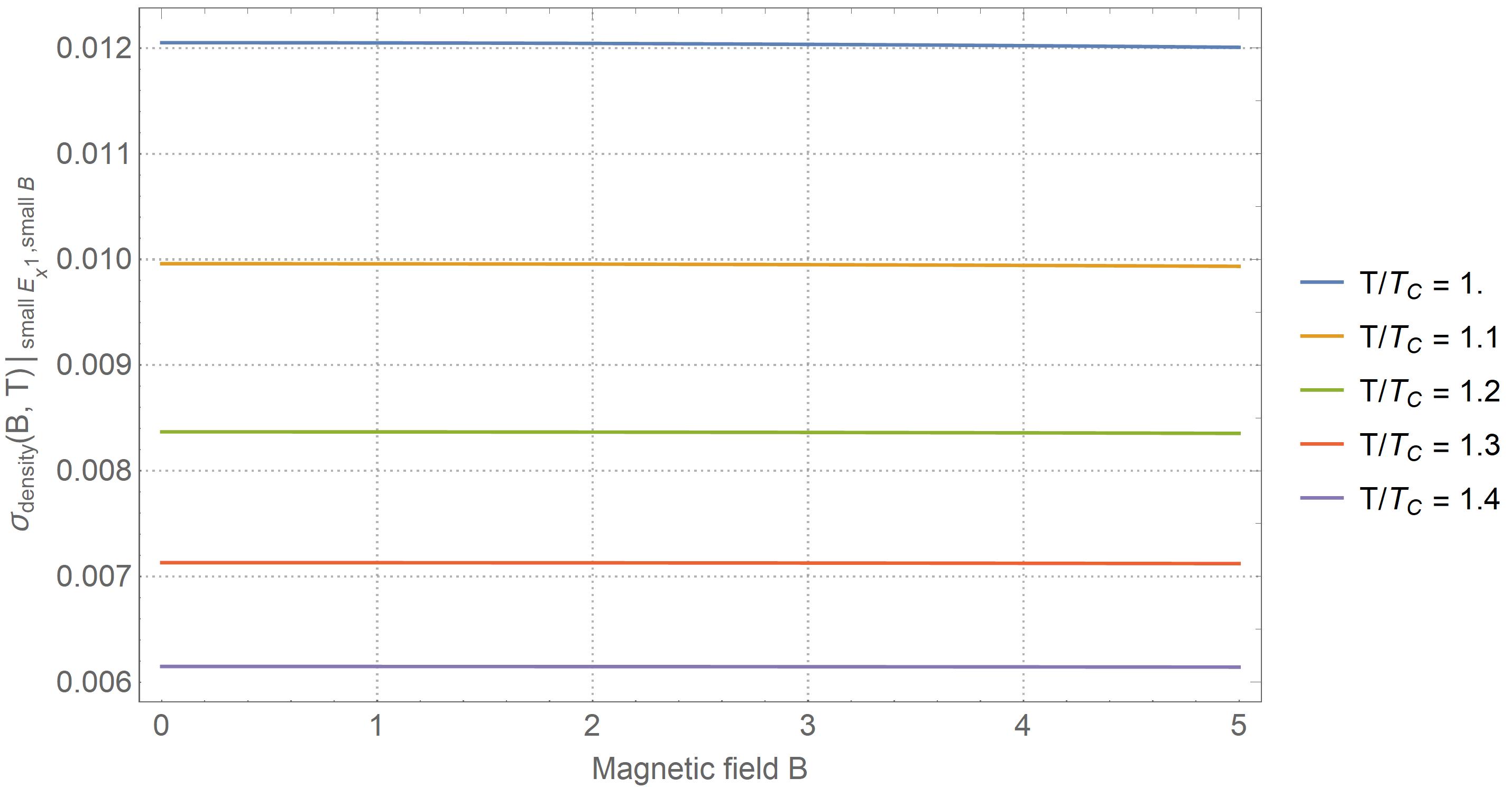}
        \label{fig:density_vs_T}
    }
    \hfill
    \subfigure[]{
        \includegraphics[width=0.45\textwidth]{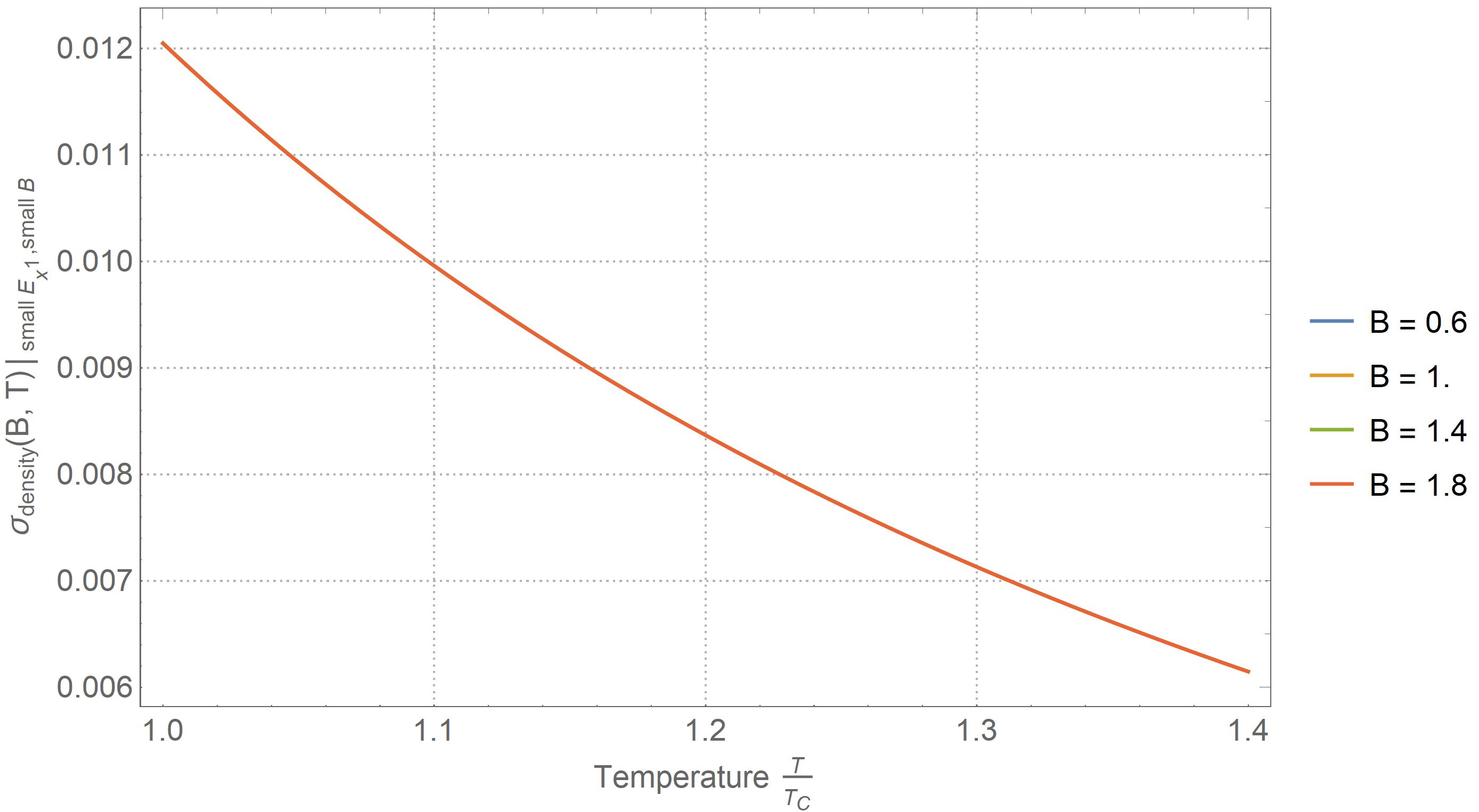}
        \label{fig:density_vs_T}
    }
    \caption{(a) Changes in the density-dependent conductivity $\sigma_{\text{density}}$ at constant temperature and fixed $E_x1=0.5$ with the external magnetic field $B$. The comparatively flat profiles show that in this regime, electromagnetic effects on charge transport are subdominant. 
(b) $\sigma_{\text{density}}$'s temperature dependency at a constant electric field $E_x1 = 0.5$. As $\sigma_{\text{density}} \propto T^{-2}$, the power law describes the decrease in conductivity, which is qualitatively typical of a Drude-like fluid where thermal fluctuations increase momentum dissipation. We use the word  Drude-like fluid to characterise the inverse relationship between conductivity and temperature, despite the fact that this scaling is different from the traditional Drude scaling. The magnetic field has very little effect in this limit, which causes the trajectories for different values of $B$ to significantly overlap.}
\end{figure}
\item \textbf{Hall Conductivity:}
As shown in the previous computations, the Hall conductivity turns out to be,
\begin{equation}
    \sigma_{xy}=\frac{B J_{t}^{(0)}}{B^2+(f_{11}^{(0)})^2}
\end{equation}
using the metric components up to $\mathcal{O}(\beta^{0})$, one gets the following expression for the Hall conductivity in the weak electric field limit,
\begin{equation}
\label{HallEx0arbB}
\sigma_{xy}^{\beta^{0}}=\frac{4 B  J_{t}^{(0)}}{4 B^2+9 \pi ^5 g_{s} N T^4}-\frac{144 \pi ^8 B E_{x^{1}}^2 g_{s}^4  J_{t}^{(0)} N^4
   T^4}{\left(4 \pi  B^2 g_{s} N+9 \pi ^6 g_{s}^2 N^2 T^4\right)^3}
\end{equation}

In the weak magnetic field, the Hall conductivity becomes
\begin{equation}
    \sigma_{xy}^{\beta^{0}}=\frac{4 B J_{t}^{(0)}}{9 \pi ^5 g_{s} N T^4}
\end{equation}

In the small magnetic-field regime, the Hall conductivity is found to be directly proportional to the magnetic field, which is the characteristic signature of the Hall effect. In contrast, in the weak electric field and large magnetic field limit, the behavior changes qualitatively. Here we define the large magnetic-field scale as $B \approx \frac{3\pi^{5/2}\sqrt{g_s N}\,T^{2}}{2}$, which represents a parametrically large magnetic field in the holographic setup. Numerically, at the de-confinement temperature $T=T_c$, this corresponds to $B \approx 83\,T_c^{2}$ (for $g_s=0.1,N=100$) indicating an extremely strong magnetic field scale within the model. In this extremely large magnetic field regime, we have $ \sigma_{xy}=\frac{J_{t}^{0}}{B}$  which is inversely proportional to B, analogous to the classical Hall effect (i.e. $\sigma_{xy} = \frac{nq}{e B}$, in the units of $e=1$) which mimics the zero temperature limit explored in \cite{Bergman:2010gm}. Such large field strengths ($B \approx 83\,T_c^{2}$) exceed those typically realized in astrophysical environments, including strongly magnetized compact objects such as neutron stars and magnetars. Consequently, we do not attempt a detailed phenomenological interpretation of this extremely large magnetic-field regime. Instead, in the remainder of this work we focus on magnetic-field strengths in the phenomenologically relevant range near the deconfinement temperature, corresponding to values accessible in heavy-ion collision experiments such as those at RHIC, as well as in astrophysical environments including strongly magnetized compact objects such as neutron stars.

\begin{figure}[h]
    \centering
    \subfigure[]{
        \includegraphics[width=0.45\textwidth]{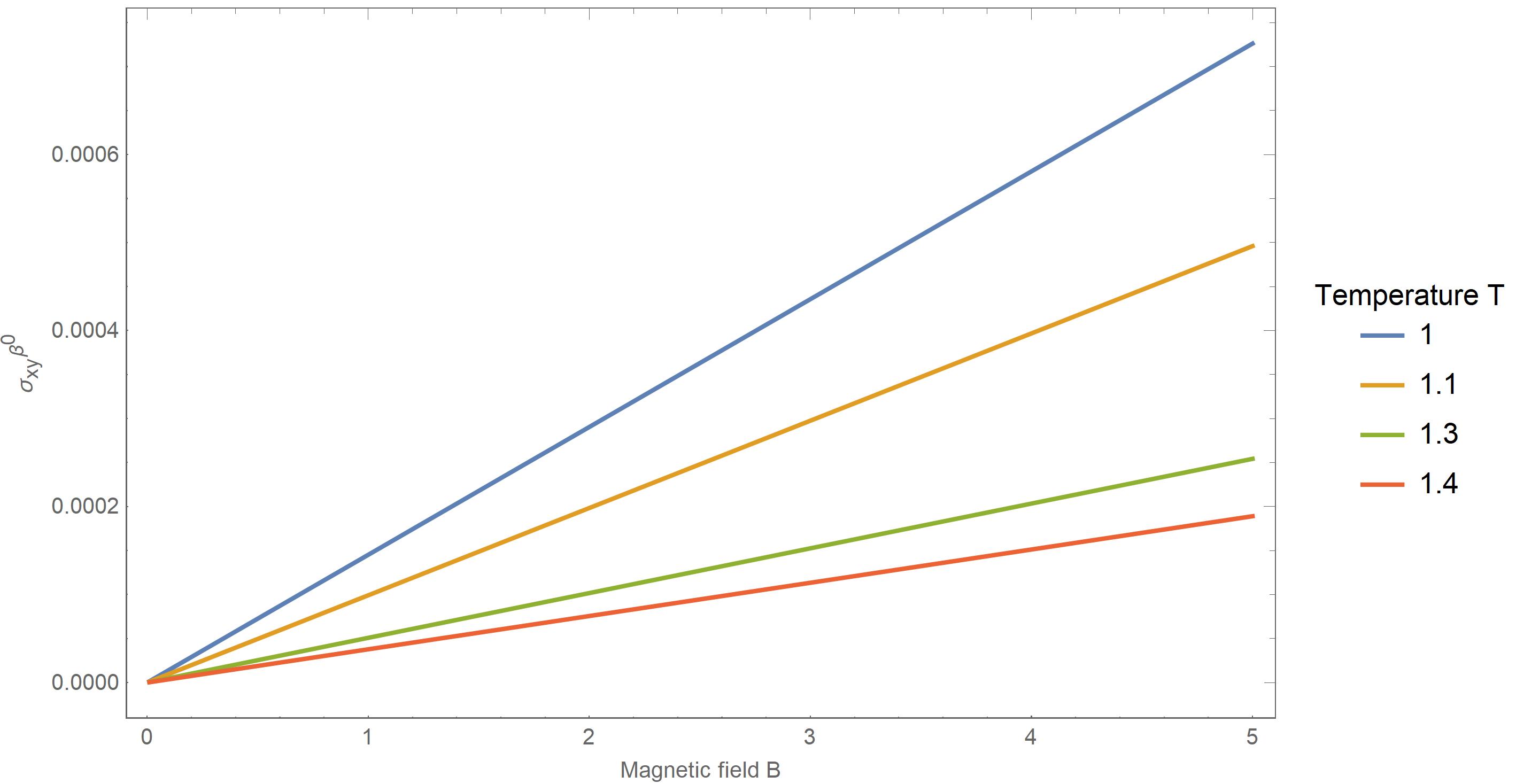}
        \label{fig:HAll_vs_B}
    }
    \hfill
    \subfigure[]{
        \includegraphics[width=0.45\textwidth]{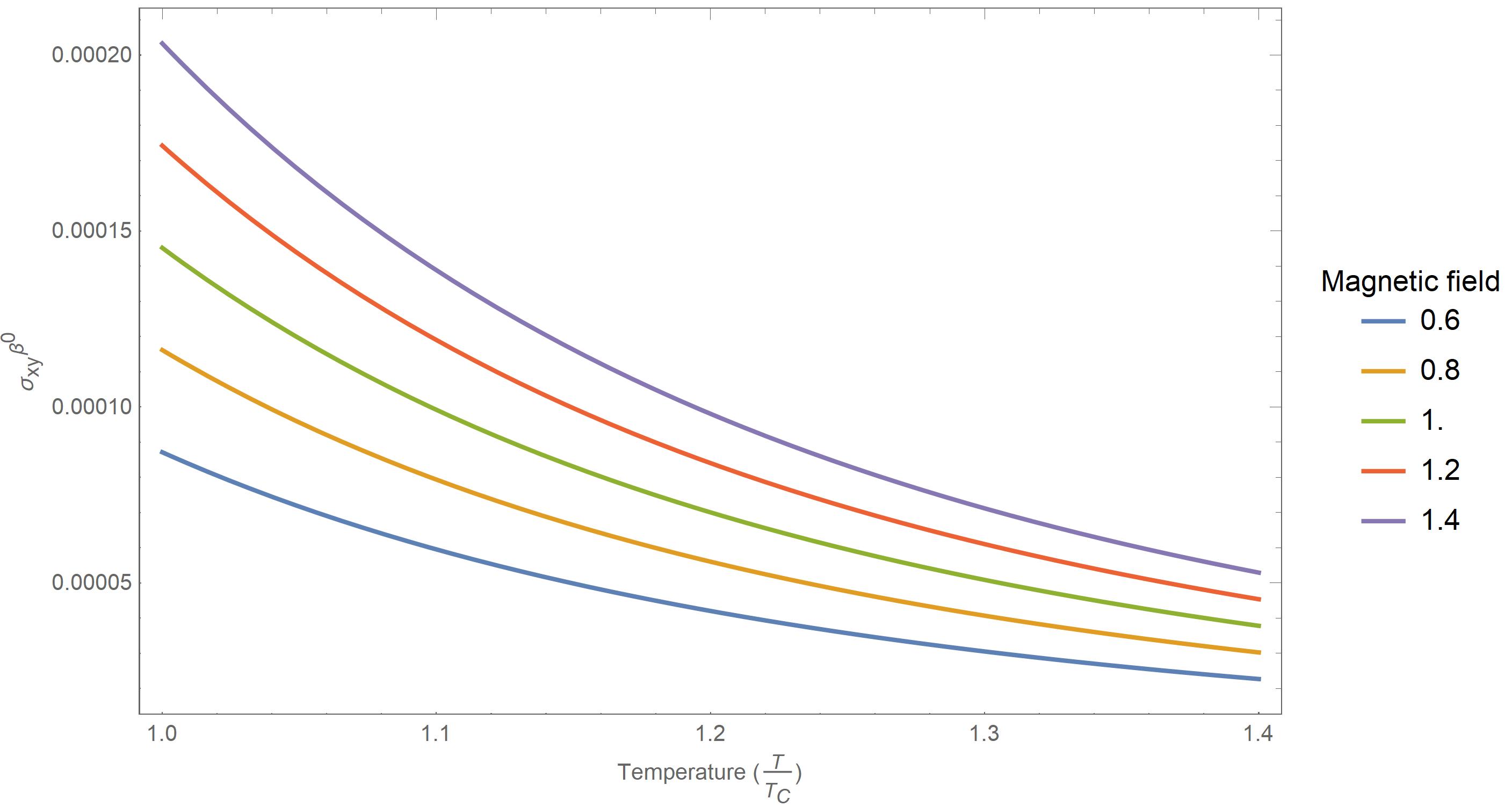}
        \label{fig:Hall_vs_T}
    }
    \caption{(a) Hall conductivity at a fixed electric field $E_x^{1}=0.5$ as a function of the magnetic field $B$. The Hall effect, in which the transverse response grows proportionately with the Lorentz force applied to the charge carriers, is characterised by a linear dependency on $B$.
(b) The Hall conductivity's temperature dependence. With increasing temperature, the results demonstrate a rapid suppression (as $\sigma_{xy}\propto \frac{1}{T^{4}}$), declining even more quickly than the density-dependent conductivity, where $\sigma_{\text{density}}\propto \frac{1}{T^{2}}$ (see leading contribution in eq(\ref{U1num}). This implies that in the weak-field regime of the dual field theory, thermal fluctuations considerably disturb the transverse momentum balance, resulting in a quicker restoration of isotropy.}
\end{figure}

For $\mathcal{O}(\beta)$, using Eqs.~(\ref{rstarbeta}) and (\ref{HallEx0arbB}), 
the Hall conductivity at $\mathcal{O}(\beta)$ takes the form
\begin{equation}
\label{Hallbeta}
\sigma_{xy}^{(\beta)}=
-\frac{32 \beta B^3 E_{x^{1}}^2 J_{t}^{(0)}}
{243 \sqrt{3}\, \pi^{27/2}\, g_{s}^{5/2} N^{5/2} T^{11}} \, .
\end{equation}
The above expression is explicitly suppressed in the large-$N$ limit, scaling as  $\sigma_{xy}^{(\beta)} \sim N^{-5/2}$. Consequently, the HD correction to the Hall conductivity is subleading in the large-$N$ expansion and does not significantly modify the transport behavior in the planar limit. We therefore conclude that, up to $\mathcal{O}(\beta)$, the Hall conductivity remains unchanged at leading order in $N$, receiving only parametrically suppressed corrections.

\item \textbf{Thermal Pair-Production Conductivity}: In this part, we investigate the transport properties arising from the pair-production mechanism induced by finite temperature in the deconfined phase ($T>T_c $). In this phase, thermal excitation of charge carrier pairs gives rise to a finite contribution to the electrical conductivity. The corresponding pair-production conductivity is given by:

\begin{equation}
\label{sigmapp}
\sigma_{tpp} = \frac{N_{f} T_{D6} \mathcal{I}}{\mathcal{Z}} \frac{e^{-\phi} f_{11}^{(0)3/2}}{\sqrt{B^2 + f_{11}^{(0)2}}} \, ,
\end{equation}

The renormalization of the boundary current introduces a factor 
 $\mathcal{Z} $, while the angular regularization over the compact $\theta_2$ and $\psi$ directions yields the multiplicative constant 
 $ \mathcal{I} $. Combining these, the product $\mathcal{I}\mathcal{Z} \sim g_{s}N=\frac{L^{4}}{\alpha^{\prime \,2}}$. Under the standard holographic dictionary, this identifies with $L^4$. Incorporating the metric components and other relevant factors, the expression for the thermal pair-production conductivity turns out to be

\begin{equation}
\label{ppbeta0}
\sigma_{tpp}^{\beta^{0}} =
\frac{9\, \sqrt{\tfrac{3}{2}}\, \pi^{11/4}\, g_s^{3/4}\, N^{3/4}\, 
      N_f\, T_{D6}\,(L^{4}/\alpha^{\prime \,2})
      \left(\mathcal{C} - 3 N_f \log T\right)
      \left(\dfrac{E_{x^{1}}^2\, T}{4B^2 + 9\pi^5 g_s N T^4} + T\right)^{\!3}}
{8\,\sqrt{\dfrac{9}{4}\,\pi^5\, g_s\, N 
      \left(\dfrac{E_{x^{1}}^2\, T}{4B^2 + 9\pi^5 g_s N T^4} + T\right)^{\!4}
      + B^2}}
\end{equation}

Considering the small $E_{x^{1}}$, and small $B$ limit the resulting expression turns out to be

\begin{equation}
\label{ppNum}
\begin{split}
\sigma_{tpp}^{\beta^{0}} 
&= \frac{3\sqrt{3}\, \pi^{1/4}\, g_s^{1/4}\, N^{1/4}\, N_f\, T\, T_{D6}(L^{4}/\alpha^{\prime \,2})
   \left(\mathcal{C} - 3 N_f \log T\right)}
   {4\sqrt{2}}- \frac{B^2\, N_f\, T_{D6}\,(L^{4}/\alpha^{\prime \,2})
   \left(\mathcal{C} - 3 N_f \log T\right)}
   {2\sqrt{6}\, \pi^{19/4}\, g_s^{3/4}\, N^{3/4}\, T^3} \\
&\quad + \frac{E_{x^{1}}^2\, N_f\, T_{D6}\,(L^{4}/\alpha^{\prime \,2})
   \left(\mathcal{C} - 3 N_f \log T\right)}
   {4\sqrt{6}\, \pi^{19/4}\, g_s^{3/4}\, N^{3/4}\, T^3} + \frac{B^2\, E_{x^{1}}^2\, N_f\, T_{D6}\,(L^{4}/\alpha^{\prime \,2})
   \left(\mathcal{C} - 3 N_f \log T\right)}
   {18\sqrt{6}\, \pi^{39/4}\, g_s^{7/4}\, N^{7/4}\, T^7}
\end{split}
\end{equation}

It is evident from Eq.~(\ref{ppNum}) that the leading-order term, representing the linear Ohmic contribution to the thermal pair-production conductivity, is independent of the external electromagnetic fields. This term scales linearly with temperature, $T$, and includes a logarithmic correction of the form $\log T$. This logarithmic dependence is a direct consequence of the non-trivial Type IIA dilaton profile, serving as a holographic signature of the running coupling constant and the breaking of conformal invariance in the thermal pair-production sector. The subsequent sub-leading terms are suppressed in both the large-$N$ and high-temperature limits. This hierarchy implies that the non-linear transport effects induced by the external electromagnetic fields decouple as the system approaches the strict planar limit or the high-temperature regime of the Quark-Gluon Plasma (QGP).

\begin{figure}[h!]
    \centering
    \subfigure[]{
        \includegraphics[width=0.45\textwidth]{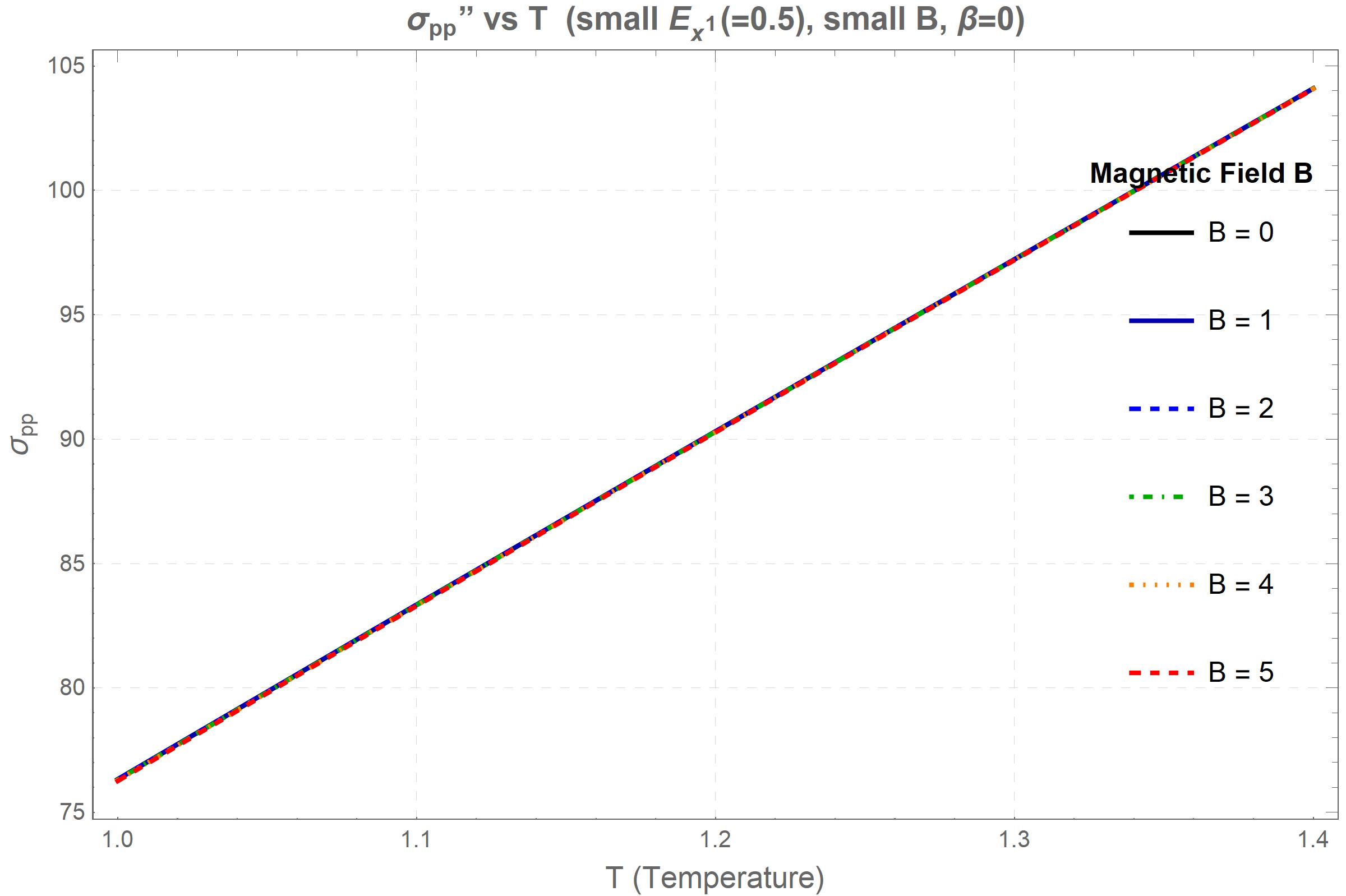}
        \label{fig:pp_vs_T}
    }
    \hfill
    \subfigure[]{
        \includegraphics[width=0.45\textwidth]{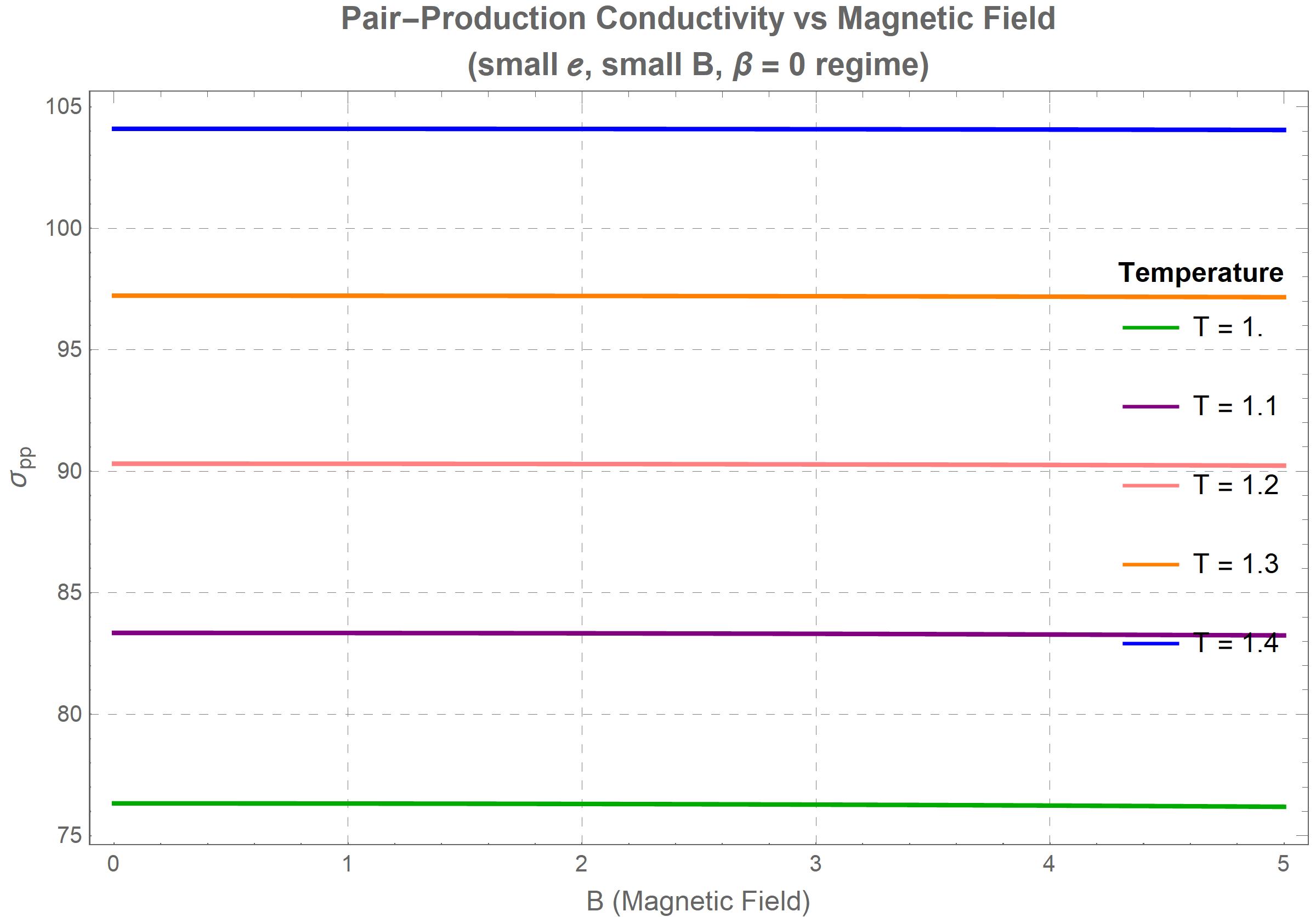}
        \label{fig:pp_vs_B}
    }
    \caption{(a) \textbf{Thermal Pair-production conductivity} $\sigma_{tpp}$ as a function of the magnetic field $B$ at a constant electric field $E_{x^{1}}=0.5$. As derived in Eq.~(\ref{ppNum}), the electromagnetic fields enter the pair-production sector only at sub-leading order. In the weak-field regime, their influence is negligible, resulting in a nearly constant profile dominated by the linear Ohmic contribution.
(b) Temperature dependence of $\sigma_{tpp}$ at $E_{x^{1}}=0.5$ for magnetic fields ranging from $B=0$ to $B=4$. The conductivity exhibits a characteristic linear growth with temperature, modulated by sub-leading logarithmic corrections ($\log T$). These corrections serve as a holographic signature of the non-conformal running of the coupling constant, sourced by the non-trivial Type IIA dilaton profile in Eq.~(\ref{leadingdilaton}). Within the relevant temperature range $T/T_c \in [1, 1.4]$ for the $\mathcal{M}$-theory-inspired QGP, the logarithmic terms provide small modifications, preserving the overall linear scaling. Furthermore, the electromagnetic field effects are large-$N$ and high-$T$ suppressed, leading to the observed overlapping of trajectories for varying $B$.}
\end{figure}

We now consider the $\mathcal{O}(\beta)$ corrections to the thermal pair-production sector of the conductivity, obtained from $\sigma_{tpp}^{\beta} = \delta r_{\ast}^{\beta}\, \partial_{r} \sigma_{tpp}^{\beta^{0}}$. Using Eqs.~\eqref{rstarbeta} and \eqref{ppbeta0}, and performing an expansion in the small $E_{x^1}$ and $B$ limit, one obtains

\begin{equation}
\sigma_{tpp}^{\beta} =
\frac{\sqrt{\frac{2}{3}} B^2 \,E_{x^{1}}^{2} \,N_{f}^{2} \,T_{D6} \,(L^{4}/\alpha^{\prime \,2})}{3 \pi ^{39/4} \,g_{s}^{7/4} N^{7/4} T^7}-\frac{\sqrt{\frac{2}{3}} B^2 E_{x^{1}}^{2} \,N_{f} \,T_{D6}\,(L^{4}/\alpha^{\prime \,2})(\mathcal{C}-3 \,{N_f} \log
   (T))}{9 \pi ^{39/4} \,{g_s}^{7/4} N^{7/4} T^7}
\end{equation}

This expression is parametrically suppressed in the large-$N$ limit, scaling as $N^{-7/4}$, and is further suppressed at high temperature as $T^{-7}$. Consequently, the higher-derivative contributions to the $\sigma_{tpp}^{\beta}$ are negligible. Thus, similar to the density-dependent conductivity $\sigma_{\text{density}}$, the thermal pair-production sector does not receive any effective higher-derivative corrections. This indicates that the result is effectively non-renormalized up to $\mathcal{O}(R^4)$.

\item \textbf{Full DC conductivity:}

We now perform a numerical analysis of the total DC conductivity, which is given by $\sigma_{\mathrm{DC}} = \sqrt{\sigma_{tpp}^2 + \sigma_{\mathrm{density}}^2}$. Using the small-$E_{x^{1}}$ and small-$B$ expansions of $\sigma_{tpp}$ and 
$\sigma_{\mathrm{density}}$ from Eqs.~\eqref{ppNum} and \eqref{U1num}, respectively, we obtain the following expressions.

The density-dependent contribution is
\begin{equation}
\sigma_{\mathrm{density}}^2 =
\frac{4 J_{t}^{(0)2}}{9 \pi^5 g_s N T^4}-\frac{32 B^2 J_{t}^{(0)2}}{81 \pi^{10} g_s^2 N^2 T^8}
-\frac{16 E_{x^{1}}^2 J_{t}^{(0)2}}{81 \pi^{10} g_s^2 N^2 T^8}+\mathcal{O}(E_{x^{1}}^{2}B^{2})
.
\end{equation}

The thermal pair-production contribution is

\begin{equation}
\begin{split}
\sigma_{tpp}^{2} 
&= \frac{27}{32} \sqrt{\pi g_s N}\, N_f^2\, T^2\, T_{D6}^2 \,(L^{4}/\alpha^{\prime\,2})^{2}
   \left(\mathcal{C} - 3 N_f \log T\right)^2  
   - \frac{3 B^2\, N_f^2\, T_{D6}^2 \,(L^{4}/\alpha^{\prime \,2})^{2}
   \left(\mathcal{C} - 3 N_f \log T\right)^2}
   {8 \pi^{9/2} \sqrt{g_s N}\, T^2} \\
&\qquad
   + \frac{3 E_{x^{1}}^2\, N_f^2\, T_{D6}^2 \,,(L^{4}/\alpha^{\prime\,2})^{2}\,
   \left(\mathcal{C} - 3 N_f \log T\right)^2}
   {16 \pi^{9/2} \sqrt{g_s N}\, T^2}+\mathcal{O}(E_{x^{1}}^{2}B^{2})
\end{split}
\end{equation}

from where the dominating contribution turns out to be,

\begin{equation}
\label{leadingDC}
\sigma_{\mathrm{DC}} =
\sqrt{
  \frac{27\, N^2\, N_f^2\,(L/\sqrt{\alpha^{\prime}})^{2}\, T^2 
  \left(\mathcal{C} - 3 N_f \log T\right)^2}
  {128\, \pi^{7/2}}
  +
  \frac{4\, J_{t}^{(0)2}}
  {9\, \pi^5\, g_s\, N\, T^4}
}
\end{equation}
where, after the Sieberg-like duality cascade, $N$ is identified with $M$, representing the number of colors in dual theory.

Considering beyond the leading order corrections at $\mathcal{O}(\beta^{(0)})$, one can incorporate more sub-leading corrections by modifying the warp factor, including the back-reaction from fractional($M$) and flavor branes($N_{f}$) as:
\begin{equation}
\label{warpfactor}
h = \frac{L^{4}}{r^{4}} \left[ 1 + \frac{3 g_{s} M^{2}}{2\pi N} \log r \left( 1 + \frac{3 g_{s} N_{f}}{2\pi} \log r \right) \right]\end{equation}

At $\mathcal{O}(\beta^{0})$, the metric component is defined as $f_{11}^{(0)} = 1/\sqrt{h}$. Incorporating the non-conformal backreaction, this component scales as $f_{11}^{(0)} \sim \frac{r^{2}}{L^{2}} \left[ 1 - \frac{3 g_{s} M^{2}}{4\pi N} \log r \left( 1 + \frac{3 g_{s} N_{f}}{2\pi} \log r \right) \right]$. When evaluating the DC conductivity in the weak-electromagnetic field limit, the pair-production term $\sigma_{tpp}^{(0)}$ follows the relation $\bar{\mathcal{N}} e^{-\phi} \sqrt{f_{11}^{(0)}}$, where $\bar{\mathcal{N}} = N_{f} T_{D6} \mathcal{I} \mathcal{Z}^{-1}$. Since the charge-density contribution is sub-leading compared to the dominant term from $\sigma_{tpp}$, we neglect the backreaction from the fractional $D5$-branes that would arise from the fully backreacted warp factor $h$ (see Eq.~\ref{warpfactor}). Consequently, we consider the $\sigma_{\text{density}}$ contribution only up to the leading order. Following this approach, the total DC conductivity is found to be:

\begin{equation}
\begin{split}
\sigma_{\rm DC}^{(0)} = \Biggl[
&\frac{27\, (L/\sqrt{\alpha^{\prime}})^{2}\, N^2 N_f^2 T^2}{128\, \pi^{7/2}} 
\Biggl(\mathcal{C}^2 - 6\, \mathcal{C}\, N_f \log(T)\,+ 9\, N_f^2 \log^2(T)
- \frac{27\, g_s\, M^2\, N_f^2 \log^4(T)}{8\, \pi^2\, N} \\
&\quad - \frac{27\, g_s\, M^2\, N_f^2 \log(N) \log^3(T)}{16\, \pi^2\, N}
\Biggr)+ \frac{4\, J_t^{(0)2}}{9\, \pi^5\, g_s\, N\, T^4}
\Biggr]^{1/2}
\end{split}
\end{equation}
where, the coefficient of $(\log T)^{3}$ term can be dropped since $\frac{\log N}{N}\approx 0$ in the large-$N$, resultingly one obtains: 
{\footnotesize
\begin{equation}
\sigma_{\rm DC}^{(0)} = \sqrt{
\frac{27\, (L/\sqrt{\alpha^{\prime}})^{2}\, N^2\, N_f^2\, T^2}{128\, \pi^{7/2}}
\left[
\mathcal{C}^2 
- 6\, \mathcal{C}\, N_f\, \log T 
+ 9\, N_f^2 (\log\,T)^{2}
- \frac{27\, g_s\, M^2\, N_f^2}{8\, \pi^2\, N}\, 
(\log T)^{4})
\right]
+ \frac{4\, J_t^{(0)2}}{9\, \pi^5\, g_s\, N\, T^4}
}
\end{equation}
}

The appearance of the logarithmic corrections reflects the breaking of conformal invariance. Notably, these $\log T$ corrections arise from two different sources:
\begin{itemize}
\item from flavor backreated type-IIA dilaton, see Eq(\ref{leadingdilaton}).
\item from subleading corrections in warp factor, see Eq(\ref{warpfactor}).
\end{itemize}
It turns out that the logarithmic corrections from the warp factor are subleading compared to the flavor-backreacted type-IIA dilaton.

Here we are working within the framework of a trivial Ouyang embedding, which corresponds to the massless quarks in the dual gauge theory; the total DC conductivity in Eq(\ref{leadingDC}) demonstrates that $\sigma_{tpp}$ consistently dominates over the density-dependent component $\sigma_{density}$ in this massless QGP regime. Consequently, the total conductivity($\sigma_{\text{DC}}$) exhibits dominantly the linear growth with temperature($T$). We further observe that non-conformal corrections, appearing as $\log T$ terms, are subleading and result in a slight departure from strict linearity. This dominant linear $T$-dependence, where NLO, NNLO, and higher-order terms are large-$N$ suppressed, turns out to be in good agreement with the results reported in \cite{Hoyos:2021njg}. Within the valid temperature range for this $\mathcal{M}$-theoretic QGP setup (where $\mathcal{C} > N_f \log T$), Eq. \eqref{leadingDC} yields the approximation $\frac{\sigma_{DC}}{T} \approx \text{constant}$. This behavior is consistent with the lattice results represented by the red \cite{Francis:2011bt} and blue \cite{Aarts:2007wj} curves in Figure(\tcb{2}) of \cite{Aarts:2020dda}.

We now comment on the higher-derivative (HD) corrections to the total DC conductivity. As established in the preceding sections, the $\mathcal{O}(\beta)$ corrections to both the charge density contribution and the pair-production sector are significantly suppressed in the large-$N$ planar limit and the high-temperature regime. Consequently, these sectors do not receive effective HD corrections at the physical scales of interest. Given that the total DC conductivity is defined as $\sigma_{DC} = \sqrt{\sigma_{tpp}^2 + \sigma_{\text{density}}^2}$, it is evident that the total transport response remains robust against higher-order curvature effects. This suggests that the leading-order DBI results are non-renormalized by $\mathcal{O}(R^4)$ corrections due to the dominant large-$N$ suppression.

\end{itemize}

\section{Conclusion and Future Directions}
\label{Results}

In this work, we have studied the charge-based transport properties of the holographic Quark-Gluon Plasma (QGP) in the framework inspired by top-down $\mathcal{M}$-theory, inclusive of $\mathcal{O}(R^{4})$ corrections. The DC, Hall, and thermal pair-production conductivities were calculated by applying the reality condition of the Dirac-Born-Infeld (DBI) action \cite{OBannon:2007cex, Karch:2007pd}. This method is then extended to incorporate $\mathcal{O}(R^4)$ higher-derivative corrections. Our key findings are summarized as follows:

\begin{itemize}

\item \textbf{Magnetically Sourced Higher-Derivative Corrections}:

Since $B=0$ correctly yields the trivial magnetic field effective horizon Eq(\ref{effhorizon}), the sole constraint achieved at $\mathcal{O}(\beta)$ is $\xi^{\beta}$, providing the shift in the horizon resulting in HD corrections, which turns out to be magnetically sourced. As a result, the HD corrections seen in the non-trivial magnetic field have a magnetic origin.

The phenomenon where the effective temperature of the dual field theory drops as the magnetic field increases is found at $\mathcal{O}(\beta^{0})$ in the weak-field limit from Eq(\ref{IMC}). This behaviour is well-known as the \textit{Inverse Magnetic Catalysis}.

\item \textbf{Comments on Conductivity Computations:}
Our findings show that the thermal pair-production sector eq(\ref{ppNum}) is characterised by a linear temperature growth controlled by logarithmic corrections ($\log T$), but the density-dependent conductivity in Eq(\ref{U1num}) displays a Drude-like suppression as $T^{-2}$. The non-trivial Type IIA dilaton profile eq(\ref{leadingdilaton}) is the source of this leading logarithmic behaviour, while the logarithmic corrections sourced by the back-reacted warp-factor in Eq~(\ref{warpfactor}) are sub-dominant compared to the one coming from the non-trivial flavor back-reacted Type IIA dilaton profile, which provides a strong holographic trace of the non-conformal functioning of the gauge coupling. Importantly, we found that in the large-$N$, high-temperature limit, the $\mathcal{O}(\beta)$ higher-derivative adjustments to the Hall conductivity, thermal pair-production conductivity, density-part conductivity, and total DC conductivity are greatly suppressed. This implies that at leading order in $N$, the transport coefficients (such as DC, Hall, thermal pair-production, and charged-density conductivities) are non-renormalized by $\mathcal{O}(R^4)$ terms. As $\mathcal{C}>\,N_{f}\,logT$ and the charged-density part contribution is suppressed at the leading order behaviour for total DC conductivity ($\sigma_{\text{DC}}$) in eq(\ref{leadingDC}), it is evident that $\frac{\sigma_{DC}}{T}\approx \text{Constant}$, which is consistent with results of the bottom-up holographic QGP from VQCD in \cite{Hoyos:2021njg}, and lattice QGP plots displayed in the red \cite{Francis:2011bt} and blue \cite{Aarts:2007wj} curves in Figure(\tcb{2}) of \cite{Aarts:2020dda}.

Furthermore, the Hall conductivity is also suppressed in the weak electromagnetic field limit at large $N$. This suppression indicates that the system tends toward isotropic behavior, as the Hall response becomes negligible compared to the longitudinal conductivity. In particular, in this regime the density-dependent contribution dominates over the Hall component, which scales as $\sigma_{Hall} \propto T^{-4}$, due to both its temperature dependence and its large-$N$ suppression.

Several promising directions could substantially extend the current framework. A primary objective is to incorporate rotational effects, which are expected to significantly influence the QCD phase structure. Various studies suggest that finite rotation can induce nontrivial shifts in the deconfinement transition temperature ($T_{C}$), e.g. for $\mathcal{M}-$QGP setup see \cite{Yadav:2022qcl}. Incorporating rotation within the top-down holographic study of QGP-like theories, for example, through rotating black hole backgrounds, could provide valuable insight into the interplay between centrifugal effects and inverse magnetic catalysis (IMC) in the strongly non-linear regime.

Another natural extension involves the study of alternating current (AC) transport. Determining the frequency-dependent conductivity, $\sigma(\omega)$, along with the associated quasinormal mode (QNM) spectrum, would offer a more comprehensive probe of the system's dissipative dynamics. This analysis is essential for extracting characteristic relaxation timescales, understanding equilibration processes, and identifying the onset of hydrodynamic behavior in a strongly coupled magnetized plasma.

\end{itemize}

\textbf{Acknowledgments}
SSK is supported by a Senior Research Fellowship (SRF) from the Ministry of Education (formerly MHRD), Government of India. I would like to express my sincere gratitude to A.~Misra for valuable discussions and for extensive technical assistance throughout the duration of this project, and especially Appendix A. I am also grateful to Mohd.~Aariyan Khan for fruitful discussions.

\textbf{Open Access}. This article is distributed under the terms of the Creative Commons Attribution License \textcolor{blue}{(CC-BY 4.0)}, which permits any use, distribution, and reproduction in any medium, provided the original author(s) and source are credited. SCOAP3 supports the goals of the International Year of Basic Sciences for Sustainable Development.

\appendix
\section{Details of the Angular-Regularization in the DBI action for the flavor D6-brane}
Here, we discuss the process of angular regularization employed to counter the divergences that arise during the angular integration of the DBI action over $\theta_2$ and $\tilde{Y}$. These specific divergences originate during the angular integration over the $\theta_2$ coordinate. To ensure a finite physical result, we introduce a counterterm scheme that isolates the singular behaviour in the $\theta_2 \to 0$ limit, thereby regularizing the volume of the internal manifold for the top-down Type IIA background. Here, we only provide the mechanism to address the leading-order divergences. Following the angular regularization procedure, the constant multiplicative factors $\mathcal{I}$ and $\tilde{\mathcal{I}}$ are rescaled to unity to simplify the subsequent transport derivations.

The angular factor appearing at $\mathcal{O}(\beta^{0})$ and $\mathcal{O}(\beta)$ which arises from the expressions \\
$\left(\sqrt{-\det\bigl(i^{\ast}(g_{S^{2}(\theta_{2},\tilde{Y})} + B_{NS\text{-}NS})\bigr)}\right)$, and $\left(\sqrt{-f_{11}^{(1)}(\theta_{2})\det\bigl(i^{\ast}(g_{S^{2}(\theta_{2},\tilde{Y})} + B_{NS\text{-}NS})\bigr)}\right)$, the off diagonal terms of which is sourced by the NS--NS $B$-field:
\begin{equation}
\left.\sqrt{i^*B + g}\right|_{S^2} =
\sqrt{
f_{\theta_2 \theta_2}(\theta_2)\, f_{\tilde{Y}\tilde{Y}}(\theta_2)
- \left(f_{\theta_2 \tilde{Y}}(\theta_2)\right)^2
}.
\end{equation}

In the small-$\theta_2$ limit, one obtains (writing the most singular term and the finite term only),
\begin{eqnarray}
& & \left(\left.\sqrt{i^*B + g}\right|_{S^2}\delta\left(\sin\theta_1 - \sqrt{2}\sin\theta_2\right)\right)(\theta_2\sim0, \theta_1\sim N^{-\frac{1}{5}}) =
\frac{\sqrt{2}\, \pi \, (g_s N)^{1/4}}{3\, N^{4/5}}
\left(
\frac{1}{\theta_2^4} + \frac{187}{90}
\right),\nonumber\\
& & \left(\sqrt{{\rm det}({\rm Ricci}_{S^2})}\delta\left(\sin\theta_1 - \sqrt{2}\sin\theta_2\right)\right)(\theta_2\sim0, \theta_1\sim N^{-\frac{1}{5}})=\frac{2}{ {\pi}^{1/4} {g_s}^{1/4} {N}^{1/4}}\left(\frac{1}{\theta_{2}^4} + \frac{1}{45}\right),\nonumber\\
& & \sqrt{-g_{S^1\times\mathbb{R}^3}}(r=r_h(1+\delta))=\frac{r_h^4\sqrt{\delta}}{4\pi g_s N}.
\end{eqnarray} 
One sees that in the IR, ${\cal L}_{\rm DBI}^{D6,\ \beta^0}\sim\left.\sqrt{i^*B + g}\right|_{S^2}F(r=r_h)$ as in the IR $r =r_h\left(1 + {\cal O}\left(\frac{1}{N}\right)\right)$.
The $\frac{1}{\theta_2^4}$ pole in $\left.\sqrt{i^*B + g}\right|_{S^2}$ is therefore cancelled by the counterterm ($L\equiv \left(4\pi g_s N\right)^{1/4}$)
\begin{equation}
{\cal L}_{\rm ct}^{\beta^0} = -\frac{F(r_h)N^{4/5}}{L\left(\frac{r_h^4\sqrt{\delta}}{L^4}\right)}\sqrt{-g_{S^1 \times \mathbb{R}^3}}\,
\sqrt{{\rm det}({\rm Ricci})_{S^2}}\delta\left(\sin\theta_1 - \sqrt{2}\sin\theta_2\right)\Big|_{r=r_h}.
\end{equation}

At $\mathcal{O}(\beta)$, ${\cal L}_{\rm DBI}^{D6,\ \beta}\sim{\cal F}(\theta_2)\tilde{F}(r=r_h)$, where,
\begin{equation}
\mathcal{F}(\theta_2) =
f_{11}^{(\beta)}(\theta_2)\,\left.\sqrt{i^*B + g}\right|_{S^2}.
\end{equation}
Noting,
\begin{equation}
\left.{\rm Pfaffian}(B)\right|_{S^2}(\theta_2\sim0) = (\pi g_s N)^{1/4}\left(\frac{1}{27\theta_2^3} - \frac{2}{3^{5/2}}\right),
\end{equation}
To regularize the angular integration over $\theta_2$, consider the following expressions (writing the most singular term and the finite term only):
\begin{equation}
\mathcal{F}(\theta_2\sim0) =
\frac{16\, \sqrt{g_s}  \pi^{5/4}}{9 \cdot 3^{1/3} }
\left(
\frac{10\sqrt{3}}{\theta_2^3} - 4428
\right),
\end{equation}

%
%
%
%

The leading-order pole in $\mathcal{F}$ is $\sim \frac{1}{\theta_2^3}$ and is therefore cancelled by the counterterm
\begin{equation}
{\cal L}_{\rm ct}^{\beta^0} = -\frac{F_1(r_h)}{L\left(\frac{r_h^4\sqrt{\delta}}{L^4}\right)}\,\sqrt{g_{S^1 \times \mathbb{R}^3}}\,\sqrt{\text{Pfaffian}(B)_{S^2}}\Big|_{r=r_h}.
\end{equation} 
  
Further, noting that,
\begin{eqnarray}
& &\left| \frac{\left.\left.\sqrt{i^*B + g}\right|_{S^2}\delta\left(\sin\theta_1 - \sqrt{2}\sin\theta_2\right)(\theta_2\sim0, \theta_1\sim N^{-\frac{1}{5}})\right|_{\rm finite\ as\ \theta_2\rightarrow0}}{\left.{\cal L}_{\rm ct}^{\beta^0}\right|_{\rm finite\ as\ \theta_2\rightarrow0}}\right|\gg1,\nonumber\\
& &\left| \frac{\left.{\cal F}(\theta_2)\delta\left(\sin\theta_1 - \sqrt{2}\sin\theta_2\right)(\theta_2\sim0, \theta_1\sim N^{-\frac{1}{5}})\right|_{\rm finite\ as\ \theta_2\rightarrow0}}{\left.{\cal L}_{\rm ct}^{\beta}\right|_{\rm finite\ as\ \theta_2\rightarrow0}}\right|\gg1,\nonumber\\
\end{eqnarray}  
We will be disregarding $\left.{\cal L}_{\rm ct}^{\beta^0}\right|_{\rm finite\ as\ \theta_2\rightarrow0}$ as compared to \\ $\left.\left.\sqrt{i^*B + g}\right|_{S^2}\delta\left(\sin\theta_1 - \sqrt{2}\sin\theta_2\right)(\theta_2\sim0, \theta_1\sim N^{-\frac{1}{5}})\right|_{\rm finite\ as\ \theta_2\rightarrow0}$, and $\left.{\cal L}_{\rm ct}^{\beta}\right|_{\rm finite\ as\ \theta_2\rightarrow0}$ as compared to  $\left.{\cal F}(\theta_2)\delta\left(\sin\theta_1 - \sqrt{2}\sin\theta_2\right)(\theta_2\sim0, \theta_1\sim N^{-\frac{1}{5}})\right|_{\rm finite\ as\ \theta_2\rightarrow0}$.

\section{Metric Components}

The temporal component of the metric is given by:
\begin{eqnarray}
f_{tt}^{(0)}(r) &=& -\frac{r^2 \left(1-\frac{r_h^4}{r^4}\right)}{2 \sqrt{\pi}\, \sqrt{g_s}\, \sqrt{N}} \nonumber\\[6pt]
f_{tt}^{(1)}(r) &=& \frac{27\, b^{10} \left(9 b^2+1\right)^4 M\, r^3 \left(6 a^2+r_h^2\right) (r-2 r_h) \left(1-\frac{r_h^4}{r^4}\right) \log^3(r_h)}{4 \pi^{3/2} \left(3 b^2-1\right)^5 \left(6 b^2+1\right)^4 \sqrt{g_s}\, N^{7/4}\, (\log N)^4\, N_f\, r_h^4 \left(9 a^2+r_h^2\right)} \nonumber\\[6pt]
f_{tt}^{(1)}(\theta_2) &=& \frac{-19683 \sqrt{6}\, \alpha_{\theta_1}^6 - 6642\, \alpha_{\theta_2}^2 \alpha_{\theta_1}^3 + 40 \sqrt{6}\, \alpha_{\theta_2}^4}{\alpha_{\theta_2}^3}
\end{eqnarray}

similarly the spatial part of the metric component is $g_{x^{1}x^{1}}=g_{x^{2}x^{2}}=g_{x^{3}x^{3}}$ is,
\begin{eqnarray}
f_{11}^{(0)}(r) &=& \frac{r^2}{2 \sqrt{\pi}\, \sqrt{g_s}\, \sqrt{N}} \nonumber\\[6pt]
f_{11}^{(1)}(r) &=& \frac{27\, b^{10} \left(9 b^2+1\right)^4 M\, r^3 \left(6 a^2+r_h^2\right) (r-2 r_h)\, \log^3(r_h)}{4 \pi^{3/2} \left(3 b^2-1\right)^5 \left(6 b^2+1\right)^4 \sqrt{g_s}\, N^{7/4}\, (\log N)^4\, N_f\, r_h^4 \left(9 a^2+r_h^2\right)} \nonumber\\[6pt]
f_{11}^{(1)}(\theta_2) &=& f_{tt}^{(1)}(\theta_2)
\end{eqnarray}

For the radial part, we have:
\begin{eqnarray}
f_{rr}^{(0)}(r) &=& 
\frac{2 \sqrt{\pi}\, \sqrt{g_s}\, \sqrt{N}\, r^2 \left(6 b^2+\frac{r^2}{r_h^2}\right)}
{r_h^4 \left(\frac{r^4}{r_h^4}-1\right)\left(9 b^2+\frac{r^2}{r_h^2}\right)}
\nonumber\\[8pt]
f_{rr}^{(1)}(r) &=& 
\frac{2 \sqrt{\pi}\, \sqrt{g_s}\, \sqrt{N}\, \left(6 b^2 r_h^2 + r^2\right)\left(C_{1010}(1) - 2 C_{610}(1) + 2 C_{68}(1)\right)}
{r^2 \left(1-\frac{r_h^4}{r^4}\right)\left(9 b^2 r_h^2 + r^2\right)}
\end{eqnarray}

The angular components sourced by the NS--NS $B$-field are:
\begin{eqnarray}
f_{\theta_2 \theta_2} &=& 
\frac{\sqrt{\pi}\, \sqrt{g_s}\, \sqrt{N}\, \left(\alpha_{\theta_2}^2 + N^{1/5}\, \alpha_{\theta_1}^2\right)}
{3^{1/3}\, \alpha_{\theta_2}^2}
\nonumber\\[8pt]
f_{\theta_2 \tilde{Y}} &=& 
\frac{f_{\theta_2}(r)}{\alpha_{\theta_2}} 
+ \frac{\pi^{1/4}\, g_s^{1/4}\, N^{3/20}\, \alpha_{\theta_2}}
{\sqrt{6}\, \alpha_{\theta_1}}
\nonumber\\[8pt]
f_{\tilde{Y}\tilde{Y}} &=& 
\frac{2 \left(2 \cdot 3^{2/3} N - 9 \sqrt{2}\, 3^{1/6} N^{4/5}\, \alpha_{\theta_1}\right)}
{27\, \alpha_{\theta_1}^2\, \alpha_{\theta_2}^2}
\end{eqnarray}

where
\begin{equation}
f_{\theta_2}(r) =
\frac{27\, g_s^{7/4} M\, N^{1/20}\, N_f\, \log(r)\, \left(36 b^2 r_h^2 \log(r) + r\right)}
{16 \pi^{5/4}\, r}.
\end{equation}

The non-trivial type-IIA dilaton profile used in the computation is

\begin{equation}
e^{-\Phi_{\mathrm{IIA}}}
=
\frac{3}{8\pi}
\left(
\frac{8\pi}{g_s}
+ 2 N_f \log N
- 6 N_f \log r
+ 4 N_f \log 4
\right)+\frac{18 a^2 g_s M^2 N_f \left(c_1 + c_2 \log r_h \right)}
{N \left(r^2 + 3 r_h^2\right)}
.
\end{equation}

Here we have considered only the leading-order behavior using Eq.~\eqref{effhorizon}, with $r_h = \sqrt{3}\, \pi^{3/2} \sqrt{g_s N}\, T$ from \cite{Yadav:2022qcl}, the dominant contribution in the small $E_{x^{1}}$ limit is then

\begin{equation}
\label{leadingdilaton}
e^{-\Phi_{\mathrm{IIA}}}
=
\frac{3 \left(\mathcal{C}_1 - 3  N_f\, \log T \right)}
{4\pi},
\end{equation}
where,
\begin{equation}
 \mathcal{C} \equiv \dfrac{4\pi}{g_s} 
+ N_f \log\!\left(\dfrac{16}{3\sqrt{3}\,\pi^{9/2}\,g_s^{3/2}\,N^{1/2}}\right)
\end{equation}

\section{Relevant Quantities subjected to $\mathcal{O}(R^{4})$-corrections:}

 There are only two types of fields which mainly participate in the $\mathcal{M}-$theory dynamics, which are the metric and the three-form potential, which receives HD corrections after solving the EOMs for the metric and three-form potential order-by-order. Hence, whether type-IIA fields such as NS-NS B-field, RR 1-form, dilaton field, and the embedding structure, whether receiving HD corrections or not, can be answered in terms of the metric components and three-form $\mathcal{M}-$theoretic $3-$form potential. The answer is summarized in multiple subparts as(for more details on mathematical computation, see\cite{Yadav:2020tyo}):

\begin{itemize}
    \item \textbf{Suppressed HD corrections in type-IIA dilaton profile.}

Metric for $\mathcal{M}$-theory uplift of the type IIB dual of thermal QCD-like theories at high temperatures, i.e., for $T>T_c$  \cite{MQGP,OR4}, is given as:
\begin{equation}
\begin{aligned}
ds_{11}^{2}
&= e^{-\frac{2\phi^{\mathrm{IIA}}}{3}}
\Bigg[
\frac{1}{\sqrt{h(r,\theta_{1,2})}}
\left(
-g(r)\,dt^{2}
+(dx^{1})^{2}
+(dx^{2})^{2}
+(dx^{3})^{2}
\right) \\
&\qquad\qquad
+\sqrt{h(r,\theta_{1,2})}
\left(
\frac{dr^{2}}{g(r)}
+ds_{\mathrm{IIA}}^{2}(r,\theta_{1,2},\phi_{1,2},\psi)
\right)
\Bigg] \\
&\quad
+e^{\frac{4\phi^{\mathrm{IIA}}}{3}}
\left(
dx^{11}
+A_{\mathrm{IIA}}^{F_{1}^{\mathrm{IIB}}
+F_{3}^{\mathrm{IIB}}
+F_{5}^{\mathrm{IIB}}}
\right)^{2}.
\end{aligned}
\label{eq:Mtheoryuplift}
\end{equation}
one obtains the dilaton as, $e^{\Phi_{IIA}}=(G^{\mathcal{M}}_{x^{10}x^{10}})^{3/4}$, where $G^{\mathcal{M}}_{x^{10}x^{10}}$ this is associated with the size of the $\mathcal{M}-$theory circle. Using this one obtains,

{\footnotesize
\begin{eqnarray}
\Phi_{\rm IIA} &=& -\log\left[\frac{9\left(\dfrac{8\pi}{g_s}+2\log N\, N_f-N_f\log\!\left(9a^{2}r^{4}+r^{6}\right)\right)^{2}}{64\pi^{2}}\right] \nonumber\\[2mm]
&&+\ \frac{96\pi\, a^{2}\, g_s^{2}\, M^{2}\, N_f\,\big(c_1+c_2\log(r_h)\big)}
{N\left(9a^{2}+r^{2}\right)\Big(8\pi+2\, g_s\log N\, N_f-g_s N_f\log\!\left(9a^{2}r^{4}+r^{6}\right)\Big)} \nonumber\\[2mm]
&& \hskip-2cm +\ \frac{243\,\beta\, b^{10}\left(9b^{2}+1\right)^{4} M\, N^{-5/4}\, r\,(r-2r_h)\left(6a^{2}+r_h^{2}\right)\left(19683\sqrt{6}\,\alpha_{\theta_1}^{6}+6642\,\alpha_{\theta_1}^{3}\alpha_{\theta_2}^{2}-40\sqrt{6}\,\alpha_{\theta_2}^{4}\right)\log^{3}(r_h)}
{4\pi\,\alpha_{\theta_2}^{3}\, r_h^{4}\, N_f\,(\log N)^{4}\left(3b^{2}-1\right)^{5}\left(6b^{2}+1\right)^{4}\left(9a^{2}+r_h^{2}\right)} \nonumber\\
\end{eqnarray}
}

Consequently, the $\mathcal{O}(\beta)$ correction in the dilaton field behaves as $\mathcal{O}(\beta)\sim \beta\,N^{-5/4}$, and is therefore parametrically suppressed in the intermediate MQGP limit.


\item \textbf{No HD corrections in the NS-NS $B$-field and RR three-form:}
This follows from the type IIA-to-$\mathcal{M}-$theory uplift dictionary, under which components of the $M$-theory three-form potential $C_{MNP}(=C^{\mathcal{M}}_{MNP})$ with a leg along the $M$-theory circle $x^{10}$ give the type IIA NS-NS two-form,
\begin{equation}
B_{MN}^{\rm NS\text{-}NS,\, IIA} = C^{\mathcal{M}}_{MN\,x^{10}},
\end{equation}
while components with no leg along $x^{10}$ give the type IIA RR three-form,
\begin{equation}
C_{(3)\,MNP}^{\rm RR,\,IIA} = C^{\mathcal{M}}_{MNP}.
\end{equation}
Thus, the question of whether the type IIA $B$-field and RR three-form receive $\mathcal{O}(R^4)$ corrections reduces entirely to the question of whether the $\mathcal{M}$-theory three-form $C_{MNP}$ itself is corrected at $\mathcal{O}(\beta)$.

The $\mathcal{O}(l_p^6)$-corrected flux equation of motion is:
\begin{equation}
d\star G = \frac{1}{2}\,G\wedge G + 3^2\cdot 2^{13}(2\pi)^4\,\beta\,X_8.
\end{equation}
Making the ansatz
\begin{equation}
g_{MN} = g^{(0)}_{MN} + \beta\, g^{(1)}_{MN}, \qquad
C_{MNP} = C^{(0)}_{MNP} + \beta\, C^{(1)}_{MNP},
\end{equation}
one finds 
\begin{equation}
\beta\,\partial\!\left(\sqrt{-g}^{\,(0)}\,\partial C^{(1)}\right)
+ \beta\,\partial\!\left[\left(\sqrt{-g}\right)^{(1)}\partial C^{(0)}\right]
+ \beta\,\epsilon_{11}\,\partial C^{(0)}\partial C^{(1)}
= \mathcal{O}(\beta^2) \sim 0 \quad [\text{up to } \mathcal{O}(\beta)].
\end{equation}

\textbf{Lemma 3} of \cite{Yadav:2020tyo} establishes that $C^{(1)}_{MNP} = 0$ up to $\mathcal{O}(\beta)$ is a consistent solution of this equation. The key input is topological: since Pontryagin classes of the internal manifold $p_1^2(M_{11}) = p_2(M_{11}) = 0$ up to $\mathcal{O}(\beta^0)$ (as shown in \cite{MQGP}), one has $X_8 = 0$ up to $\mathcal{O}(\beta^0)$, so there is no leading-order source term forcing $C^{(1)}_{MNP}$ away from zero. Working near the $\psi = 2n\pi$, $n=0,1,2$ branches with the delocalized Ouyang embedding, the linearized flux equation above reduces to a small set of algebraic conditions on the constants of integration $C^{(1)}_{MN}$ that already appear in the solutions of the $\mathcal{O}(\beta)$ metric-perturbation EOMs (Lemma 2). These conditions take the explicit form
\begin{equation}
C^{(1)}_{\theta_1 x} = 0, \qquad C^{(1)}_{zz} = 2\,C^{(1)}_{\theta_1 z}, \qquad C^{(1)}_{\theta_2 z} = 0,
\end{equation}
 where these $C_{MN}^{(1)}$ are the integration constants appearing in the solutions to the equations of motion for the $\mathcal{O}(R^4)$-corrected $\mathcal{M}$-QGP metric $g_{MN}^{\mathrm{M}}$. Consistently, one obtains
\begin{equation}
C^{(1)}_{MNP} = 0 \quad \text{up to } \mathcal{O}(\beta).
\end{equation}
Hence the $M$-theoretic three-form potential receives no $\mathcal{O}(R^4)$ correction, and consequently -- via the uplift relations above -- neither does the type IIA NS-NS $B$-field nor the RR three-form $C_3$. All the $\mathcal{O}(R^4)/\mathcal{O}(l_p^6)$ information is instead carried entirely by the metric perturbations $f_{MN}(r)$.

\item \textbf{No independent HD corrections to the RR one-form:}
The type-IIA RR one-form arises from the Kaluza-Klein reduction of the eleven-dimensional metric according to
\[
A^{(1)}_{\mu}=\frac{G^{\mathcal{M}}_{x^{10}\mu}}{G^{\mathcal{M}}_{x^{10}x^{10}}}.
\]
Working in the coordinate patches $\psi=2n\pi$ ($n=0,1,2$) employed throughout this work, the off-diagonal metric components satisfy $G^{\mathcal{M}}_{x^{10}N}=0$ for $N\neq x^{10}$ and $G^{\mathcal{M}}_{rM}=0$ for $M\neq r$, implying that the background RR one-form vanishes identically. Furthermore, the $\mathcal{O}(\beta)$ metric corrections are introduced through the perturbative ansatz
\[
\delta G_{MN}=G_{MN}^{\rm MQGP}\,f_{MN},
\]
where $G_{MN}^{\rm MQGP}$ denotes the leading-order ($\mathcal{O}(\beta^0)$) $\mathcal{M}$QGP metric. Since the off-diagonal components $G^{\rm MQGP}_{x^{10}N}$ vanish identically in these patches, their $\mathcal{O}(\beta)$ perturbations remain unsourced, and consequently no independent higher-derivative correction to the RR one-form is generated. As shown in \cite{Yadav:2020tyo}, the same conclusion continues to hold even away from the $\psi=2n\pi$ patches, where a complete analysis of the higher-derivative-corrected metric likewise yields no $\mathcal{O}(\beta)$ correction to the RR one-form.

\item \textbf{No HD corrections in the brane embedding:}
Note that, following \cite{Yadav:2017bbe}, the flavor $D6$-branes in our setup arise as the SYZ mirror (constructed via the delocalized SYZ prescription) of the Ouyang-embedded flavor $D7$-branes in type IIB. Under the SYZ mirror transformation, $(x,y,z)$ (of type IIB) $\longrightarrow$ ($\tilde{x},\tilde{y},\tilde{z}$) (of type-IIA),
the type IIB $D7$-brane embedding is mapped to the corresponding type IIA $D6$-brane embedding. In particular, the embedding is characterized by the profile $\tilde{z}(r)$. As shown in \cite{Yadav:2017bbe}, the pullback determinant
governing the $D6$-brane DBI action takes the form
\begin{equation}
\det\big(i^*(g+B)\big)=\Sigma_0(r;g_s,N_f,N,M)+\Sigma_1(r;g_s,N_f,N,M)\,\big(\tilde{z}^{\prime}(r)\big)^2,
\end{equation}
where $\Sigma_0$ is the embedding function obtained from the pulled-back metric and $B$-field components, and $\Sigma_1$ is the coefficient of $(\tilde{z}^{\prime})^2$ generated by the $\tilde{z}^{\prime}$-dependent piece of the pullback; both are pure functions of $r$ (and the background parameters $g_s,N_f,N,M $), with no explicit $\tilde z$-dependence. Since the DBI Lagrangian depends on $\tilde{z}$ only through $\tilde{z}^{\prime}$, the coordinate $\tilde z$ is cyclic, and the resulting equation of motion is homogeneous of degree one in $\tilde{z}^{\prime}$: it is built entirely out of $\Sigma_0$, $\Sigma_1$, and $\tilde{z}^{\prime}$ itself, with no term independent of $\tilde{z}^{\prime}$.

It implies that $\tilde{z}^{\prime}=0$, i.e. $\tilde{z}=\text{constant}$, solves the equation of motion \emph{exactly}, regardless of the detailed functional form of $\Sigma_0$ and $\Sigma_1$: setting $\tilde{z}^{\prime}=0$ trivially satisfies any equation built purely from terms proportional to $\tilde{z}^{\prime}$ or its derivatives. This is not a fine-tuned cancellation special to a particular order in the large-$N$ expansion, but a direct consequence of the algebraic structure of the DBI Lagrangian.

This is precisely why the constant embedding survives the further extension of $\Sigma_0,\Sigma_1$ to their NLO-in-$N$ forms, as carried out in \cite{Yadav:2017bbe}: substituting the NLO expressions only reshapes the coefficients appearing in the equation of motion, while $\tilde z'=0$ continues to solve the resulting equation trivially. The same argument extends to any further correction to the background, including higher-derivative corrections to the M-theory/type IIA background: such corrections enter the $D6$-brane action only through the pulled-back metric and $B$-field, i.e.through $\Sigma_0,\Sigma_1\to\Sigma_0^{\rm HD},\Sigma_1^{\rm HD}$, without introducing any term in the equation of motion that is independent of $\tilde z'$. Consequently, the constant embedding $\tilde{z}=\pm C\frac{\pi}{2}$ (the antipodal $D6/\overline{D6}$ locus) remains an exact solution to all orders, rather than being corrected order by order.

\end{itemize}
Considering the discussion above regarding the higher-derivative corrections to the various fields of the theory, namely the eleven-dimensional metric, the three-form potential, the type-IIA dilaton, the NS--NS two-form, and the RR one-form, we conclude that, within the consistent truncation and to $\mathcal{O}(\beta)$ adopted in the present work, the only non-trivial higher-derivative contribution entering the probe D6-brane dynamics arises through the corrected background metric, while the $\mathcal{O}(\beta)$ correction to the worldvolume $U(1)$ gauge field is induced solely by this corrected geometry. Consequently, the determinant appearing in the DBI action,
\[
\mathcal{D}=\det\!\left(\iota^{*}(G_{ab}+B_{ab})+F_{ab}\right),
\]
is systematically expanded up to $\mathcal{O}(\beta)$ by decomposing the induced metric and the worldvolume gauge field as
\[
G_{ab}=G_{ab}^{(0)}+\beta G_{ab}^{(1)},\qquad
A_a=A_a^{(0)}+\beta A_a^{(1)},
\]
with the $U(1)$ field strength given by $F=dA$. Thus, to the order considered in this work, no independent higher-derivative corrections from the remaining bulk fields contribute explicitly to the probe D6-brane DBI action.


\begin{thebibliography}{99}
\bibitem{OBannon:2007cex}
A.~O'Bannon,
Phys. Rev. D \textbf{76}, 086007 (2007)
doi:10.1103/PhysRevD.76.086007
[arXiv:0708.1994 [hep-th]].
\bibitem{Karch:2007pd}
A.~Karch and A.~O'Bannon,
JHEP \textbf{09}, 024 (2007)
doi:10.1088/1126-6708/2007/09/024
[arXiv:0705.3870 [hep-th]].
\bibitem{Maldacena:1997re}
J.~M.~Maldacena,
Adv. Theor. Math. Phys. \textbf{2}, 231-252 (1998)
doi:10.4310/ATMP.1998.v2.n2.a1
[arXiv:hep-th/9711200 [hep-th]].

\bibitem{Witten:1998zw}
E.~Witten,
Adv. Theor. Math. Phys. \textbf{2}, 505-532 (1998)
doi:10.4310/ATMP.1998.v2.n3.a3
[arXiv:hep-th/9803131 [hep-th]].


\bibitem{Kovtun:2004de}
P.~Kovtun, D.~T.~Son and A.~O.~Starinets,
Phys. Rev. Lett. \textbf{94}, 111601 (2005)
doi:10.1103/PhysRevLett.94.111601
[arXiv:hep-th/0405231 [hep-th]].


\bibitem{Policastro:2001yc}
G.~Policastro, D.~T.~Son and A.~O.~Starinets,
Phys. Rev. Lett. \textbf{87}, 081601 (2001)
doi:10.1103/PhysRevLett.87.081601
[arXiv:hep-th/0104066 [hep-th]].


\bibitem{Meyer:2007ic}
H.~B.~Meyer,
Phys. Rev. D \textbf{76}, 101701 (2007)
doi:10.1103/PhysRevD.76.101701
[arXiv:0704.1801 [hep-lat]].


\bibitem{Astrakhantsev:2017nrs}
N.~Astrakhantsev, V.~Braguta and A.~Kotov,
JHEP \textbf{04}, 101 (2017)
doi:10.1007/JHEP04(2017)101
[arXiv:1701.02266 [hep-lat]].


\bibitem{Arnold:2003zc}
P.~B.~Arnold, G.~D.~Moore and L.~G.~Yaffe,
JHEP \textbf{05}, 051 (2003)
doi:10.1088/1126-6708/2003/05/051
[arXiv:hep-ph/0302165 [hep-ph]].


\bibitem{Bulk-Viscosity-McGill-IIT-Roorkee} A.~Czajka, K.~Dasgupta, C.~Gale, S.~Jeon, A.~Misra, M.~Richard and K.~Sil, {\it Bulk Viscosity at Extreme Limits: From Kinetic Theory to Strings}, JHEP {\bf 07}, 145 (2019)[arXiv:1807.04713 [hep-th]].

\bibitem{Kushwah:2024ngr}
S.~S.~Kushwah and A.~Misra,
Phys. Rev. D \textbf{110}, no.12, 126010 (2024)
doi:10.1103/PhysRevD.110.126010
[arXiv:2403.10541 [hep-th]].

\bibitem{Meyer:2008sn}
H.~B.~Meyer,
PoS \textbf{LATTICE2008}, 017 (2008)
doi:10.22323/1.066.0017
[arXiv:0809.5202 [hep-lat]].

\bibitem{Bluhm:2011xu}
M.~Bluhm, B.~Kampfer and K.~Redlich,
Phys. Lett. B \textbf{709}, 77-81 (2012)
doi:10.1016/j.physletb.2012.01.069
[arXiv:1101.3072 [hep-ph]].

\bibitem{Kadam:2014cua}
G.~P.~Kadam and H.~Mishra,
Nucl. Phys. A \textbf{934}, 133-147 (2014)
doi:10.1016/j.nuclphysa.2014.12.004
[arXiv:1408.6329 [hep-ph]].

\bibitem{Yaresko:2014fia}
R.~Yaresko and B.~Kampfer,
Acta Phys. Polon. Supp. \textbf{7}, no.1, 137-144 (2014)
doi:10.5506/APhysPolBSupp.7.137
[arXiv:1403.3581 [hep-ph]].

\bibitem{Gupta:2003zh}
S.~Gupta,
Phys. Lett. B \textbf{597}, 57-62 (2004)
doi:10.1016/j.physletb.2004.05.079
[arXiv:hep-lat/0301006 [hep-lat]].

\bibitem{Amato:2013naa}
A.~Amato, G.~Aarts, C.~Allton, P.~Giudice, S.~Hands and J.~I.~Skullerud,
Phys. Rev. Lett. \textbf{111}, no.17, 172001 (2013)
doi:10.1103/PhysRevLett.111.172001
[arXiv:1307.6763 [hep-lat]].

\bibitem{Skokov:2009qp}
V.~Skokov, A.~Y.~Illarionov and V.~Toneev,
Int. J. Mod. Phys. A \textbf{24}, 5925-5932 (2009)
doi:10.1142/S0217751X09047570
[arXiv:0907.1396 [nucl-th]].

\bibitem{Caron-Huot:2006pee}
S.~Caron-Huot, P.~Kovtun, G.~D.~Moore, A.~Starinets and L.~G.~Yaffe,
JHEP \textbf{12}, 015 (2006)
doi:10.1088/1126-6708/2006/12/015
[arXiv:hep-th/0607237 [hep-th]].

\bibitem{Kushwah:2025ymb}
S.~S.~Kushwah and A.~Misra,
Fortsch. Phys. \textbf{73}, no.12, e70044 (2025)
doi:10.1002/prop.70044
[arXiv:2503.07732 [hep-th]].

\bibitem{Iqbal:2008by}
N.~Iqbal and H.~Liu,
Phys. Rev. D \textbf{79}, 025023 (2009)
doi:10.1103/PhysRevD.79.025023
[arXiv:0809.3808 [hep-th]].



\bibitem{metrics} M.~Mia, K.~Dasgupta, C.~Gale and S.~Jeon, {\it Five Easy Pieces: The Dynamics of Quarks in Strongly Coupled Plasmas}, Nucl.\ Phys.\ B {\bf 839}, 187 (2010) [arXiv:hep-th/0902.1540].


\bibitem{OR4} A.~Misra and V.~Yadav, {\it On ${\cal M}$-Theory Dual of Large-$N$ Thermal QCD-Like Theories up to ${\cal O}(R^4)$ and $G$-Structure Classification of Underlying Non-Supersymmetric Geometries}, Adv. Theor. Math. Phys. 26, 10 (2022) [arXiv:2004.07259 [hep-th]].


\bibitem{MQGP} M.~Dhuria and A.~Misra, {\it Towards MQGP}, JHEP 1311 (2013) 001 [arXiv:hep-th/1306.4339].

\bibitem{Ali-Akbari:2010xwz}
M.~Ali-Akbari and K.~B.~Fadafan,
Nucl. Phys. B \textbf{844} (2011), 397-408
doi:10.1016/j.nuclphysb.2010.10.028
[arXiv:1008.2430 [hep-th]].





\bibitem{Kim:2011zd}
K.~Y.~Kim and D.~W.~Pang,
JHEP \textbf{09} (2011), 051
doi:10.1007/JHEP09(2011)051
[arXiv:1108.3791 [hep-th]].

\bibitem{Lee:2010uy}
B.~H.~Lee and D.~W.~Pang,
Phys. Rev. D \textbf{82} (2010), 104011
doi:10.1103/PhysRevD.82.104011
[arXiv:1006.4915 [hep-th]].



\bibitem{ACMS} A.~Misra and G.~Yadav, {\it (Almost) Contact (3) (Metric) Structure(s) and Transverse $SU(3)$ Structures Associated with ${\cal M}$-Theory Dual of Thermal QCD at Intermediate Coupling}, [arXiv:2211.13186[hep-th]].


\bibitem{Vikas+Gopal+Aalok} V.~Yadav, G.~Yadav and A.~Misra,{\it (Phenomenology/Lattice-Compatible) $SU(3)$ M$\chi$PT HD up to ${\cal O}(p^4)$ and the ${\cal O}\left(R^4\right)$-Large-$N$ Connection}, JHEP 2108 (2021) 151 [arXiv:2011.04660 [hep-th]].
\bibitem{NPB} K.~Sil and A.~Misra, {\it On Aspects of Holographic Thermal QCD at Finite Coupling},
  Nucl.\ Phys.\ B {\bf 910}, 754 (2016) [arXiv:1507.02692 [hep-th]].

\bibitem{SYZ-free-delocalization}M.~Becker, K.~Dasgupta, A.~Knauf and R.~Tatar, {\it Geometric transitions, flops and nonKahler manifolds. I.},  Nucl.\ Phys.\ B {\bf 702}, 207 (2004) [hep-th/0403288].

\bibitem{VA-Glueball-decay} V.~Yadav and A.~Misra, {\it M-Theory Exotic Scalar Glueball Decays to Mesons at Finite Coupling}, JHEP {\bf 09}, 133 (2018) [arXiv:1808.01182 [hep-th]].
\bibitem{Kruczenski et al-2003} M.~Kruczenski, D.~Mateos, R.~C.~Myers and D.~J.~Winters, {\it Towards a holographic dual of large-$N_c$ QCD}, JHEP 0405 (2004) 041 [arXiv:hep-th/0311270].
\bibitem{DM-transport-2014} M.~Dhuria and A.~Misra, {\it Transport Coefficients of Black MQGP M3-Branes}, Eur. Phys. J. C {\bf 75}, 16 (2015) [arXiv:1406.6076[hep-th]].


\bibitem{Armoni et al-2020} R.~Argurio, A.~Armoni, M.~Bertolini, F.~Mignosa and P.~Niro, {\it Vacuum structure of large N QCD$_3$ from holography}, JHEP {\bf 07}, 134 (2020) [arXiv:2006.01755[hep-th]].
\bibitem{Becker-sisters-O(R^4)}K.~Becker and M.~Becker, {\it Supersymmetry Breaking, M-Theory and Fluxes}, JHEP {\bf 07}, 038 (2001) [arXiv:hep-th/0107044].


\bibitem{zeta_IITR+McGill} K.~Dasgupta, A.~Misra, S.S.~Kushwah, C.~Gale, S.~Jeon, to appear.  



\bibitem{Yadav:2022qcl}
G.~Yadav,
Phys. Lett. B \textbf{841}, 137925 (2023)
doi:10.1016/j.physletb.2023.137925
[arXiv:2203.11959 [hep-th]].

\bibitem{Bergman:2010gm}
O.~Bergman, N.~Jokela, G.~Lifschytz and M.~Lippert,
JHEP \textbf{10}, 063 (2010)
doi:10.1007/JHEP10(2010)063
[arXiv:1003.4965 [hep-th]].

\bibitem{Hoyos:2021njg}
C.~Hoyos, N.~Jokela, M.~J{\"a}rvinen, J.~G.~Subils, J.~Tarrio and A.~Vuorinen,
Phys. Rev. D \textbf{105}, no.6, 066014 (2022)
doi:10.1103/PhysRevD.105.066014
[arXiv:2109.12122 [hep-th]].

\bibitem{Aarts:2007wj}
G.~Aarts, C.~Allton, J.~Foley, S.~Hands and S.~Kim,
Phys. Rev. Lett. \textbf{99}, 022002 (2007)
doi:10.1103/PhysRevLett.99.022002
[arXiv:hep-lat/0703008 [hep-lat]].

\bibitem{Francis:2011bt}
A.~Francis and O.~Kaczmarek,
Prog. Part. Nucl. Phys. \textbf{67}, 212-217 (2012)
doi:10.1016/j.ppnp.2011.12.020
[arXiv:1112.4802 [hep-lat]].

\bibitem{Aarts:2020dda}
G.~Aarts and A.~Nikolaev,
Eur. Phys. J. A \textbf{57}, no.4, 118 (2021)
doi:10.1140/epja/s10050-021-00436-5
[arXiv:2008.12326 [hep-lat]].



\bibitem{Yadav:2020tyo}
V.~Yadav and A.~Misra,
Adv. Theor. Math. Phys. \textbf{26} (2022) no.10, 3801-3894
doi:10.4310/ATMP.2022.v26.n10.a11
[arXiv:2004.07259 [hep-th]].

\bibitem{Yadav:2017bbe}
V.~Yadav, A.~Misra and K.~Sil,
Eur. Phys. J. C \textbf{77} (2017) no.10, 656
doi:10.1140/epjc/s10052-017-5219-5
[arXiv:1707.02818 [hep-th]].
  
\end{thebibliography}
\end{document}